\documentclass[reprint,aps,pra,10pt,nofootinbib,english,longbibliography,nosuperscriptaddress]{revtex4-2}

\usepackage{graphicx}
\usepackage[shortlabels]{enumitem}
\usepackage[normalem]{ulem}
\usepackage{bm,amssymb,amsmath,mathrsfs,amstext,latexsym,physics}

\usepackage[usenames,dvipsnames]{xcolor}
\usepackage[colorlinks=true,citecolor=blue,urlcolor=blue,linkcolor=blue]{hyperref}

\usepackage{tikz,ytableau}
\usetikzlibrary{decorations.pathreplacing,
                calligraphy,% must be loaded after decorations.pathreplacing
                matrix}

\usepackage{comment}

\newcommand{\der}[2]{\frac{d #1}{d #2}}

\newcommand{\phdagger}[0]{{\phantom{\dagger}}}

\let\tilde\relax
\newcommand{\tilde}[1]{\widetilde{#1}}

\definecolor{cornflower}{RGB}{120, 159, 255}

\begin{document}

\title{The deep Hilbert space of all-to-all interacting SU(3) atoms:\\ from quantum to classical}

\author{Federico Balducci}
\email{fbalducci@pks.mpg.de}
\affiliation{Max Planck Institute for the Physics of Complex Systems, N\"othnitzer Str.\ 38, 01187 Dresden, Germany}

\author{Aleksandra A.\ Ziolkowska}
\email{aaz1@st-andrews.ac.uk}
\affiliation{Institut f\"ur Physik, Johannes Gutenberg-Universit\"at Mainz, D-55099 Mainz, Germany}
\affiliation{School of Physics and Astronomy, University of St Andrews, St Andrews KY16 9SS, United Kingdom}

\date{\today}

%###########################################--ABSTRACT--###########################################

\begin{abstract}
    We study the emergence of chaos in multilevel atoms with all-to-all interactions, inspired by cavity QED. Focusing on a 3-level Tavis-Cummings model in a far detuned limit, we detail its deep Hilbert space structure---i.e.\ we enumerate all distinct dynamical sectors, beyond the totally symmetric subspace---by using the Schur-Weyl duality, which is applicable thanks to the permutation symmetry in the all-to-all Hamiltonian. Strong Hilbert space fragmentation ensues from the non-abelian nature of the symmetry, with some sectors displaying regular dynamics and others being chaotic. We uncover that many permutation symmetry sectors contribute to the dynamics in the classical limit, in addition to the commonly studied totally symmetric subspace. To elucidate the dynamical responses in each of the symmetry sectors, we propose a semiclassical description in terms of spin coherent states, which is also able to explain the origin of chaotic or regular dynamics with a simple geometrical argument. Our work contributes to the study of the quantum-classical correspondence in chaotic systems, and uncovers a rich structure in multilevel all-to-all interacting models.
\end{abstract}

\maketitle

%--------------------------------------------------------------------------------------------------
%--------------------------------------------------------------------------------------------------
\section{Introduction}

One of the most intriguing aspects of all-to-all interacting systems is their ability to explore the boundary between quantum and classical behavior, for instance, in the context of quantum signatures of chaos~\cite{Haake2010Quantum}. They provide an ideal arena for exploring such correspondences, since their semiclassical limits are well defined, yet their quantum Hilbert spaces are finite and accessible to direct computation~\cite{Sciolla2011Dynamical,LewisSwan2019Unifying}. Experimentally, such models are  most commonly realised in cavity quantum electrodynamics (QED), which is a versatile, high control platform for exploring the interaction between quantized light and matter~\cite{Walther2006Cavity,Mivehvar2021Cavity,Nussmann2005Submicron,Baumann2010Dicke,Zhang2012Collective,Thompson2013Coupling,Begley2016Optimized,Neuzner2016Interference,Davis2018Painting,Norcia2018Cavity,Davis2019Photon,Davis2020Protecting,Yan2023Superradiant,Sauerwein2023Engineering,Orsi2024Cavity}. By confining electromagnetic fields within high-finesse cavities, it becomes possible to reach regimes where the coherent coupling between cavity modes and atoms induces collective interactions, leading to interesting dynamical regimes like superradiance and squeezing~\cite{Leroux2010Implementation}. Paradigmatic examples of systems modeling the physics within a single-mode cavity are the Dicke and Tavis–Cummings Hamiltonians, where classical trajectories typically exhibit chaotic (in the former)~\cite{Emary2003Quantum,*Emary2003Chaos} or regular (in the latter)~\cite{Bogoliubov1996Exact} motion. In the far detuned limit, the cavity field can be eliminated from the dynamics of those models, rendering effective all-to-all interactions studied in this work.

In all-to-all interacting systems, the collective nature of interactions makes the thermodynamic limit coincide both with a mean-field approximation and a classical limit. Conveniently, quantum effects can be incorporated order by order in perturbation theory on top of the mean-field solution~\cite{my2PI}. However, the details of how the approximations are taken reveal diametrically different dynamics. Typically, only mean-field states that are fully symmetric under permutation of the atoms are considered. Correspondingly, usually only the so-called totally symmetric subspace is studied, which represents but a vanishing fraction of the total Hilbert space. This realization has led some authors to study the \emph{deep Hilbert space} of fully-connected spin models~\cite{Qi2023Surprises}---by definition, all other sectors beyond the totally symmetric one---showing that the range of dynamical responses can be very different, depending on the symmetry sector the mean-field approximation is considered in.

In this work, we study the emergence of chaos in a multilevel Tavis-Cummings model with adiabatically eliminated photon field~\cite{Pineiro2022Emergent,Chu2023Photon} from the perspective of the deep Hilbert space. The choice of the model hinges upon its experimental feasibility~\cite{ValenciaTortora2023Crafting}, and the richness of its dynamical responses already uncovered. The presence of chaos in the model was indeed already noticed within the traditional mean-field framework, showing that a $3$-level [also referred to as SU(3)] Tavis-Cummings Hamiltonian can show chaotic behavior in the far-detuned limit~\cite{ValenciaTortora2023Crafting}, while its $2$-level counterpart is always regular. Inspired by these preliminary findings, we perform a thorough analysis of the model, in particular from the perspective of the deep Hilbert space, and show how the deep Hilbert space becomes relevant in the thermodynamic limit. While the question of the origin of chaos in one particular model is the \emph{casus belli} for the whole analysis, our results uncover a very general structure and are applicable to a broader class of permutationally symmetric models.

The first significant finding is the use of the Schur-Weyl duality~\cite{Fulton1991Representation} to describe the deep Hilbert space of the model. This well-established algebraic relation uncovers two special unitary structures at play in the $3$-level model: SU(3), which is the symmetry group of the $3$-level atoms---that we therefore refer to as SU$(3)_\mathrm{atoms}$---, and SU($d$) with $d=(n+1)(n+2)/2$, arising when $n$ $3$-level atoms are grouped together~\cite{ValenciaTortora2023Crafting}. We refer to this second structure as SU$(d)_\mathrm{local}$, as it describes the local Hilbert space structure. 

The interplay of these two structures, together with the permutation symmetry entailed by the all-to-all nature of interactions, gives rise to a \emph{strongly fragmented} Hilbert space~\cite{Sala2020Ergodicity,Moudgalya2022Quantum}. Namely, the Hamiltonian acquires a direct sum structure over many disconnected dynamical sectors, which can be either chaotic or regular on their own, and which are all asymptotically smaller than the total Hilbert space. Only considering them separately, the question of whether the model is overall chaotic can be answered. By means of numerical exact diagonalization, we show how the totally symmetric subspace---the one most easily accessible with mean-field methods---displays in general regular dynamics, while the deep Hilbert space can host chaos. This first set of results points at the origin of the chaotic behaviour observed in Ref.~\cite{ValenciaTortora2023Crafting}, and furthermore calls for a more careful use of mean-field methods to address the question in general.

The second notable result we obtain is the explanation of chaos \emph{via a semiclassical approach as well}. We argue that most mean-field/semiclassical methods employ U($1$) coherent states, which do not respect the symmetries of the model under study---or of its many cousins used in cavity QED settings. In practice, this means that the manifold of classical states accessed via mean-field is composed by a mixture of the deep-Hilbert-space dynamically-disconnected sectors, \emph{which can be either chaotic or regular independently of each other}. This leads to a mixed dynamical response, which is difficult to frame as either regular or chaotic. We propose the use of SU(3) coherent states as a viable alternative, which is able to distinguish much better the nature of the dynamics. To be specific, we argue that while only SU$(d)_\mathrm{local}$ coherent states fully respect the structure of the deep Hilbert space, in practice the use of SU$(3)_\mathrm{atoms}$ is enough. Hence, we avoid the challenges of constructing SU($d$) spin-coherent states for large $d$ by developing a framework for projecting the representations of $\mathrm{S}_L$ onto the effective SU$(3)_\mathrm{atoms}$ representations and deriving the explicit formulas for that mapping.

The use of SU(3) coherent states further provides a natural geometrical interpretation of the emergence of chaos in the model, based on known results in other SU(3) models of chaos~\cite{Gnutzmann2000Quantum}. In practice, the totally symmetric subspace tends to show regular behavior since it corresponds to a one-dimensional classical system (in the thermodynamic limit, having fixed all the symmetry numbers), while other sectors in the deep Hilbert space have a classical limit that is higher-dimensional. Thus, by leveraging the quantum-classical correspondence of chaos, we provide a justification for some of the observed chaoticity.

Our results represent a stepping stone for the analytical description of fully-connected models of atoms in a cavity. They call for the use of more sophisticated analytical methods in the study of quantum chaos, and, by extension, in the study of dynamical responses in general. They furthermore highlight that the mathematical structure of commutant algebras~\cite{Moudgalya2022Hilbert}, at the basis of the Schur-Weyl duality and used recently in the context of quantum state preparation~\cite{Li2025Highly}, naturally arises in the context of cavity QED models. This suggests yet another way to use multilevel cavity QED as a quantum simulation platform.

The manuscript is organized as follows. In Sec.~\ref{sec:SU2} we start by recalling the most commonly studied models of $2$-level atoms in a single-mode cavity, and why one expects regular or chaotic motion in that setting.
In Sec.~\ref{sec:model}, we introduce the $3$-level Tavis-Cummings model, and the all-to-all interacting model of $3$-level atoms resulting from adiabatically eliminating the photon field. In Sec.~\ref{sec:symmetries}, we present the symmetries of the all-to-all interacting model and discuss their impact on the Hilbert space structure, leading e.g.\ to fragmentation. In order to help with the algebraic tools introduced, we provide some explicit examples in the subsequent Sec.~\ref{sec:examples}. In Sec.~\ref{sec:numerics_quantum} we provide the numerical analysis of chaotic vs regular behavior in the deep Hilbert space by means of level spacing indicators. 
Section \ref{sec:classical} describes how to take a classical limit in the model under consideration, by using SU(3) coherent states. Later, Sec.~\ref{sec:classical_chaos} makes use of the introduced coherent states to study the emergence of chaos in the model. We take our conclusions and discuss further directions in Sec.~\ref{sec:conclusions}.

%--------------------------------------------------------------------------------------------------
%--------------------------------------------------------------------------------------------------
\section{Warm-up: cavity QED with SU(2) atoms}
\label{sec:SU2}

In this Section, we review some basic facts about $2$-level cavity QED models. In Sec.~\ref{sec:SU2_models} we introduce the standard Dicke and Tavis-Cummings Hamiltonians and the process of adiabatic elimination of the photon. In Sec.~\ref{sec:SU2_deep} we comment on the deep Hilbert space structure of such models, and in Sec.~\ref{sec:SU2_chaos} we briefly summarize the known results about quantum chaos in them.

%--------------------------------------------------------------------------------------------------
\subsection{Dicke and Tavis-Cummings models}
\label{sec:SU2_models}

Some of the most widely studied models within the cavity QED framework are the \emph{$2$-level Tavis-Cummings model}~\cite{Tavis1968Exact} and the \emph{$2$-level Dicke model}~\cite{Dicke1054Coherence}; for a review see e.g.\ Ref.~\cite{Kirton2019Introduction}. They describe an ensemble of SU($2$) spin-$1/2$'s $\hat{S}_j^\mu$, $\mu = x,y,z$, interacting with a single cavity mode $\hat{a}$, under the conditions of the rotating-wave approximation (the former) or not (the latter). The Hamiltonian reads
\begin{multline}
    \label{eq:H_Dicke}
    \hat{H} = \omega_0 \hat{a}^\dagger \hat{a} + h \sum_{j=1}^L \hat{S}_j^z\\
    +\frac{g}{\sqrt{L}} \sum_{j=1}^L \big( \hat{S}_j^+ \hat{a} + \hat{S}_j^-\hat{a}^\dagger\big) + \frac{\lambda}{\sqrt{L}} \sum_{j=1}^L \big( \hat{S}_j^+ \hat{a}^\dagger + \hat{S}_j^- \hat{a}\big)  \ ,
\end{multline}
where $\lambda=g$ corresponds to the Dicke model while $\lambda=0$ recovers the Tavis-Cummings model.

In case of a large detuning between the cavity mode frequency and the transition energies---an approximation valid in many single-mode cavity QED experiments---only the absorption and emission of virtual photons are possible. It is then justified to adiabatically eliminate the photon field, leading to a permutation-symmetric effective Hamiltonian
\begin{multline}
    \label{eq:H_LMG}
    \hat{H} =h\sum_{j=1}^L \hat{S}^z_j -\frac{g^2+\lambda^2}{\omega_0 L} \sum_{i,j=1}^L \hat{S}^-_i \hat{S}^+_j \\
    - \frac{g\lambda}{\omega_0 L}\sum_{i,j=1}^L  \big(\hat{S}^-_i \hat{S}^-_j + \hat{S}^+_i \hat{S}^+_j \big)  \ .
\end{multline}
This is the well-known Lipkin-Meshkov-Glick (LMG) model~\cite{Meshkov1965Validity}.

In order to pave the way to the understanding of the SU(3) case later on, it is convenient to perform some intermediate steps and rewrite the Hamiltonians above in another basis. We express the spin operators in terms of Schwinger bosons $\hat{b}_{j,\alpha},\hat{b}_{j,\alpha}^\dagger$, with $j=1,\dots,L$ and $\alpha=1,2$~\cite{Schuckert2018Nonequilibrium,Halimeh2018Aging}: 
\begin{equation}
\begin{gathered}    
    \hat{S}_j^x = \frac{\hat{b}_{j,1}^\dagger \hat{b}_{j,2}^\phdagger + \hat{b}_{j,2}^\dagger \hat{b}_{j,1}^\phdagger}{2}, \qquad 
    \hat{S}_j^y = \frac{\hat{b}_{j,1}^\dagger \hat{b}_{j,2}^\phdagger - \hat{b}_{j,2}^\dagger \hat{b}_{j,1}^\phdagger}{2i}, \\
    \hat{S}_j^z = \frac{\hat{b}_{j,1}^\dagger \hat{b}_{j,1}^\phdagger - \hat{b}_{j,2}^\dagger \hat{b}_{j,2}^\phdagger}{2} \, ,
\end{gathered}
\end{equation}
with the additional constraint $\hat{b}^\dagger_{j,1} \hat{b}^\phdagger_{j,1} + \hat{b}^\dagger_{j,2} \hat{b}^\phdagger_{j,2} = 1$. The collective-spin Hamiltonian becomes 
\begin{multline}
    \label{eq:H_TC_bosons_SU2}
    \hat{H} =\sum_{j=1}^L \sum_{\alpha=1}^2 h_{j,\alpha} \hat{b}^\dagger_{j,\alpha} \hat{b}^\phdagger_{j,\alpha} -\frac{g^2+\lambda^2}{\omega_0 L} \sum_{i,j=1}^L \hat{b}^\dagger_{i,2} \hat{b}^\phdagger_{i,1} \hat{b}^\dagger_{j,1} \hat{b}^\phdagger_{j,2} \\
    - \frac{g\lambda}{\omega_0 L}\sum_{i,j=1}^L  \big(\hat{b}^\dagger_{i,1} \hat{b}^\phdagger_{i,2}\hat{b}^\dagger_{j,1} \hat{b}^\phdagger_{j,2} + \mathrm{h.c.}\big)  \ .
\end{multline}
The above rewriting shifts the attention from the spins on each site $\hat{S}_j$ to the mode occupancy: ``up'' ($\hat{b}^\dagger_{j,1} \hat{b}^\phdagger_{j,1}$) and ``down'' ($\hat{b}^\dagger_{j,2} \hat{b}^\phdagger_{j,2}$). Since there are two modes to be occupied, it is usually said that the cavity contains SU(2) atoms.

The situation can be generalized by assuming that each spatial site $j = 1,\dots,L$ can host more than one atom: this is achieved in the Schwinger boson formalism via the constraint $\hat{b}^\dagger_{j,1} \hat{b}^\phdagger_{j,1} + \hat{b}^\dagger_{j,2} \hat{b}^\phdagger_{j,2} = n_j$, where $n_j$ is an integer in general greater than 1. The Hamiltonian remains the same as in Eq.~\eqref{eq:H_TC_bosons_SU2}. The same effect can be obtained via Eq.~\eqref{eq:H_Dicke} by enlarging the spin representation on each site $j$, from $S=1/2$ to general $S$. Notice, however, that the transformation group is still dictated by SU(2), since there are only two distinct levels per atom.

%--------------------------------------------------------------------------------------------------
\subsection{The deep Hilbert space of SU($2$) atoms with collective interactions}
\label{sec:SU2_deep}

In this work, we study a SU(3) Tavis-Commings model in a far detuned regime from the perspective of the \emph{deep Hilbert space}~\cite{Qi2023Surprises}. To understand the meaning of the term, it is convenient to first rewrite the SU(3) LMG Hamiltonian, Eq.~\eqref{eq:H_LMG}, in terms of collective spin operators $\hat{S}^\mu \equiv \sum_i \hat{S}_i^\mu$, $\mu=x,y,z$: 
\begin{equation}
    \label{eq:H_LMG_collective}
    \hat{H} = h \hat{S}^z -\frac{g^2+\lambda^2}{\omega_0 L} \hat{S}^- \hat{S}^+ - \frac{g\lambda}{\omega_0 L} \big(\hat{S}^- \hat{S}^- + \hat{S}^+ \hat{S}^+ \big)  \ .
\end{equation}
The collective spins $\hat{S}^\mu$ come from the sum of $L$ spins-1/2. In algebraic terms, they are the tensor product of $L$ separate $S=1/2$ representations of SU(2), an operation that does not yield an irreducible representation (irrep) of SU(2). Rather, the operators $\hat{S}^\mu$ can be decomposed  via Clebsch-Gordan formulae in the sum of irreducible representations, each indexed by the value assumed by the total spin $\hat{S}^2$. Consequently, the Hamiltonian $\hat{H}$ can be decomposed itself into the sum of irreps of the total spin, each one representing a different dynamical sector. Now, typically in cavity QED settings only the totally symmetric irrep, associated to the maximal value of $\hat{S}^2$, is considered. The deep Hilbert space, by definition, is composed by all the other irreps, which can have rather different dynamical behaviors~\cite{Qi2023Surprises}.

We provide a more detailed description of the deep Hilbert space of SU(2) models later on in Sec.~\ref{sec:SU2_Schur-Weyl}, after having introduced the necessary analytical tools.

%--------------------------------------------------------------------------------------------------
\subsection{Chaos in SU(2) collective models}
\label{sec:SU2_chaos}

Both the Dicke and Tavis-Cummings model have been extensively studied from the perspective of quantum chaos. In the classical limit ($L \to \infty$), the Dicke model reduces to two matter degrees of freedom, and is therefore generically chaotic. The Tavis-Cummings model, conserves the number of excitations, i.e.\ it has a U(1) symmetry that, added to the energy conservation, makes the classical limit effectively one-dimensional and therefore integrable. 

The chaos properties of the two models have been investigated also in the quantum regime, i.e.\ at finite $L$. Again, it is found that the Tavis-Cummings model is integrable for all $L$ and all spin representations $S \geq 1/2$~\cite{Bogoliubov1996Exact}. The Dicke model displays instead a richer behavior, with a large literature investigating its dynamical phase diagram~\cite{Emary2003Quantum,Emary2003Chaos,Lambert2009Quantum,Altland2012Quantum,Bakemeier2013Dynamics}.

Finally, in the case of a far-detuned photon, where the physics is described by the LMG model, Eq.~\eqref{eq:H_LMG}, an extensive number of Gaudin-type conservation laws guarantees that the model is integrable for any $L$, irrespective of the values of $g$ and $\lambda$~\cite{Bentsen2019Integrable}.

In the following, we consider the chaos properties of a multilevel Tavis-Cummings model with a far-detuned photon, i.e.\ the SU(3) generalization of the LMG model without counter-rotating terms. The presence of a larger symmetry group for the ``spins'' modifies the dimensionality-counting arguments for integrability or chaos, leading to the presence of chaos also in the case of U(1) excitation number conservation, usually associated with integrability. In particular, we study the emergence of chaos keeping in mind the richness of the structure of the deep Hilbert space.

%--------------------------------------------------------------------------------------------------
%--------------------------------------------------------------------------------------------------
\section{Model: cavity QED with SU(3) atoms}
\label{sec:model}

In this Section, we introduce the model studied in the paper: an ensemble of $3$-level atoms, commonly referred to as SU(3) atoms, interacting with a single photon mode in a cavity. We first briefly describe how its Hamiltonian arises in experimentally relevant settings (Sec.~\ref{sec:microscopic_origin}), and then express it in terms of the commonly-used collective operators (Sec.~\ref{sec:collective_spin}).

%--------------------------------------------------------------------------------------------------
\subsection{Microscopic origin}
\label{sec:microscopic_origin}

We consider a generalization of the Tavis-Cummings introduced above in Sec.~\ref{sec:SU2}. We assume that in the cavity there is a single photonic mode ($\hat{a},\hat{a}^\dagger$) interacting with $\alpha=1,2,\dots,N$ bosonic modes ($\hat{b}^\phdagger_{j,\alpha},\hat{b}^\dagger_{j,\alpha}$) on each site $j=1,2,\dots,L$:
\begin{multline}
    \label{eq:H_ab}
    \hat{H} = \omega_0 \hat{a}^\dagger \hat{a} + \sum_{j=1}^L \sum_{\alpha=1}^N h_{j,\alpha} \hat{b}^\dagger_{j,\alpha} \hat{b}^\phdagger_{j,\alpha} \\
    + \sum_{j=1}^L \sum_{\alpha=1}^{N-1}  \frac{g_\alpha}{\sqrt{L}} \big( \hat{b}^\dagger_{j,\alpha} \hat{b}^\phdagger_{j,\alpha+1} a + \mathrm{h.c.}\big)  \ ,
\end{multline}
where we are using a convention designating $1$ as the highest energy state, and the states with increasing labels having decreasing energy. This is usually referred to as \emph{$N$-level Tavis-Cummings model}.

\begin{figure}
    \centering
    \includegraphics[width=\linewidth]{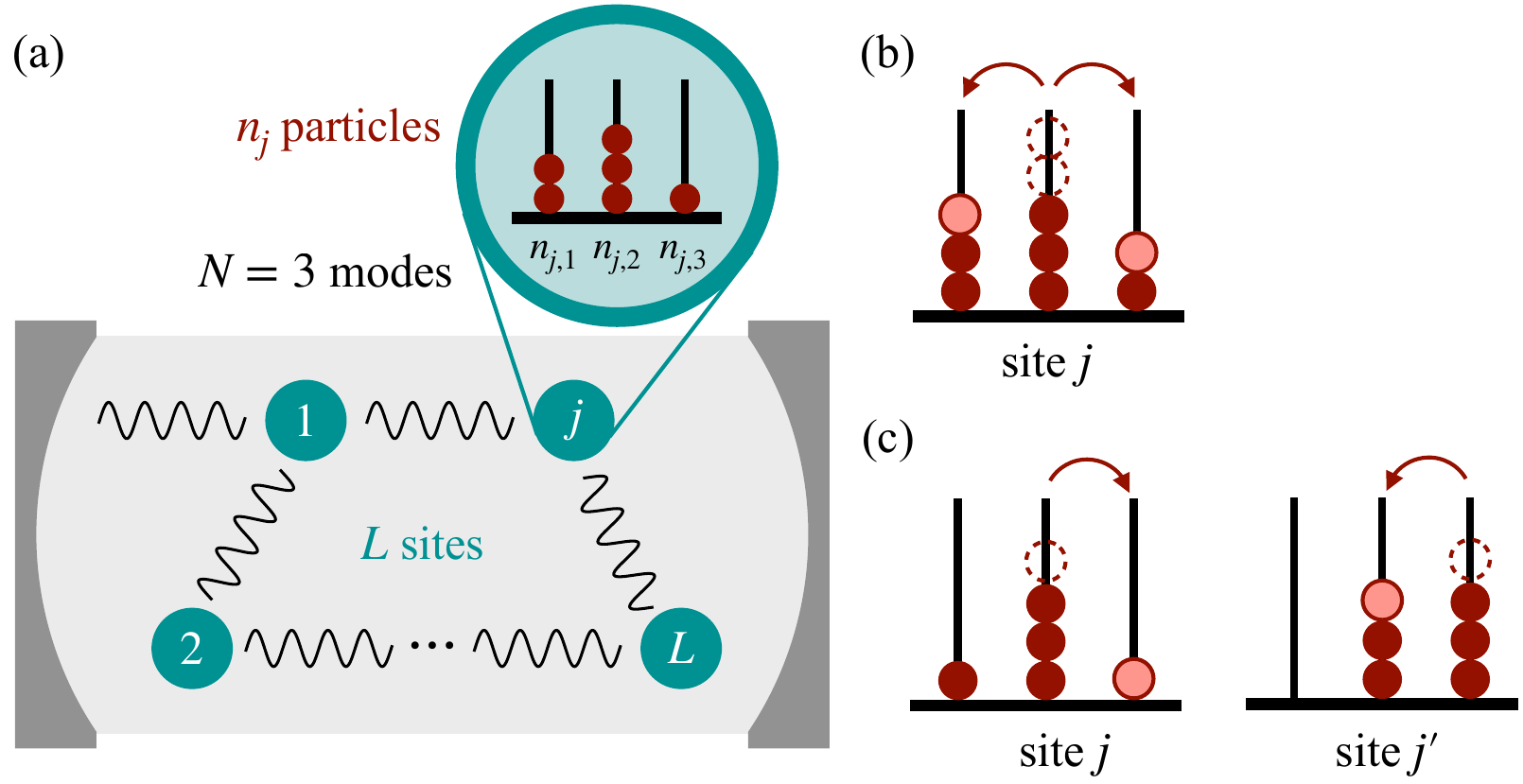}
    \caption{(a) Sketch of the experimental setup. Bosons are placed on $L$ sites in an optical cavity. Each site has $n_j$ particles, which can be in $N=3$ different modes, visualized as a Hanoi tower. The allowed transitions involve two particles at a time, and respect the total magnetization, defined as the number of particles on all columns 1 minus the ones on columns 3, Eq.~\eqref{eq:M}. The transitions can be either intra-site (b) or inter-site (c).}
    \label{fig:model}
\end{figure}

Again, if the frequency of the photonic mode is far detuned from the Raman resonance, the photon can be adiabatically eliminated, as in Eq.~\eqref{eq:H_LMG}. Then, an effective Hamiltonian for the bosons is to the lowest order,
\begin{multline}
    \label{eq:H_b}
    \hat{H} = \sum_{j=1}^L \sum_{\alpha=1}^N h_{j,\alpha} \hat{b}^\dagger_{j,\alpha} \hat{b}^\phdagger_{j,\alpha} \\
    - \sum_{\alpha,\beta=1}^{N-1} \frac{g_\alpha g_\beta}{\omega_0 L} \sum_{i,j=1}^L \hat{b}^\dagger_{i,\alpha+1} \hat{b}^\phdagger_{i,\alpha} \hat{b}^\dagger_{j,\beta} \hat{b}^\phdagger_{j,\beta+1}.
\end{multline}
An intuitive sketch of the Hamiltonian is presented in Fig.~\ref{fig:model}. The usual far-detuned Tavis-Cummings model is recovered when the number of modes $N$ is set to back to 2. To go beyond this simple setting, one needs to set $N \geq 3$. To keep the discussion as simple as possible, we fix the number of modes to $N=3$; we briefly comment on the case $N>3$ in Sec.~\ref{sec:conclusions}. 

The model, Eqs.~\eqref{eq:H_ab}--\eqref{eq:H_b}, allows for a high degree of tunability in its couplings $g_\alpha$ and local fields $h_{j,\alpha}$. Experimentally, these parameters can be controlled with high accuracy via the spatial profile of the cavity field and the amplitude of the driving field~\cite{Davis2019Photon,Periwal2021Programmable}. For our purposes, it is sufficient to fix $h_{j,\alpha} \equiv h_\alpha$, thus suppressing the dependence on the site, and in particular $h_3-h_2 \equiv h_2-h_1 \equiv h$, i.e.\ on each site the local fields act as a standard magnetic field on a spin-1 particle. The asymmetry between modes is governed solely by the interaction coefficients $g_\alpha$. In conclusion, the Hamiltonian simplifies to
\begin{multline}
    \label{eq:H}
    \hat{H} = h \sum_{j=1}^L (\hat{n}_{j,1} - \hat{n}_{j,3}) \\
    - \sum_{\alpha,\beta=1}^{2} \frac{g_\alpha g_\beta}{L} \sum_{i,j=1}^L \hat{b}^\dagger_{i,\alpha+1} \hat{b}^\phdagger_{i,\alpha} \hat{b}^\dagger_{j,\beta} \hat{b}^\phdagger_{j,\beta+1} \, ,
\end{multline}
where
\begin{equation}
    \hat{n}_{j,\alpha} \equiv \hat{b}_{j,\alpha}^\dagger \hat{b}_{j,\alpha}^\phdagger\ ,
\end{equation}
and we absorbed $\omega_0$ into the normalization of the interactions constants. From now on, this is the only Hamiltonian considered.

Here, it is important to stress the fact that, if one further assumes $g_\alpha \equiv g$ for all modes $\alpha$, the physics becomes the same as that of the SU(2) LMG model, i.e., integrable. This is because the Hamiltonian can still be written in terms of SU(2) spin-1 operators:
\begin{equation}
\label{eq: SU2}
\begin{gathered}    
    \hat{S}_j^+ = \hat{b}_{j,1}^\dagger \hat{b}_{j,2}^\phdagger + \hat{b}_{j,2}^\dagger \hat{b}_{j,3}^\phdagger, \quad 
    \hat{S}_j^- = \hat{b}_{j,2}^\dagger \hat{b}_{j,1}^\phdagger + \hat{b}_{j,3}^\dagger \hat{b}_{j,2}^\phdagger, \\
    \hat{S}_j^z = \hat{b}_{j,1}^\dagger \hat{b}_{j,1}^\phdagger - \hat{b}_{j,3}^\dagger \hat{b}_{j,3}^\phdagger \, .
\end{gathered}
\end{equation}
An extensive number of conservation laws follows.

The physics instead changes substantially if the cavity field couples differently to each of the transitions between the spin levels due to, for instance, different detuning of each of the transitions from the cavity field. This is exactly the situation we explore in the present work.

%--------------------------------------------------------------------------------------------------
\subsection{Collective spin representation}
\label{sec:collective_spin}

Owing to the all-to-all nature of the interactions mediated by the cavity field, the Hamiltonian~\eqref{eq:H} can be expressed in a natural way in terms of global operators, like it was done for the SU(2) LMG Hamiltonian to get from Eq.~\eqref{eq:H_LMG} to Eq.~\eqref{eq:H_LMG_collective}. Here, one needs to define collective SU(3) operators
\begin{equation}
    \label{eq:def_T}
    \hat{T}_{\alpha \beta} = \sum_{j=1}^L\hat{T}^{(j)}_{\alpha \beta} = \sum_{j=1}^L \sum_{\alpha,\beta=1}^3 \hat{b}^\dagger_{j,\alpha} \hat{b}^\phdagger_{j,\beta},
\end{equation}
in terms of which Eq.~\eqref{eq:H} becomes
\begin{equation}
    \label{eq:H_SU3}
    \hat{H} = h \big(\hat{T}_{11} - \hat{T}_{33} \big) - \frac{1}{\omega_0 L} \big(g_1 \hat{T}_{21} + g_2 \hat{T}_{32}\big) \big(g_1 \hat{T}_{12} + g_2 \hat{T}_{23}\big).
\end{equation}
The $\hat{T}$ operators can be seen as fundamental generators of SU(3)---a natural group of basis transformations of a 3-mode system\footnote{In fact, a fundamental symmetry of a $N$-level system is GL($N$). It reduces to SL($N$) via the requirement of particle number conservation and eventually to SU($N$) because of the Hermicity condition~\cite{Sawicki2010Classical}.}~\cite{Fradkin1965Three}. We refer to this SU(3) structure as SU$(3)_\mathrm{atoms}$, not to confuse it with other special unitary group structures introduced later.

For later use, it is convenient to introduce the Cartan-Weyl basis, which interprets the SU$(3)_\mathrm{atoms}$ generators as generators of three \emph{dependent} SU(2) algebras, the so-called isospin algebras:
\begin{subequations}   
    \label{eq:isospin}
\begin{alignat}{4}    
    \hat{I}_+ &= \hat{T}_{12}, \quad
    \hat{I}_- &&= \hat{T}_{21}, \quad
    \hat{I}_3 &&= \frac{1}{2} \left(\hat{T}_{11}-\hat{T}_{22}\right), \\
    \hat{V}_+ &= \hat{T}_{13}, \quad
    \hat{V}_- &&= \hat{T}_{31}, \quad
    \hat{V}_3 &&= \frac{1}{2} \left(\hat{T}_{11} - \hat{T}_{33}\right),
     \\
    \hat{U}_+ &= \hat{T}_{23}, \quad
    \hat{U}_- &&= \hat{T}_{32}, \quad
    \hat{U}_3 &&= \frac{1}{2} \left(\hat{T}_{22} - \hat{T}_{33}\right).
\end{alignat}
\end{subequations}
Equation~\eqref{eq:H} can be then expressed as
\begin{equation}
    \label{eq:H_SU3_Cartan}
    \hat{H} = 2h \hat{V}_3 - \frac{1}{\omega_0 L}  \big(g_1 \hat{I}_- + g_2 \hat{U}_-\big) \big(g_1 \hat{I}_+ + g_2 \hat{U}_+\big) .
\end{equation}
The equation above is the equivalent to the macroscopic spin representation of Eq.~\eqref{eq:H_LMG_collective}, where SU$(2)_\mathrm{atoms}$ has been replaced with the larger group of SU$(3)_\mathrm{atoms}$. For later convenience, we also define two hypercharge operators commonly used to characterize the representations of SU(3)
\begin{equation}
    \begin{split}
        Y&=\frac{1}{3}\left(\hat{T}_{11}+\hat{T}_{22}-2\hat{T}_{33}\right)\ ,\\
        Y'&=\frac{1}{3}\left(2\hat{T}_{22}-\hat{T}_{11}-\hat{T}_{33}\right)\ .
    \end{split}
\end{equation}

Equations~\eqref{eq:H_SU3}--\eqref{eq:H_SU3_Cartan} may suggest that the Hamiltonian, and thus the Hilbert space, should be partitioned according to the irreducible representations (irreps) of SU$(3)_\mathrm{atoms}$. This is indeed the case if we consider a single bosonic particle (i.e., a single SU(3) spin) per site. This setup is equivalent to an atomic cloud trapped in a cavity, with each atom constituting its own ``site'' in the absence of the externally imposed spatial structure.
However, for atoms trapped in an optical lattice with multiple occupancies allowed, the fact that there are $n>1$ particles per site becomes relevant. In this latter case, the absence of an explicit SU(3) \emph{symmetry} in the Hamiltonian implies that the Hilbert space is partitioned based on irreps of higher SU($d$) groups, with $d \geq 3$ indicated by bosonic occupations. A detailed discussion of this fact is presented in the following section.

%--------------------------------------------------------------------------------------------------
%--------------------------------------------------------------------------------------------------
\section{Symmetries, deep Hilbert space structure, and fragmentation}
\label{sec:symmetries}

In order to construct the deep Hilbert space of the model, Eq.~\eqref{eq:H}, we need to discuss in detail its symmetries. Indeed, the all-to-all nature of the interactions gives a lot of structure to the model, which can be exploited both for the mathematical characterization of all the symmetry sectors, and for a more efficient classical simulation.

The Hamiltonian~\eqref{eq:H} admits three independent conservation laws:
\begin{enumerate}
    \item U(1) symmetry of the particle number conservation on each site: $[\hat{H}, \hat{n}_j] = 0$ for all
    \begin{equation}
        \hat{n}_j = \sum_{\alpha=1}^3 \hat{n}_{j,\alpha} , \qquad j=1,2,\dots,L;
    \end{equation}

    \item $\mathrm{S}_L$ permutation symmetry of the sites: $[\hat{H}, \hat{\sigma}_\pi] = 0$ for all the operators $\hat{\sigma}_\pi$ defined by
    \begin{equation}
        \label{eq:action_S_L}
        \hat{\sigma}_\pi^\dagger \hat{b}_{j,\alpha} \hat{\sigma}_\pi = \hat{b}_{\pi(j),\alpha} ,
        \quad \text{for each } \pi \in \mathrm{S}_L;
    \end{equation}

    \item U(1) symmetry of the total magnetization conservation: $[\hat{H}, \hat{M}] = 0$ for
    \begin{equation}
        \label{eq:M}
        \hat{M} = \sum_{j=1}^L \left( \hat{n}_{j,1} - \hat{n}_{j,3} \right).
    \end{equation}
\end{enumerate}
While the conservation laws 1.\ and 3.\ are hard-wired already at the level of Eq.~\eqref{eq:H_b}, the conservation law 2.\ can be broken by some choices of local fields $h_{j,\alpha}$, e.g.\ if they depend on the site index $j$. Owing to our choices of magnetic fields, viz.\ $h_1$, $h_2$ and $h_3$ being independent of the site, all conservations laws 1.--3.\ are respected, and must be taken into account.

\begin{figure}
    \centering
    \includegraphics[width=0.9\linewidth]{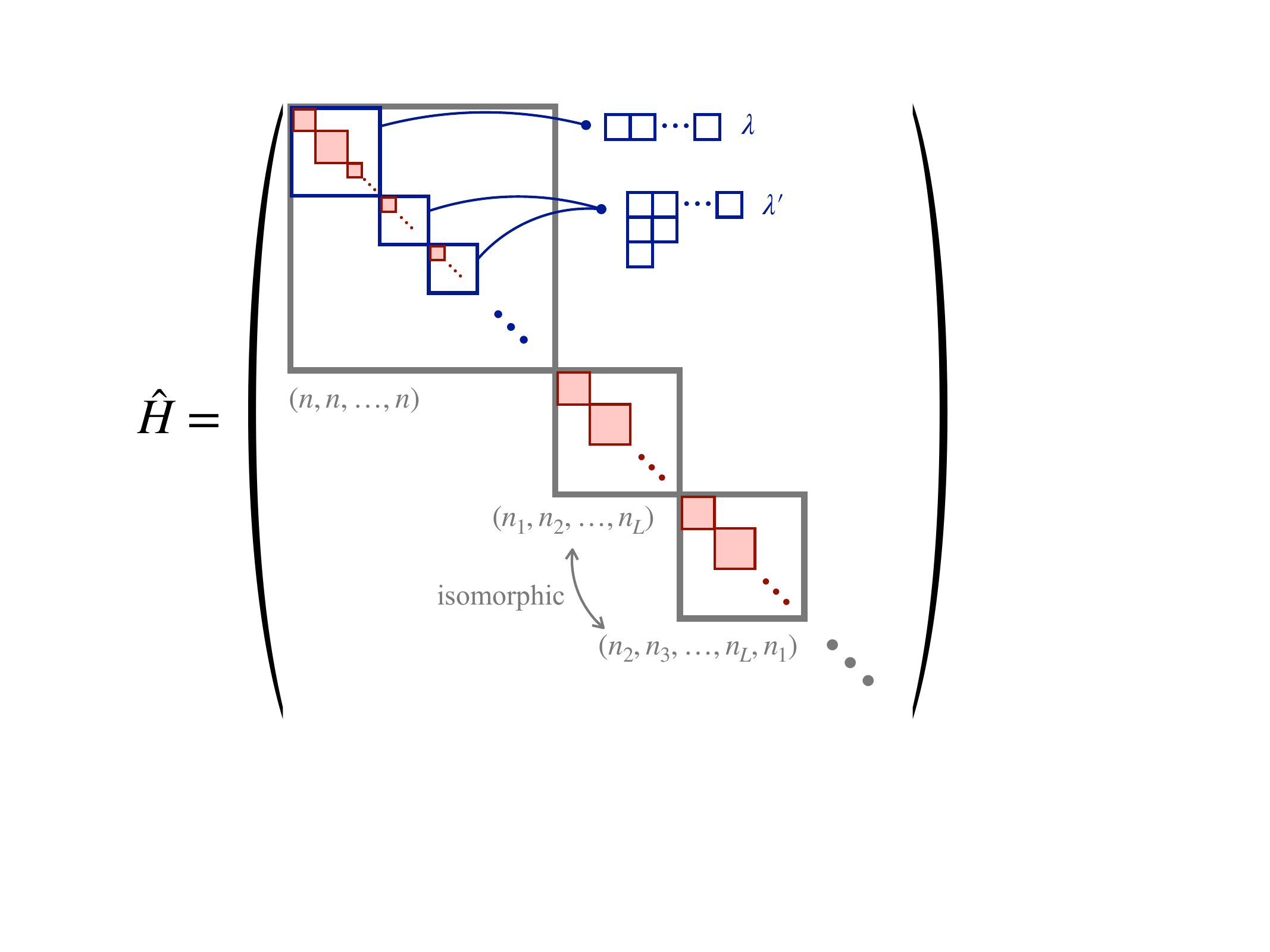}
    \caption{Fragmentation of the Hamiltonian into block-diagonal sectors, according to the successive application of the symmetries: on-site particle number conservation (gray), permutations (blue) and total magnetization (red). If the number of particles $n_j$ on each site is the same (top left gray box), then the Hilbert space is decomposed according to the irreps of the symmetric group $\mathrm{S}_L$, Eq.~\eqref{eq:Schur-Weyl}, that are labeled by Young diagrams $\lambda$. Finally, the magnetization constraint splits every intermediate (blue) sector into finer (red) sectors according to Eq.~\eqref{eq:decomposition_Hilbert_perm_mag}. If instead the number of particles is different on each site (other two gray boxes), only the magnetization constraint needs to be applied, Eq.~\eqref{eq:decomposition_Hilbert_noperm}. The permutation symmetry, not respected by the states, remains manifest in the isomorphicity between sectors that have the same occupation numbers, but arranged in a different order. }
    \label{fig:fragmentation}
\end{figure}

For future use, it is important also to notice that if the number of particles on each site, $n_j$, is different for different sites, then permutation symmetry is broken at the level of the sector explored dynamically, while being respected overall at the level of the Hamiltonian---see also Fig.~\ref{fig:fragmentation} below\footnote{More precisely, permutation symmetry is restricted to those groups of sites that share both the same $n_j$ and the same $h_{j,\alpha}$ pattern.}. In this paper, this is the only case considered in which the permutation symmetry is not fully respected.

We now construct the Hilbert space of this model by imposing progressively the conservation laws 1.--3. We consider here the permutational symmetric case, while we leave to App.~\ref{app:sec:no_permutation_symmetry} the case of $n_j$ different from site to site. The final result is Eq.~\eqref{eq:decomposition_Hilbert_perm_mag}, to which we refer the reader that is not interested in the group-theory arguments. 

Let us start from the conservation law 1., and denote as $n$ the particle number on each site, i.e.\ $n_j \equiv n$. Since a single site with $n$ particles can host 
\begin{equation}
    d = \binom{n+2}{2} = \frac{(n+2)(n+1)}{2}
\end{equation}
different states, the local Hilbert space is isomorphic to $\mathbb{C}^d$. From now on, $d$ will be fixed to the value above, if not otherwise stated. It follows that the total Hilbert space $\mathcal{H}$ can be taken to be
\begin{equation}
    \label{eq:Hilbert_space_tensor_product}
    \mathcal{H} = \bigotimes_{j=1}^L \mathbb{C}^d.
\end{equation}

According to Maschke's theorem~\cite{Fulton1991Representation}, the Hamiltonian can be decomposed into blocks (and thus the Hilbert space into sectors), according to the irreducible representations (irreps) of the site-permutation group $\mathrm{S}_L$---i.e.\ conservation law 2. While we leave all details to App.~\ref{app:sec:decomposition_sectors}, here we summarize only the main results. The action of the symmetric group on the Hilbert space, Eq.~\eqref{eq:action_S_L}, entails that the latter can be partitioned as
\begin{equation}
    \label{eq:decomposition_Hilbert_perm}
    \mathcal{H} = \bigoplus_{\lambda} \bigoplus_{k=1}^{m_\lambda} \mathcal{H}_{\lambda,h}, 
\end{equation}
where $\lambda$ labels the irreps of $\mathrm{S}_L$, that can appear with multiplicity $m_\lambda$ in the decomposition.

It is a well known fact (briefly explained in App.~\ref{app:sec:irreps_S_L}) that the irreps of $\mathrm{S}_L$ can be labeled by the \emph{integer partitions} of $L$: by definition, they are the tuples of integers $\lambda = (\lambda_1,\dots, \lambda_l)$ such that $L = \lambda_1 + \cdots + \lambda_l$ and $\lambda_1 \geq\cdots \geq \lambda_l$. They can be represented graphically as \emph{Young diagrams}, i.e.\ collections of $L$ boxes arranged in rows of length $\lambda_1,\lambda_2,\dots \lambda_L$. For example, for $L=3$ all the admissible partitions and corresponding diagrams are
\begin{equation}
    \ytableausetup{centertableaux,boxsize=1.2em}
    (1,1,1) = \ydiagram{1,1,1} \; ,\;
    (2,1,0) = \ydiagram{2,1} \; ,\;
    (3,0,0) = \ydiagram{3} \; .  
\end{equation}
To the Young diagram $\lambda$ with $L$ boxes, there corresponds an irrep of $\mathrm{S}_L$, denoted by $\mathcal{S}^\lambda$. The irreps $\mathcal{S}^\lambda$ are usually referred to as \emph{Specht modules}. Let us introduce also the irreps of the special unitary group SU$(d)_\mathrm{local}$, related to the fact that each site hosts a $d$-dimensional local Hilbert space. We denote the irreps as $\mathcal{U}^\lambda$. Also the representations of SU($d$) can be specified by Young diagrams of at most $d-1$ rows (more in App.~\ref{app:sec:irreps_SU(d)}). Notice that SU$(3)_\mathrm{atoms}$ (introduced in Sec.~\ref{sec:collective_spin}) and SU$(d)_\mathrm{local}$ (introduced here) are conceptually distinct special unitary structures on the same many-body Hilbert space, the first related to the transformation group of a single atom, and the latter to the transformation group of a single site, occupied by one or more atoms.

In the case under consideration, the decomposition Eq.~\eqref{eq:decomposition_Hilbert_perm} can be refined through the celebrated \emph{Schur-Weyl duality}. Namely,
\begin{equation}
    \label{eq:Schur-Weyl}
    \mathcal{H} \cong \bigoplus_{\lambda \in \mathrm{Par}(L,d)} \mathcal{S}^\lambda \otimes \mathcal{U}^{\bar\lambda}
\end{equation}
where $\mathrm{Par}(L,d)$ are the partitions of the integer $L$ into at most $d$ parts, or equivalently Young diagrams with at most $d$ rows, and $\bar\lambda$ is the diagram obtained by removing from $\lambda$ eventual columns of length $d$ (thus $\bar\lambda$ has at most $d-1$ rows, as required for SU$(d)_\mathrm{local}$). The meaning of Eq.~\eqref{eq:Schur-Weyl} is that the partitioning of the Hilbert space is specified by the complimentary action of SU$(d)_\mathrm{local}$, which describes transformations acting on the single site, and $\mathrm{S}_L$, that permutes sites. In particular, it tells us that the subspaces $\mathcal{H}_{\lambda,h} \cong \mathcal{U}^{\bar\lambda}$ and the multiplicity $m_\lambda = \dim \mathcal{S}^\lambda$. In Sec.~\ref{sec:examples} we provide examples for this construction. Refer to App.~\ref{app:sec:Schur-Weyl} for a more detailed explanation and more examples.

We are now in a position to quantify the fragmentation of the Hilbert space due to permutation symmetry. We assume that the thermodynamic limit is taken by sending the number of sites $L \to \infty$, while keeping $n$ finite.
In App.~\ref{app:sec:dimensions} we detail that the number of different Young diagrams $\lambda \in \mathrm{Par}(L,d)$ figuring in the direct sum is $p_d(L) \leq L^d$, while the size of the Specht modules is exponential in $L$. This leads to the formation of exponentially many disconnected dynamical sectors, the size of which is bounded by $\dim \mathcal{H}_{\lambda,h} = \dim \mathcal{U}^{\bar\lambda} \leq L^{d^2}$. This last inequality implies that the fragmentation is \emph{strong} in the classification of Ref.~\cite{Sala2020Ergodicity}, since the largest dynamical sector is asymptotically smaller than the whole Hilbert space.

Finally, one needs to enforce the magnetization constraint 3. Fixing $\hat{M}=M$ from Eq.~\eqref{eq:M} cannot be done straightforwardly, since the sectors $\mathcal{H}_{\lambda}$ are more easily described in terms of the permutation-symmetrized basis, not the Fock basis. Notice, however, that any permutation of the sites cannot change the global magnetization, and indeed the magnetization operator is diagonal as well in the permutation-symmetrized basis. Therefore, each of the sectors $\mathcal{H}_{\lambda,h}$ is further decomposed into $2nL+1$ subsectors, according to the eigenvalues of $\hat{M}$:
\begin{equation}
    \label{eq:decomposition_Hilbert_perm_mag}
    \mathcal{H} \cong \bigoplus_n \bigoplus_{\lambda \in \mathrm{Par}(L,d)} \bigoplus_{k=1}^{m_\lambda} \bigoplus_{M=-nL}^{nL} \mathcal{H}_{\lambda,k,M},
\end{equation}
where we reintroduced the direct sum over the particle number per site, $n$, explicitly.

The take-home message from this section is that the conservation laws 1.--3.\ lead to the formation of exponentially many (in $L$) distinct dynamical sectors in the model under consideration. These can be labeled explicitly with the quantum numbers of the particle number per site, the permutation group (given in terms of Young diagrams), and of the magnetization, as entailed by Eq.~\eqref{eq:decomposition_Hilbert_perm_mag} and shown graphically in Fig.~\ref{fig:fragmentation}. The physical consequence of this fragmentation is that the chaoticity of the model is severely suppressed, as it can happen in presence of dynamical constraints. In the language of Ref.~\cite{Sala2020Ergodicity}, the fragmentation is \emph{strong} since the largest sector is asymptotically much smaller than the total Hilbert space. Moreover, the system under consideration---and in general any permutation-symmetric system---provides an example of \emph{quantum} fragmentation~\cite{Moudgalya2022Hilbert}, i.e.\ the sectors are not diagonal in the computational basis. We stress, however, that in the present case the fragmentation is only caused by symmetries, while in general the term fragmentation is used also for the breaking into distinct dynamical sectors due to non-evident, non-local symmetries~\cite{Mukherjee2021Minimal,Brighi2023Hilbert}.

%--------------------------------------------------------------------------------------------------
%--------------------------------------------------------------------------------------------------
\section{Worked-out examples}
\label{sec:examples}

In this Section, we provide explicitly worked-out examples to make the theoretical construction of the Schur-Weyl duality as clear as possible. First, in Sec.~\ref{sec:SU2_Schur-Weyl} we show how the same algebraic analysis can be applied to SU(2) atoms as well. We then consider in Sec.~\ref{sec:1particle} the case of $N=3$ modes with only $n=1$ particle per site: this way, the local Hilbert space dimension $d=3$ and is the smallest allowed. Finally, in Sec.~\ref{sec:arbitrary_particles}, we sketch the case with an arbitrary number $n$ of particles per site for SU(3) atoms.

%--------------------------------------------------------------------------------------------------
\subsection{A look back on SU(2) atoms}
\label{sec:SU2_Schur-Weyl}

The Schur-Weyl duality, reviewed in the previous section, applies to any Hilbert space of the form Eq.~\eqref{eq:Hilbert_space_tensor_product}. While for the SU(3) model under consideration the local dimension $d \geq 3$, the theorem applies equally for $d=2$, namely for the tensor product of several spins-1/2. Thus, one can write
\begin{equation}
    \left( \mathbb{C}^2 \right)^{\otimes L} \cong \bigoplus_{\lambda \in \mathrm{Par}(L,2)} \mathcal{S}^\lambda \otimes \mathcal{U}^\lambda .
\end{equation}
Now, notice that the partitions of $L$ in at most 2 parts are of the form $\lambda = (L/2 + j,\, L/2 - j)$, and thus can be indexed by a single integer $j$ (we are assuming $L$ even for simplicity). This is exactly the total spin, as $\mathcal{U}^\lambda$ is the $(2j+1)$-dimensional representation of SU(2) in this case---namely, a spin-$j$. The duality further tells that the multiplicity space $\mathcal{S}^\lambda$ carries the structure of an irrep of $\mathrm{S}_L$.

Let us give the simplest example: the sum of two spins-1/2. The Schur-Weyl duality states that
\begin{equation}
    \left(\mathbb{C}^2 \right)^{\otimes 2} = \left( \mathcal{S}_\mathrm{sym} \otimes \mathcal{U}_{j=1} \right) \oplus \left( \mathcal{S}_\mathrm{antisym} \otimes \mathcal{U}_{j=0} \right),
\end{equation}
which is sometimes written as $2 \otimes 2 = 3 \oplus 1$. This is exactly the Clebsch--Gordan decomposition in triplet (symmetric) plus singlet (antisymmetric).

%--------------------------------------------------------------------------------------------------
\subsection{SU(3): One particle per site}
\label{sec:1particle}

Let us now explain in detail the Schur-Weyl duality, Eq.~\eqref{eq:Schur-Weyl}, in the case of SU(3) atoms with one particle per site: $n_j=1$ for all $j=1,\dots,L$. The local dimension of the Hilbert space is $d=3$, therefore in the decomposition Eq.~\eqref{eq:Schur-Weyl} one needs to consider the irreps of SU$(d=3)_\mathrm{local}$ and $\mathrm{S}_L$. Notice that this case is special, since the SU$(d)_\mathrm{local}$ structure on each site, needed for the duality, coincides with the SU$(3)_\mathrm{atoms}$ structure of the collective spin representation of the Hamiltonian, see Sec.~\ref{sec:collective_spin}. In group theory language, this is because on each site the local Hilbert space transforms as the fundamental representation of SU$(3)_\mathrm{atoms}$.

As detailed in App.~\ref{app:sec:irreps_S_L}, the irreps of $\mathrm{S}_L$ can be labeled by Young diagrams composed of $L$ boxes, each one representing a site. Boxes within a given row correspond to the mutually symmetrized sites, whereas the sites in different rows are antisymmetric under exchange. A ($d=3$)-dimensional local Hilbert space limits to a maximum of 3 rows, see Eq.~\eqref{eq:Schur-Weyl}. Hence, the diagrams can be labeled by 3 integers $(p,q,r)$, which correspond to the number of columns with one, two, and three rows, respectively:
\begin{equation}
    \begin{tabular}{c}
        $\hphantom{\lambda = }\overbrace{\hspace{3.6em}}^{\displaystyle r} \overbrace{\hspace{2.4em}}^{\displaystyle q} \overbrace{\hspace{2.4em}}^{\displaystyle p}  \hspace{1em} $ \\[-3pt]
        $ \lambda = \ydiagram{7,5,3} $ \, .
    \end{tabular}
\end{equation} 
The corresponding SU(3) diagram $\bar{\lambda}$ is constructed by removing all the ``$r$'' columns from $\lambda$. Indeed, SU(3) representations have to be parametrized by diagrams with at most 2 rows, and are usually denoted as $D(p,q)$:
\begin{equation}
    \begin{tabular}{c}
        $\hphantom{\lambda = } \overbrace{\hspace{2.4em}}^{\displaystyle q} \overbrace{\hspace{2.4em}}^{\displaystyle p} \hspace{1em}$ \\[-3pt]
        $ \bar{\lambda} = \ydiagram{4,2} $ \, ,
    \end{tabular}
\end{equation} 
see also App.~\ref{app:sec:irreps_SU(d)}. The labels $(p,q)$ of the SU(3) representations $D(p,q)$ correspond to the eigenvalues of the Cartan subalgebra generators $I_3$ and $U_3$, respectively---defined in Eqs.~\eqref{eq:isospin}. The representation $D(p,q)$ has dimension $d_{p,q}=(p+1)(q+1)(p+q+2)/2$.

The states can be constructed explicitly by using a Young tableau. One starts by applying a labeling as
\begin{equation}
    \label{eq:Yt_labeling}
    \begin{tabular}{c}
        $\overbrace{\hspace{2.8em}}^{\displaystyle r} \overbrace{\hspace{4.2em}}^{\displaystyle q} \overbrace{\hspace{2.8em}}^{\displaystyle p}  \hspace{0.5em} $ \\[-3pt]
        $ \ytableausetup{centertableaux,boxsize=1.4em}
        \begin{ytableau} 
            i_1 & i_2 & i_3 & i_4 & i_5 & i_6 & i_7\\
            j_1 & j_2 & j_3 & j_4 & j_5 & \none & \none\\
            k_1 & k_2 & \none & \none & \none & \none & \none
        \end{ytableau} \;, $
    \end{tabular}
\end{equation}
i.e.\ one assigns the state ``1'' to all sites with indices in the first row (indexed by $i_1, \dots,i_{p+q+r}$), the state ``2'' to all sites with indices in the second row ($j_1, \dots,j_{q+r}$), and the state ``3'' to the sites with indices in the third row ($k_1, \dots,k_r$)---indeed, the local Hilbert space is composed of only $d=3$ states. We call this reference state $\ket{\phi}$. Then, the highest weight state in a given representation is obtained by symmetrizing the sites in every row and antisymmetrizing the sites in every column. This can be done in terms of the $S, A$ operators, i.e.\ the symmetric and antisymmetric projection operators, respectively, see App.~\ref{app:sec:irreps_S_L}. Explicitly, the highest weight state corresponding to the tableau in Eq.~\eqref{eq:Yt_labeling} is
\begin{multline}
\ket{\mu}=
A_{i_1,j_1,k_1}A_{i_2,j_2,k_2}A_{i_3,j_3}A_{i_4,j_4}A_{i_5,j_5}\\
S_{i_1,i_2,i_3,i_4,i_5,i_6,i_7}S_{j_1,j_2,j_3,j_4,j_5}S_{k_1,k_2}\ket{\phi}\ .  
\end{multline}
The symmetry projection operators are defined in terms of all possible combinations of 2-cycles, $P_{ab}\equiv (a,b)$, so that $S_{ab}=1+P_{ab}$ and $A_{ab}=1-P_{ab}$. For $n$ indices, the projector contains $n!$ terms.

The remaining states are generated then by acting with the global annihilation operators $\{\hat{T}_{21}, \hat{T}_{31}, \hat{T}_{32}\}$. In the language of high-energy physics, these are the lowering operators of the $\hat{I}$,$\hat{U}$,$\hat{V}$ isospins, see Eqs.~\eqref{eq:isospin}. The procedure for constructing the SU(3) irreps is outlined in more detail in App.~\ref{app:sec:irreps_SU(d)}.

\begin{figure}
    \centering
    \includegraphics[width=0.7\linewidth]{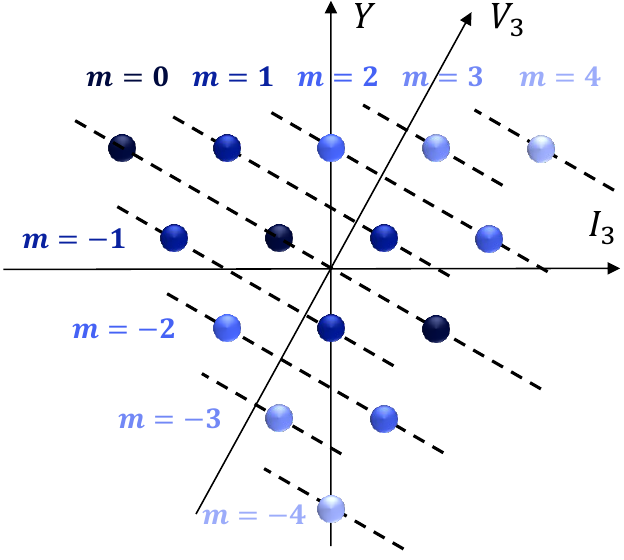}
    \caption{An example fully symmetric SU$(3)_\mathrm{atoms}$ representation for $L=4$ sites with $n=1$ particle each. The partition into magnetization sectors corresponds to fixing the isospin quantum number $V_3$. }
    \label{fig:su3_rep}
\end{figure}

The last step consists in imposing the magnetization conservation, corresponding to a U$(1)$ symmetry. This breaks the representation into smaller multiplets, as exemplified in Fig.~\ref{fig:su3_rep}. In the high-energy notation, these correspond to eigenstates of $\hat{V}_3$, the third component of the $\hat{V}$ isospin [defined in Eqs.~\eqref{eq:isospin}].

\begin{figure*}
    \centering
    \includegraphics[width=\textwidth]{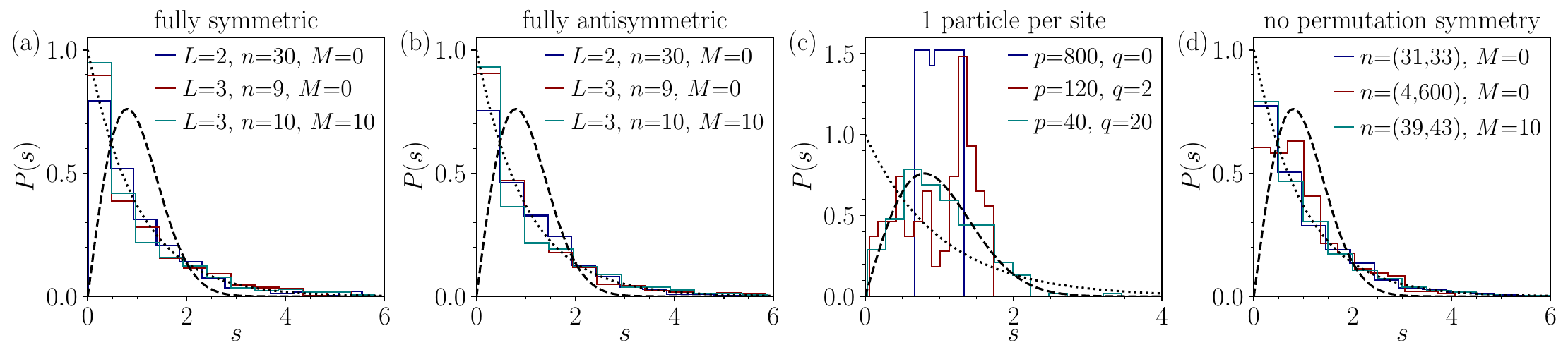}
    \caption{Comparison of Wigner-Dyson (dashed black lines) vs Poissonian (dotted black lines) level statistics in the shallow and deep portions of the Hilbert space.
    (a) Poissonian statistics, associated with non-chaotic behavior, is found when considering systems with $n_j \equiv n >1$ particles on each site, in the permutation fully-symmetric sector (one-row Young diagram).
    (b) The same happens for the permutation fully-antisymmetric sectors, irrespective of the choice of sites, occupation number, and magnetization. 
    (c) Fixing one particle per site $n_j \equiv 1$ and the magnetization $M=0$, chaos appears only for non-planar representations of SU(3), i.e.\ $D(p,q)$ with both $p>0$ and $q>0$. This result can be understood from the quantum-classical correspondence, as detailed in Sec.~\ref{sec:classical}. The spectral statistics for $q=0$ is not Poissonian, but displays the rigidity typical of harmonic oscillators (rather than chaotic behaviour).
    (d) We generically find the absence of chaotic behaviour also when the permutation symmetry is broken by a choice of site-dependent number of bosons $n_j$. 
    In all panels, the parameters of the Hamiltonian, Eq.~\eqref{eq:H_b}, read $g_1=1.7$, $g_2=1$ and $h=1$.}
    \label{fig:level_stat}
\end{figure*}

Within each multiple of magnetization $M$, there are states $\ket{x}, x=0,\dots,x^*$, where $x$ quantifies the number of \textit{pairs} of particles displaced from zero, and $x^*=\lfloor (L-M)/2\rfloor$ is the maximal value.
The largest multiple corresponds to magnetization $M=0$, and the multiples of decreasing size have an increasing absolute value of magnetization.

%--------------------------------------------------------------------------------------------------
\subsection{SU(3): Arbitrary number of particles per site}
\label{sec:arbitrary_particles}

The explicit construction of states transforming under SU$(d)_\mathrm{local}$ irreps, i.e.\ with more than one particle per site, follows an analogous procedure to the one outlined in the case of SU$(3)_\mathrm{local}$. Here, however, to each row one needs to assign a state label ranging from $1$ to $d$: label ``1'' to the first row, label ``2'' to the second row, and so on. After constructing such a reference state $\ket{\phi}$, the highest weight state is obtained by applying a combination of the symmetric/antisymmetric projection operators dictated by the Young tableau. All the remaining states in the representation are constructed by the application of the $d(d-1)/2$ annihilation operators of SU$(d)_\mathrm{local}$, see App.~\ref{app:sec:irreps_SU(d)}. The ordering of the Fock states, in terms of mode occupation, into the consecutive labels from $1$ to $d$, depends on the corresponding eigenvalues of the $(d-1)$ Cartan subalgebra generators of SU$(d)_\mathrm{local}$.

%--------------------------------------------------------------------------------------------------
\section{Integrability vs chaos in the deep Hilbert space}
\label{sec:numerics_quantum}

A standard way to detect the presence of chaos in a quantum system is spectral statistics. Quantum energy levels repel in presence of chaos, as conjectured by Bohigas, Giannoni and Schmit~\cite{Bohigas1984Characterization}, while they behave as independent random variables for integrable systems, as predicted by Berry and Tabor~\cite{Berry1977Level}. Correspondingly, the normalized nearest-neighbour spacing distribution of energy levels is described by that of gaussian ensembles of random matrices, $P(s) = \frac{\pi s}{2} e^{-\pi s^2/4}$, for chaotic systems with time-reversal invariance (i.e.\ the Gaussian Orthogonal Ensemble, GOE), and by the Poisson (exponential) distribution, $P(s) = e^{-s}$, for non-chaotic ones. These two are not the only possibilities; for instance, systems as the harmonic oscillator are integrable but have a non-Poisson, regular spacing structure, leading to the so-called ``spectral stiffness''~\cite{Haake2010Quantum,Gnutzmann2000Quantum}. Notice also that the observation of a Wigner-Dyson level spacing is not a sufficient condition to guarantee the chaoticity of a system; a typical counterexample is the presence of quantum scars~\cite{Moudgalya2022Quantum}. Nevertheless, in the following we take the stance that a Wigner-Dyson statistics indicates the presence of at least \emph{some} chaos in the system, and that the absence of level repulsion indicates lack of chaos.

Let us move to the numerical results. First, we fix the same number of bosons on each site, $n_j \equiv n >1$. In Fig.~\ref{fig:level_stat}(a), we consider the totally symmetric subspace (i.e.\ $\lambda$ is one-row Young diagram). One can see that in general the spectral statistics is Poisson-like, independently from the number of sites, occupation number, and magnetization. From Fig.~\ref{fig:level_stat}(b), it can also be seen that Poisson statistics is associated to the fully antisymmetric permutation sector (i.e.\ $\lambda$ is one-column Young diagram)\footnote{We refrain from investigating mixed-permutation-symmetry sectors with $n>1$ in the present work, since the problem becomes numerically much more demanding.}.

In Fig.~\ref{fig:level_stat}(c), we consider the case of one particle per site, $n_j \equiv 1$. Since in this case the local Hilbert space is 3-dimensional ($d=3$), we label the different permutation sectors with the tuple $(p,q)$, as explained in Sec.~\ref{sec:1particle}. The fully symmetric sector is found to present spectral stiffness, as exemplified by the choice $(p,q) = (800,0)$. Level repulsion arises instead for generic Young diagrams, with a clear chaotic behaviour. These results are explained from a semiclassical perspective in the following sections.

Finally, in Fig.~\ref{fig:level_stat}(d) we consider the case of permutation symmetry explicitly broken by the choice of a number of bosons that is different on each site. Here, the number of particles on each site $n_j$ and the magnetization $M$ completely define the symmetry sector. We find numerically that such a sector is not chaotic in general, as the spectral statistics is Poissonian. We do not have at present an analytical argument to explain these findings, which surely deserve a deeper investigation in future work.

%--------------------------------------------------------------------------------------------------
%--------------------------------------------------------------------------------------------------
\section{Classical limit}
\label{sec:classical}

In the previous sections, we have described in detail the deep Hilbert space structure of the model under study, Eq.~\eqref{eq:H}. The question was somehow overlooked in the literature since the model is fully connected, and the solution for its dynamics can be obtained by means of the classical (mean-field) limit $\hbar_\mathrm{eff} =1/L \to 0$, which becomes exact for thermodynamically large systems. However, there are some conceptual issues with taking a ``naive'' classical limit of the model, as we now detail.

The main point concerns the nature of the manifold of classical states $\mathcal{A}$, that is explored when taking the classical limit. In fact, some mean-field approaches are restricted to the (quantum) totally symmetric subspace, which limits the analysis to a shallow portion of the Hilbert space. On the other hand, and crucially, the U(1) coherent states $\ket{\Psi}$, commonly used for taking the classical limit $O_\mathrm{cl}(t) = \lim_{\hbar_\mathrm{eff}\to0}\ev*{\hat{O}}{\Psi(t)}$, \emph{mix a large number of permutation symmetry sectors}. In this case, the classical phase space manifold $\mathcal{A}$ does not respect at all the decomposition Eq.~\eqref{eq:decomposition_Hilbert_perm_mag}, and the totally symmetric subspace forms only an exponentially small part of $\mathcal{A}$. Consequently, the dynamical responses of different symmetry sectors may obscure each other, preventing the understanding of the full spectrum of the dynamics.

To disentangle different contributions to the classical dynamics, we study the latter projected onto each permutation symmetry sector separately. A response measured in a physical system initialized in a state not respecting the symmetries of the Hamiltonian will be a mix of the responses we uncover.

We split the discussion in several sections. First, in Sec.~\ref{sec:coherent_states} we argue for the use of spin-coherent states, instead of the U(1) coherent states. In Sec.~\ref{sec:SU3_coherent_states} we describe in detail how SU(3) coherent states work, and in Sec.~\ref{sec:Hamiltonian_struc} we use them to perform the classical limit on the Hamiltonian. Finally, in Sec.~\ref{sec:chaos_from_dimensionality} we argue that chaos or regular motion can follow from a geometrical dimension-counting procedure.

%--------------------------------------------------------------------------------------------------
\subsection{On the choice of coherent states depending on the symmetries}
\label{sec:coherent_states}

In cavity QED settings, it is common to prepare the bosons $\hat{b}_{j,\alpha}$ in a U(1) coherent state (commonly referred to as simply a coherent state), i.e.
\begin{equation}
    \ket{\Psi}_{\vec{\gamma}} = \exp[\sum_{j=1}^L \sum_{\alpha=1}^3 \left( \gamma^*_{j,\alpha} \hat{b}_{j,\alpha} +  \gamma_{j,\alpha} \hat{b}^\dagger_{j,\alpha} \right)] \ket{0},
\end{equation}
where $\gamma_{j,\alpha}$ are the coefficients parametrizing the coherent state, and $\ket{0}$ is the bosonic vacuum. Such states are eigenstates of the annihilation operator $\hat{b}_{j,\alpha}$, form an overcomplete set and, crucially, \emph{do not have a fixed number of particles}. From this last point in particular, one can evince that the chaos properties of the Hamiltonian, Eq.~\eqref{eq:H} are difficult to be extracted if $\hat{H}$ is projected onto such a coherent state basis---since the latter does not respect the particle-conservation symmetry of the Hamiltonian. 

Spin coherent states offer an alternative that instead respects as much as possible the fragmented structure of the Hamiltonian, shown in Fig.~\ref{fig:fragmentation}. Guided by the Schur-Weyl duality, one understands that each permutation symmetry sector can be parametrized by SU$(d)_\mathrm{local}$ coherent states. These states can be defined for each irreducible representation as follows. Take the highest weight state $\ket{\mu}$, defined to be the state annihilated by all creation operators $\hat{T}_{\alpha\beta}^\mathrm{local}$, $\alpha<\beta$, of the $\mathfrak{gl}(d,\mathbb{C})_\mathrm{local}$ algebra---see also Eq.~\eqref{eq:annihilation}. A coherent superposition of all states within the representation is then generated by the exponentiation of all annihilation operators $\hat{T}_{\alpha\beta}^\mathrm{local}$, $\alpha>\beta$, acting on the highest weight state $\ket{\mu}$\footnote{This approach introduces degeneracies if any of the representation defining integers $(p,q)$ is zero. To avoid those degeneracies, one can remove certain annihilation operators from the definition of a coherent state along the lines of Ref.~\cite{Gnutzmann2000Quantum}.}:
\begin{equation}
    \ket{\Vec{\gamma}}=\mathcal{N}_{\Vec{\gamma}} \ \exp \bigg( \sum_{\alpha=2}^{d} \sum_{\beta<\alpha} \gamma_{\alpha\beta} \hat{T}_{\alpha\beta}^\mathrm{local} \bigg) \ket{\mu}\, ,
\end{equation}
where $\gamma_{\alpha\beta}$ are new coefficients parametrizing the coherent state, and $\mathcal{N}_{\Vec{\gamma}}$ is a normalization constant. Notice that we are using the same notation, $\hat{T}_{\alpha\beta}$, for both the $\mathfrak{gl}(3,\mathbb{C})_\mathrm{atoms}$ and $\mathfrak{gl}(d,\mathbb{C})_\mathrm{local}$ algebras, since they are both instances of special unitary algebras. When necessary, we distinguish them via the sub/superscripts ``atoms'' and ``local''.

To understand the significance of the coherent states above, let us briefly review some general results valid for Hamiltonians expressible in terms of the generators of SU($d$). Consider any quantum Hamiltonian which is at most quadratic when decomposed in the $d^2-1$ generators of SU($d$), i.e.
\begin{equation}
    \hat{H}=\sum_{\alpha,\beta=1}^{d}c^{(1)}_{\alpha\beta} \hat{T}_{\alpha\beta} + \sum_{\alpha,\beta,\mu,\nu=1}^{d}c^{(2)}_{\alpha\beta,\mu\nu} \hat{T}_{\alpha\beta} \hat{T}_{\mu\nu}\ ,
\end{equation}
where $c^{(1)}_{\alpha\beta}, c^{(2)}_{\alpha\beta,\mu\nu}$ are arbitrary coefficients. For such a Hamiltonian, exact formulae for the classical dynamics can be derived in the basis of the \emph{equivalent} SU($d$) coherent states~\cite{Davis2019Photon,Dahlbom2022Langevin}. Let us define $G_{\alpha\beta} \equiv \ev*{\hat{T}_{\alpha\beta}}$ as the expectation value of a SU($d$) generator on a SU($d$) coherent state. In the classical limit, the Hamiltonian takes the form
\begin{equation}
\label{eq: gen_mf}H=\sum_{\alpha,\beta=1}^{d}c^{(1)}_{\alpha\beta} G_{\alpha\beta} + \sum_{\alpha,\beta,\mu,\nu=1}^{d}c^{(2)}_{\alpha\beta,\mu\nu} G_{\alpha\beta} G_{\mu\nu}\ +\mathcal{O}(\hbar_\mathrm{eff})\ ,
\end{equation}
and the corresponding classical equations of motion are dictated by the Poisson-Lie structure on the manifold $\mathcal{A}$
\begin{equation}
    \der{}{t}G_{\alpha\beta} = f_{\alpha\beta;\gamma\delta;\epsilon\zeta}\ \frac{\partial H}{\partial G_{\gamma\delta}}G_{\epsilon\zeta}\ ,
\end{equation}
where $f_{\alpha\beta;\gamma\delta;\epsilon\zeta}$ are the SU($d$) structure constants. 

The formula \eqref{eq: gen_mf} implies that, depending on the choice of coherent states, different order of fluctuations is captured by the classical limit. Let us provide two simple examples to illustrate this point.

When one considers $n=1$ particle per site, then the local Hilbert space dimension is $d=3$, and SU$(d)_\mathrm{local}$ coincides with SU$(3)_\mathrm{atoms}$. The generators of SU$(3)_\mathrm{local}$ are bilinear in bosonic creation/annihilation operators. As a result, when taking a classical limit in the basis of SU(3) coherent states, the expectation values of these generators capture, at most, dipolar fluctuations on the ground state. 

Conversely, let us take $n=2$ particles per site, leading to a local Hilbert space dimension $d=6$, and a structure SU$(6)_\mathrm{local}$ needed for the Schur-Weyl duality. The generators of SU$(6)_\mathrm{local}$, unlike the previous example, can also contain \emph{four} bosonic creation/annihilation operators. Thus, taking a classical limit in the SU$(6)_\mathrm{local}$ coherent state basis also preserves the quadrupolar fluctuations, which do not factorize via Wick's theorem. While the single-particle case is described exactly in the basis of SU$(3)_\mathrm{atoms}$ coherent states, the two-particle case is fully captured only in the basis of the SU$(6)_\mathrm{local}$ coherent states.

On the other hand, for large representations, the multipolar effects can still be captured by the states based on \emph{lower} symmetry groups, as they can be described as a continuum of multiple spin waves~\cite{Zhang2021Classical}. We use this fact to extract the classical dynamical responses of the model by using a SU$(3)_\mathrm{local}$ coherent state basis, even when the local Hilbert space has dimension $d>3$. This is also convenient, since any explicit calculations on SU($d$) coherent states become progressively harder with increasing $d$, due to the increasing dimensionality of the corresponding parameter space. The price to pay is the factorization of higher-point correlation functions in terms of lower-order ones.

As a last comment, we stress that spin-coherent states have been experimentally implemented~\cite{Gross2010Nonlinear,Leroux2010Implementation,Fernholz2008Spin,Hamley2012Spin} and the SU(3) coherent states were used elsewhere for computing experimentally relevant quantities~\cite{Corre2015Spin}.

%--------------------------------------------------------------------------------------------------
\subsection{SU(3) coherent states}
\label{sec:SU3_coherent_states}

We describe here in detail the SU(3) coherent state basis. This will be needed in the following to study the dynamics of the model in Eq.~\eqref{eq:H_b}, with a higher resolution than the known mean-field results, obtained with U(1) coherent states~\cite{ValenciaTortora2023Crafting}. 

%--------------------------------------------------------------------------------------------------
\subsubsection*{One particle per site}

Let us consider first the permutation symmetry sectors in the case of one particle per site. As previously stated, for $L=p+2q+3r$ the size of the SU$(3)_\mathrm{local}$ representation is $d_{p,q}=(p+1)(q+1)(p+q+2)/2$. Taking a classical limit implies taking the size of the system and the number of blocks within the Young diagram to infinity, $L\rightarrow\infty$, together with the total number of particles, $n_\mathrm{tot} \to \infty$, while keeping the number of particles per site $n_i \equiv n = n_\mathrm{tot}/L$ constant. This ensures that the dimension of the local Hilbert space remains unchanged. Consequently, for Young diagrams with both $p$ and $q$ scaling with $L$, the size of the representations grows as $L^3$ for large system sizes. For only one parameter between $p$ and $q$ scaling with $L$, the size of the representations grows as $L^2$. If only $r$ scales with $L$, the representation remains finite even for large system sizes. This means that, from the perspective of the quantum Hamiltonian, the only permutation symmetry sectors contributing to the classical dynamics correspond to the thermodynamically large representations $p,q \to \infty$.

The SU$(3)_\mathrm{local}$ coherent states parametrizing the classical manifold $\mathcal{A}$ are explicitly~\cite{Gnutzmann1998Coherent}
\begin{equation}
    \label{eq:coherent_states}
    \ket{\Vec{\gamma}}_{p,q} = \frac{1}{\sqrt{A_1^pA_2^q}}e^{\gamma_3 \hat{T}_{31}}e^{\gamma_1 \hat{T}_{21}}e^{\gamma_2 \hat{T}_{32}}\ket{\mu}_{p,q}\ ,
\end{equation}
where $\gamma_1,\gamma_2,\gamma_3$ are complex parameters defining the state, and the normalization constants are
\begin{equation}
    \begin{split}
        A_1=& 1 + |\gamma_1|^2 + |\gamma_3|^2\ ,\\
        A_2=& 1 + |\gamma_2|^2 + |\gamma_3 -\gamma_1\gamma_2|^2\ . 
    \end{split}
\end{equation}
The $\ket{\Vec{\gamma}}_{p,q}$ coherent states form an overcomplete basis for the Hilbert space of the $D(p,q)$ representation of SU$(3)_\mathrm{local}$, with a resolution of identity
\begin{equation}
    \mathbb{I}=d_{p,q}\int \frac{d^2\gamma_1 \ d^2\gamma_2 \ d^2\gamma_3}{\pi^3/2}\frac{1}{A_1^2A_2^2}\ketbra{\Vec{\gamma}}\ .
\end{equation}
This definition generates degeneracies in the cases where either $p=0$ or $q=0$. To remove them, Eq.~\eqref{eq:coherent_states} needs to be restricted to the case of $\gamma_2=0$ for $p=0$, or $\gamma_1=0$ for $q=0$.

The explicit formulae for the expectation values of the boson bilinears in these states are known~\cite{Gnutzmann2000Quantum} and quoted explicitly in App.~\ref{app:exp_vals}. What is important to remark is that expectation values depend linearly on the representation parameters $(p,q)$. This means that, in the thermodynamic limit, one or both of these integers need to scale with the system size for the dynamics in the classical limit to be non-trivial.

%--------------------------------------------------------------------------------------------------
\subsubsection*{Multiple particles per site}
\label{sec:multi_particle}

As stated above, we treat the cases with $n\neq 1$ particles per site by still using a SU$(3)_\mathrm{local}$ basis. This way, the classical limit remains analytically treatable, at the price of neglecting some dynamical correlations. Here, we describe how to chose the correct SU$(3)_\mathrm{local}$ basis, when the local Hilbert space is $d$-dimensional.

The first step is to define an effective SU$(3)_\mathrm{local}$ representation $D(p^*,q^*)$, that parametrizes a subspace of the SU$(d)_\mathrm{local}$ representation, while preserving the required permutation symmetry under which the states transform. Here, there is no simple relationship between the Young diagram of the permutation group $\mathrm{S}_L$ and the desired effective SU$(3)_\mathrm{local}$ representation. Nevertheless, by constructing a hierarchy of $n$-particle states, we obtained analytical expressions for the relations between the set of integers defining an irrep of $\mathrm{S}_L$, i.e.\ the Cartan eigenvalues $(h_1,h_2,\dots,h_{\alpha_{max}})$ [see Eq.~\eqref{app:eq:Cartan}] and the effective SU$(3)_\mathrm{local}$ irrep. The rules for the construction follow the same procedure as presented for a single particle in Sec.~\ref{sec:1particle}, and involve assigning to each consecutive row of the $\mathrm{S}_L$ Young diagram particle states of decreasing order in the hierarchy of states. Their ordering is governed by two factors:
\begin{enumerate}
    \item maximize the eigenvalue of the $\hat{Y}$ hypercharge;
    \item among the states with the largest hypercharge, maximize the third component of the isospin $\hat{I}_3$.
\end{enumerate}

In essence, we start with a Young tableau defining an irrep of $\mathrm{S}_L$ through Cartan eigenvalues $(h_1,h_2,\dots,h_{\alpha_{max}})$ with all $h_\beta=0$ for $\beta>\alpha_{max}$. The number of available blocks to arrange within the tableau is $L$, which imposes the constraint
\begin{equation}
L=\sum_{\alpha=1}^{\alpha_{max}}\alpha h_\alpha\ .
\end{equation}
Due to the underlying SU$(3)_\mathrm{atoms}$ structure, the maximal integer is constrained by the number of distinct states per site
\begin{equation}
    \mathrm{max}(\alpha_{max})=\frac{(n+1)(n+2)}{2}\ .
\end{equation}

Whereas the $\mathrm{S}_L$ Young diagram governs the permutation of sites, the particle content of those sites is reflected in the corresponding Young diagram of an effective U(3) irrep. It is defined by only three integers $(p^*,q^*, r^*)$  and consists of $nL$ blocks such that
\begin{equation}
    p^*+2q^*+ 3 r^*=nL\ ,
\end{equation}
meaning that each block within this new diagram corresponds to a single bosonic state rather than a site. It is important to notice that the mapping is \emph{surjective}. By further following the standard procedure of ``cutting'' the columns with three rows, we eventually obtain a Young diagram corresponding to the effective SU(3) irrep $D(p^*,q^*)$. The permutation of sites, however, is still governed by the original $\mathrm{S}_L$ Young diagram. The effective SU(3) corresponds to constructing the highest weight diagram using the $\mathrm{S}_L$ representation and applying the SU(3) generators upon it.

Following the construction outlined above, we determined that the integers defining the effective U(3), and, in accordance with the Schur-Weyl formula, also the SU(3) irrep are
\begin{subequations}
\begin{align}
    p^* &=\sum_{t=0}^{n}\sum_{\alpha=1}^{n+1-t}\alpha (n-\alpha-t+1)h_{\varphi_t+\alpha}\ ,\\
    q^* &= \frac{1}{2}\sum_{t=0}^{n}\sum_{\alpha=1}^{n+1-t} \left[m_t +\alpha (\alpha-1-2t)\right]h_{\varphi_t+\alpha}\ ,\\
    r^* &= \frac{1}{3}\sum_{t=0}^{n}\sum_{\alpha=1}^{n+1-t} (n\varphi_t-m_t+3\alpha t)h_{\varphi_t+\alpha}\ ,
\end{align}
\end{subequations}
where $\varphi_t=nt-\frac{1}{2}t(t-3)$ and $m_t=t(n+1-t)(n+2-t)$. This definition correctly reproduces $p^*=h_1$, $q^*=h_2$ and $r^*=h_3$ in the case of a single particle per site. In the simplest case of $n\geq (\alpha_{max}-1)$ (i.e.\ $t=0$), these reduce to
\begin{subequations}
\begin{align}
    p^* &=\sum_{\alpha=1}^{\alpha_{max}}\alpha (n-\alpha+1)h_\alpha\ ,\\
    q^* &=\frac{1}{2}\sum_{\alpha=2}^{\alpha_{max}}\alpha (\alpha-1)h_\alpha\ ,\\
    r^* &= 0\ .
\end{align}
\end{subequations}

Note, that the formulae in this section do not depend on the thermodynamic limit and are valid for any system size. In principle, these could be used to simplify the construction of finite size Hilbert spaces for a $3$-level model with many-particles per site, albeit with a proportionally larger information loss.

%--------------------------------------------------------------------------------------------------
\subsubsection*{Fixed magnetization sectors}

\begin{figure}
    \centering
    \includegraphics[width=0.6\linewidth]{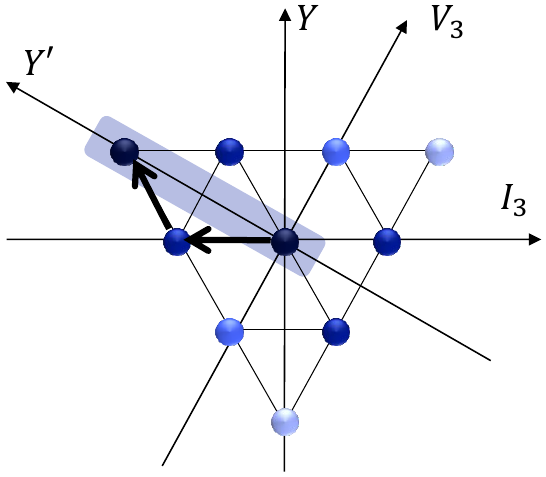}
    \caption{Example of a non-degenerate SU(3) representation D(3,0), with a magnetization sector highlighted. Each ``link'' between two states corresponds to the application of an annihilation/creation operator, which for SU(3) is a bosonic bilinear. Any two neighboring states in a fixed magnetization sector are connected by two such links. This sketch showcases how states within each magnetization sector have a fixed value of the $V_3$ isospin and different values of the $Y'$ hypercharge.}
    \label{fig:4point}
\end{figure}

To get the irreducible blocks of the quantum Hamiltonian, one needs to fix the magnetization within each of the permutation symmetry sectors, according to Eq.~\eqref{eq:decomposition_Hilbert_perm_mag}. Any two states within the sector, regardless of representation, are connected by the application of operators containing \emph{four} bosonic creation/annihilation operators, as shown in Fig.~\ref{fig:4point}. The classical limit in the basis of SU(3) coherent states does not resolve the magnetization sector and, thus, mixes their dynamical responses. However, we observe that this level of mixing does not obscure the underlying, symmetry-dictated mean-field dynamics.

\begin{figure*}
    \centering
    \includegraphics[width=1.\linewidth]{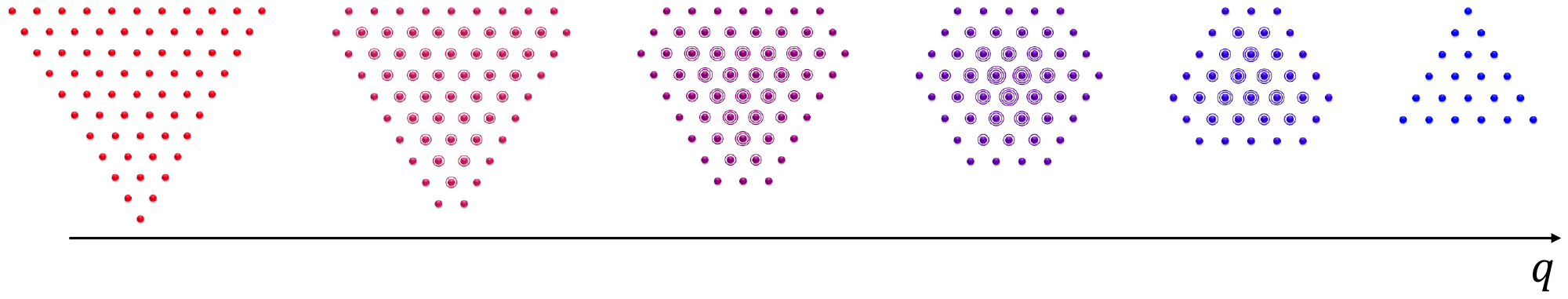}
    \caption{SU(3) irreps for $L=10$ and one particle per site ($n=1$). Each dot symbolizes a state transforming under the given representation, and rings around the dots indicate degenerate states. The representations, from left to right, are $D(10,0)$ (red, fully symmetric), $D(8,1), D(6,2), D(4,3)$ (most degenerate states, with up to four-fold degeneracy), $D(2,4), D(0,5)$ (blue, fully antisymmetric). The colors of the irreps match the corresponding curves for trajectory divergence in Fig.~\ref{fig:n1_n100}. }
    \label{fig:su3_q}
\end{figure*}

%--------------------------------------------------------------------------------------------------
\subsection{Hamiltonian structure}
\label{sec:Hamiltonian_struc}

The local bosonic bilinears, $\hat{T}^{(j)}_{\alpha\beta}= \hat{b}^\dagger_{\alpha,j} \hat{b}_{\beta,j}^\phdagger$, follow the $\mathfrak{gl}(3,\mathbb{C})$ algebra on each site:
\begin{equation}
    \comm{T^{(i)}_{\alpha\beta}}{T^{(j)}_{\mu\nu}}=\delta^{ij}\left(\delta_{\beta\mu}T^{(j)}_{\alpha\nu} - \delta_{\alpha\nu}T^{(j)}_{\mu\beta}\right)\ .
\end{equation}
Computing explicitly the classical limit in the SU$(3)_\mathrm{local}$ coherent state basis, the Hamiltonian \eqref{eq:H_SU3} can be re-expressed in terms of the expectation values of these bilinears $G_{\alpha\beta}^{(j)} \equiv \ev*{\hat{T}_{\alpha\beta}^{(j)}}$:
\begin{multline}
    \label{eq:Hamiltonian_order_L0}
    H = -\left(g_1G_{21} + g_2G_{32}\right) \left(g_1G_{12} + g_2G_{23}\right)\\
    +h\left(G_{11} - G_{33}\right) + \mathcal{O}(1/L)\ ,
\end{multline}
where $H \equiv \lim_{L\rightarrow\infty}\ev*{\hat{H}}/L$ and $G_{\alpha\beta} \equiv \frac{1}{L}\sum_{j=1}^L G_{\alpha\beta}^{(j)}$ are the averages across the system. Looking at Eq.~\eqref{eq:Hamiltonian_order_L0}, one can see that higher-point correlation functions are higher order in $1/L$ and do not contribute to the dynamics in the thermodynamic limit---here synonymous with the classical limit. Hence, the classical phase space $\mathcal{A}$ is spanned by the non-anomalous\footnote{We stick to the nomenclature by which $\ev*{b^\dagger b}$ is a non-anomalous correlation function, while $\ev*{b^\dagger b^\dagger}$ and $\ev*{bb}$ are anomalous correlation functions.} connected two-point correlation functions for every site. Here, we used the fact that the anomalous two-point correlators and two-point correlators linking different sites are incompatible with the symmetries of each sector considered.

The equations of motion for the two-point functions admit a Hamiltonian structure on the manifold of classical states $\mathcal{A}$:
\begin{equation}
    -i\der{}{t}G^{(j)}_{\alpha\beta}=\left\{G^{(j)}_{\alpha\beta},H \right\}_{G}\ ,
\end{equation}
where the bilinear map is defined in terms of local degrees of freedom
\begin{equation}
    \{\ ,\ \}_{G} \equiv \sum_{j=1}^L\sum_{\alpha,\beta,\gamma=1}^3 G^{(j)}_{\alpha\gamma}\left[ \frac{\overleftarrow{\partial}}{\partial G^{(j)}_{\alpha\beta}}\frac{\overrightarrow{\partial}}{\partial G^{(j)}_{\gamma\beta}} - \frac{\overleftarrow{\partial}}{\partial G^{(j)}_{\gamma\beta}}\frac{\overrightarrow{\partial}}{\partial G^{(j)}_{\alpha\beta}}\right]\ .
\end{equation}
This bracket is manifestly skew-symmetric, as required. The rescaling factor of the bilinear form is reminiscent of Poisson brackets of hydrodynamic type: effectively, the correlation functions provide a ``metric'' to the flows on the manifold $\mathcal{A}$, that is defined by themselves. Under this Poisson bracket, the classical phase space degrees of freedom admit again a $\mathfrak{gl}(3,\mathbb{C})$ algebra
\begin{equation}
     \left\{G^{(i)}_{ab},G^{(j)}_{cd} \right\}_G=\delta^{ij}\left(G^{(j)}_{ad}\delta_{bc}-G^{(j)}_{cb}\delta_{ad}\right)\ .
\end{equation}

\begin{figure*}
    \centering
    \includegraphics[width=\linewidth]{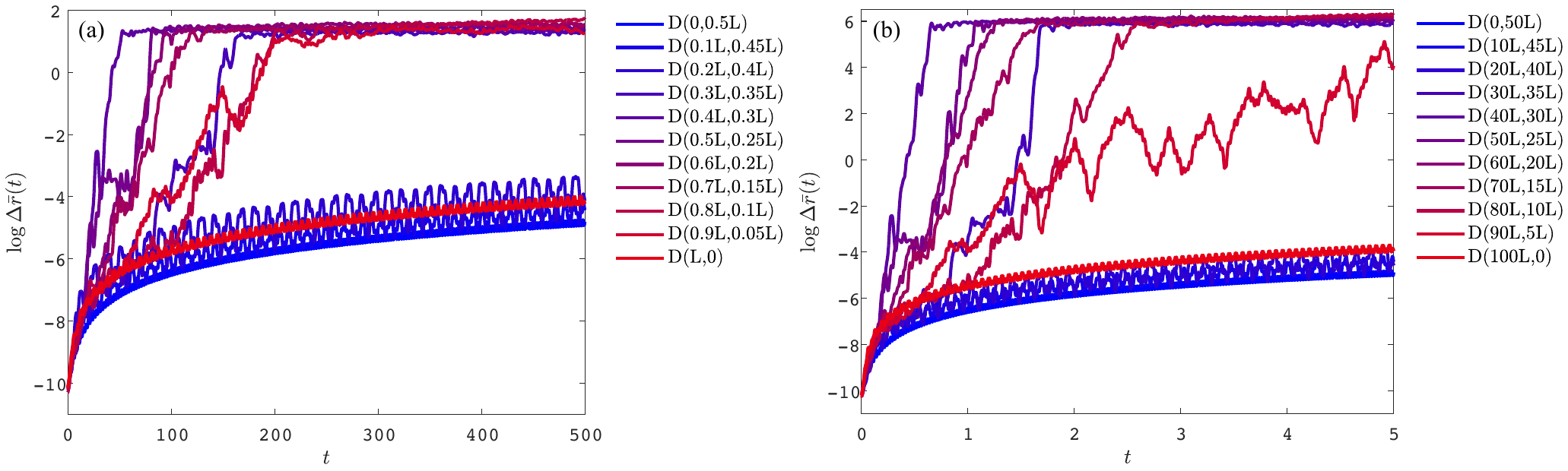}
    \caption{Chaotic vs regular dynamics in the classical limit $L\to\infty$, taken on SU(3) coherent states. Dynamics is regular when the divergence of nearby trajectories, quantified by $\Delta \bar{r}$ given in Eq.~\eqref{eq:Delta_r}, is slower than exponential; it is chaotic when the divergence is instead exponential. 
    (a) For $n=1$ particle per site, the use of SU(3) coherent states fully captures the classical limit. Regular dynamics takes place for non-degenerate representations of SU(3), i.e.\ for $D(p,q)$ with either $p$ or $q$ remaining finite as $L \to \infty$, while dynamics is chaotic for significantly degenerate representations. The parameters of the Hamiltonian [Eq.~\eqref{eq:H}] are fixed to $g_1=2$, $g_2=0.4$, $h=1$. The initial spin-coherent states [Eq.~\eqref{eq:coherent_states}] have $\gamma_1=4/\sqrt{21}$, $\gamma_2=2/\sqrt{21}$, $\gamma_3=i/\sqrt{21}$. The notation $D(p=\alpha L,q=\beta L)$ indicates how $p$ and $q$ scale in the limit $L\to\infty$.
    (b) For $n=100$ particles per site, the SU(3) coherent states still identify a regular-to-chaotic transition of the same type. The parameters of the Hamiltonian [Eq.~\eqref{eq:H}] are $g_1=2$, $g_2=0.4$, $h=1$; those of the initial state [Eq.~\eqref{eq:coherent_states}] are $\gamma_1=4/\sqrt{21}$, $\gamma_2=2/\sqrt{21}$, $\gamma_3=i/\sqrt{21}$.}
    \label{fig:n1_n100}
\end{figure*}

%--------------------------------------------------------------------------------------------------
\subsection{Chaos from degree-of-freedom counting}
\label{sec:chaos_from_dimensionality}

Each permutation symmetry sector is defined by the eigenvalues of the SU(3) Casimir operators
\begin{equation}
    \hat{C}_2 = \!\!\sum_{\alpha,\beta=1}^3 \hat{T}_{\alpha\beta} \hat{T}_{\beta\alpha}\ , \quad
    \hat{C}_3 = \!\!\sum_{\alpha,\beta,\gamma=1}^3 \hat{T}_{\alpha\beta} \hat{T}_{\beta\gamma} \hat{T}_{\gamma\alpha}\ .
\end{equation}
In the classical limit, this translates to the conservation of the Casimir functions
\begin{equation}
    C_2 = \!\!\sum_{\alpha,\beta=1}^3 G_{\alpha\beta} G_{\beta\alpha}\ , \quad
    C_3 = \!\!\!\!\sum_{\alpha,\beta,\gamma=1}^3 G_{\alpha\beta} G_{\beta\gamma} G_{\gamma\alpha}\ ,
\end{equation}
which results in involution with the classical Hamiltonian. Hence, the classical phase space inherits the symmetry structure of SU(3). 

For any SU(3) representation, there are two commuting Cartan subalgebra generators, $Y$ and $I_3$---or, alternatively, $Y'$ and $V_3$---which form the coordinate basis for the representations. If the representation is degenerate, i.e.\ $p\neq 0$ and $q\neq 0$, some states have degenerate eigenvalues of $Y',V_3$, and an additional operator is needed to differentiate them. A possible choice is
\begin{equation}
    V^2=V_3(V_3+1) + V_{-}V_{+}\ ,
\end{equation}
which commutes with both $Y',V_3$. In terms of the classical manifold $\mathcal{A}$, the above discussion implies that non-degenerate representations are associated with a lower-dimensional manifold.

Taking into account the magnetization symmetry, here generated by $V_3$, the dimensionality of the classical space $\mathcal{A}$ is reduced further: the dynamics is restricted to only a single coordinate $Y'$ in the case of non-degenerate representations, and to two coordinates $Y', V^2$ otherwise, see Fig.~\ref{fig:4point}. More precisely, one considers the expectation values $q_1=\expval*{Y'}$ and $q_2=\expval*{V^2}$, that are in involution under the Poisson bracket: $\left\{q_1, q_2\right\}_G=0$. Hence, these classical variables can be treated as classical phase space positions. Together with their conjugate momenta $p_1, p_2$~\cite{Gnutzmann2000Quantum}, they form a $4$-dimensional phase space $\{q_1,p_1,q_2,p_2\}$ in the case of the degenerate representations and a $2$-dimensional phase space $\{q_1,p_1\}$ in the case of the non-degenerate representations. This geometric fact is at the basis of the chaoticity or non-chaoticity of the classical dynamics on the SU$(3)_\mathrm{local}$ coherent state basis: non-degenerate representations are non-chaotic, since their phase space becomes one-dimensional after all the symmetries have been fixed. The dependence of the state degeneracy on the $p,q$ parameters labeling the representation $D(p,q)$ is presented in Fig.~\ref{fig:su3_q}. 

%--------------------------------------------------------------------------------------------------
%--------------------------------------------------------------------------------------------------
\section{Integrability vs chaos in the classical limit}
\label{sec:classical_chaos}

In this section, we investigate numerically the emergence of chaos in the semiclassical limit, by using SU(3) coherent states as described in the previous section. We focus on the dynamics of average correlation-function trajectories. The equations of motion considered are of the form
\begin{equation}
    -i\der{}{t}G_{\alpha\beta}=\{G_{\alpha\beta},\mathcal{H}\}_{G}\ ;
\end{equation}
their explicit expressions are presented in App.~\ref{app:classical_eoms}.

We initialize the system in different permutational symmetry sectors, parametrized by different representations of SU(3). We then solve the mean-field equations of motion, and diagnose the emergence of chaos by studying the divergence of nearby classical trajectories. For each given symmetry sector, we study $R=20$ closely sampled initial conditions. We define the distance between trajectories as the Frobenious norm of the difference of the average one-body reduced density matrices, i.e.\ 
\begin{equation}
    \Delta r_{uv}(t)=\sqrt{\sum_{\alpha\beta=1}^3 \left|G_{\alpha\beta}(t;u)-G_{\alpha\beta}(t;v)\right|^2}\ ,
\end{equation}
where $u,v=1,\dots,R$ enumerate the differently perturbed initial conditions. The average distance between the trajectories is
\begin{equation}
    \label{eq:Delta_r}
    \Delta \Bar{r}(t)=\frac{1}{R(R-1)}\sum_{u>v}\Delta r_{uv}(t)\ .
\end{equation}
For a chaotic system, this measure should grow exponentially, reflecting the information-scrambling characteristic of chaos~\cite{Haake2010Quantum}. The rate of exponential growth is the so-called Lyapunov exponent $\lambda$~\cite{Haake2010Quantum}, such that
\begin{equation}
    \Delta \Bar{r}(t)\sim e^{\lambda t}\ .
\end{equation}

We chose to fix the Hamiltonian parameters $h,g_1,g_2$ so that the system is far away from the integrable point of the Hamiltonian, i.e., neither of the hopping constants is zero and $g_1\neq g_2$. We also choose initial states, through the coefficients $\gamma_1,\gamma_2, \gamma_3$, which exemplify the chaotic regime. Indeed, it is important to note that within a chaotic symmetry sector, there are also regular trajectories, and the dynamical response does depend on the parametrization of the initial state. The dependence on the initial state parametrization is not, however, the subject of this work. We instead examine the logarithm of the trajectories' divergence $\log\Delta \Bar{r}(t)$ for various values of $p,q$, as shown in Fig.~\ref{fig:n1_n100}. As predicted by the geometrical argument of Sec.~\ref{sec:chaos_from_dimensionality}, for both $n=1$ and $n=100$ particles per site, the dynamics within the manifold of SU(3) coherent states are regular for representations with either $p=0$ or $q=0$, and become chaotic only for representations with a significant degree of state degeneracy. Since the limit $L\to\infty$ is already taken when performing the mean-field time evolution, each line in Fig.~\ref{fig:n1_n100} effectively corresponds to an \emph{infinitely} large SU(3) Young diagram. Its blocks are partitioned into two rows, according to the representation $D(p,q)$: since $p$ and $q$ scale in general with $L$, we use the notation $D(\alpha L, \beta L)$ to indicate that $p=\alpha L$ and $q =\beta L$ in the $L\to\infty$ limit. 

In the case of a single particle per site, the manifold of SU(3) coherent states supports the full dynamics of the model in the limit $L\to\infty$, and the representations $D(p,q)$ are determined directly from the Schur-Weyl formula. For $n=100$, the appropriate symmetry group of the spin-coherent states is SU(5151). The projection onto the manifold of SU(3) coherent states implies that we reduced the representation of SU$(5151)_\mathrm{local}$ to an effective SU$(3)_\mathrm{local}$ representation, significantly reducing the state space, but preserving the correct permutation symmetry (states belonging to the representation of SU(3) are linear combinations of the states belonging to the large symmetry group). Hence, in Fig.~\ref{fig:n1_n100} we study representations $D(p^*,q^*)$, with the integers determined via the procedure in Sec.~\ref{sec:multi_particle}.

\begin{figure}
    \centering
    \includegraphics[width=\columnwidth]{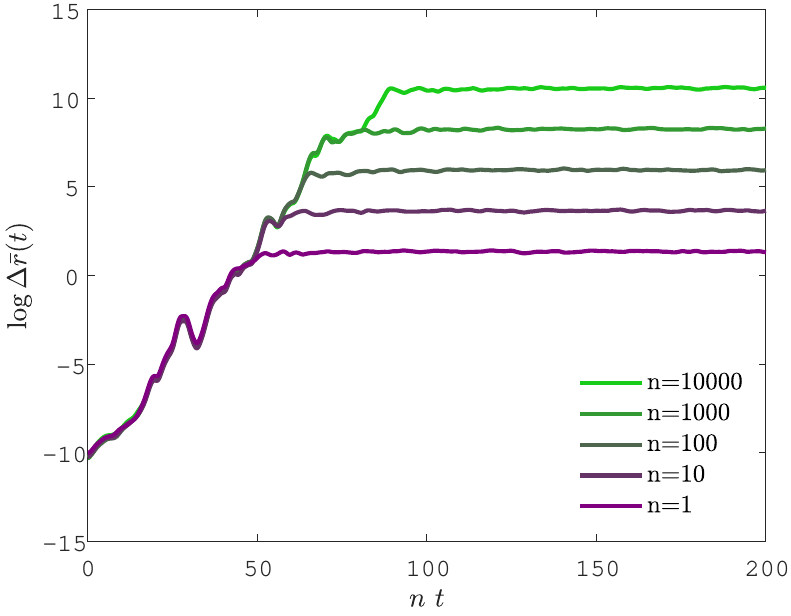}
    \caption{Divergence of nearby trajectories within the SU(3)-coherent-state manifold, varying the site occupation number $n$. The rescaling of time by $n$ shows that the Lyapunov exponent varies linearly with $n$ itself. The $p/q$ ratio was chosen to match the curves with the largest Lyapunov growth in Fig.~\ref{fig:n1_n100}, namely $p/q=4/3$. The parameters of the Hamiltonian [Eq.~\eqref{eq:H}] are $g_1=2$, $g_2=0.4$, $h=1$; the initial coherent state is characterized by $\gamma_1=4/\sqrt{21}$, $\gamma_2=2/\sqrt{21}$, $\gamma_3=i/\sqrt{21}$, see Eq.~\eqref{eq:coherent_states}.}
    \label{fig:n_comp}
\end{figure}

As shown in Fig.~\ref{fig:n_comp}, the associated Lyapunov exponent is linearly dependent on the number of particles per site: the more particles, the faster the chaotic dynamics. However, by comparing panels (a) and (b) of Fig.~\ref{fig:n1_n100}, one can see that whether the dynamics is chaotic or regular is dictated by the ratio of the integers $p/q$, rather than their absolute value, and independent of the particle number, as long as the dynamics are confined to the manifold of SU(3) coherent states.

The norm distance in Fig.~\ref{fig:n1_n100} saturates at different values depending on the chaotic or regular nature of the dynamics. The chaotic trajectories are ergodic and explore all phase space; therefore, the saturation of the distance between them indicates the time at which the trajectories have reached a typical distance in phase space (i.e.\ of the order of its diameter). As expected, the available phase-space volume increases with the increasing number of particles per site, and so does the maximal norm distance in Fig.~\ref{fig:n_comp}. On the contrary, the regular trajectories are dynamically confined to a smaller phase-space volume, i.e.\ on the tori arising from the conservation laws.

%--------------------------------------------------------------------------------------------------
%--------------------------------------------------------------------------------------------------
\section{Conclusions}
\label{sec:conclusions}

In this work, we studied chaotic and integrable behavior of the all-to-all interacting SU(3) atoms. First, with the help of the Schur-Weyl duality, we described in detail the deep Hilbert space of the model, namely, the permutation sectors beyond the totally symmetric one. This led to the conclusion that the Hilbert space is strongly fragmented, i.e.,\ the largest dynamical sector is asymptotically smaller than the total Hilbert space. Also, the model provides an instance of quantum fragmentation, meaning that the sectors are not easily identifiable from the computational basis, being instead diagonal in the basis of permutation-symmetrized states.

Second, we performed a numerical exact diagonalization of the deep Hilbert space, showing that integrable behaviour is found for the totally symmetric or antisymmetric sectors, while chaos emerges in mixed-symmetry sectors. This shows how conclusions about chaotic or regular behavior based only on the totally symmetric sector, which is widely studied by semiclassical methods, might not represent well the full range of dynamical responses displayed by the system. 

Third, we used a semiclassical description in terms of SU(3) coherent states, to explain the previous findings in a geometric way. Namely, in the simplest case of only one particle per site, chaos is associated to degenerate representations of SU(3), while non-degenerate representations are effectively lower-dimensional and display regular dynamics. We argued in detail how the use of SU(3) coherent states is both natural and required, since commonly-used U(1) coherent states obscure the contributions from the deep Hilbert space by mixing different symmetry sectors, and make it challenging to explore the dynamical phase diagram. We also used SU(3) coherent states to provide an approximate description of the general case with more than one particle per site, where one should instead use SU($d\!>\!3$) coherent states for capturing all relevant fluctuations in the classical limit. This represents a reasonable tradeoff, considering the difficulty of dealing with the representations of large unitary groups. To this end, we developed a method of projecting the permutation symmetry representations onto effective SU(3) representations and derived explicit formulas governing the projection.

Our results are naturally connected to a series of recent developments in the context of Hilbert space fragmentation that leverage the formalism of commutant algebras~\cite{Moudgalya2022Hilbert}. The system studied provides a complimentary view on the topic: while usually non-abelian U($N$) symmetries are imposed on the system, and the Schur-Weyl duality leads to a multiplicity space labeled by the irreps of the unitary group, here the system is permutation-symmetric, with the multiplicity given by the irreps of the symmetric group $\mathrm{S}_L$. A first consequence of this ``inversion'' is the presence of a stronger fragmentation~\cite{ClassenHowes2024Bipartite}. It would be interesting to study dissipative state preparation using these ideas~\cite{Li2025Highly}.

Another natural extension of our work is to consider 
SU($N$) atoms in a cavity with $N>3$. It is known that for all hopping parameters $g_\alpha\equiv g$ being equal, the Hamiltonian for SU($N$) atoms becomes integrable for all $N$ \cite{Bentsen2019Integrable}. When the full parameter freedom of the $N-1$ hopping constants is present, because of the higher dimensionality of the representations, a single U($1$) constraint of magnetization conservation does not render them automatically one-dimensional and integrable. There is, however, a possibility that the planes of constant magnetization ``cut'' the representations at such angles as to result in a one-dimensional subspace. To fully answer this question, a further investigation is required. 

To reiterate, our results point to the use of two powerful analytical tools that are deeply intertwined in the context of fully connected models: the Schur-Weyl duality and spin coherent states.  While these tools could also be applied to the most common setting of SU($2$) atoms, their merit becomes truly apparent for SU($N$) with $N \geq 3$. In the case of the SU(3) atoms considered in this work, the application of these tools allowed for the explicit constructions of the Hilbert space sectors for small system size, and the identification of distinct dynamical responses in the thermodynamic limit. The application of the framework developed in this work to a large class of permutationally symmetric models would advance the program of mapping the dynamical phases in the cavity QED setting.

%#######################################--ACKNOWLEDGMENTS--#######################################
\acknowledgments

F.B.\ thanks Giuseppe De Tomasi, Federico Roccati and Pablo Sala de Torres-Solanot for discussion.
A.Z. thanks Gianni Aupetit-Diallo for illuminating discussions.
We are deeply indebted to Jamir Marino for having spurred our interest in the topic, for guidance throughout all the stages of the project, and for a careful reading of the manuscript.
A.Z. acknowledges the support of the Alexander von Humboldt Foundation for the duration of this project.

%#######################################--APPENDIX--#######################################
\appendix

%--------------------------------------------------------------------------------------------------
%--------------------------------------------------------------------------------------------------
\section{Decomposition into sectors in absence of site-permutation symmetry}
\label{app:sec:no_permutation_symmetry}

In this Appendix, we consider the case in which the number of particles $n_j$ is different on each site. There are $\binom{3L+n_\mathrm{tot}-1}{3L}$ possible states for fixed total number of particles $n_\mathrm{tot} = \sum_j n_j$ on $L$ sites with $N=3$ modes each. Fixing all the local particle numbers $n_j$, imposes an additional U(1) symmetry for each site, which partitions the Hilbert space into much smaller blocks with $\prod_{j=1}^L \binom{n_j+2}{2}$ states each.

The on-site conservation of boson number can be imposed by labeling the Fock basis $\{\ket{n}\}$ of the system via a table of integer numbers $n_{j,\alpha}$ (notice that hats have been removed)
\begin{equation}
    \label{eq:n_table}
    \ket{n} \leftrightarrow 
    \begin{bmatrix}
        n_{1,1} &n_{1,2} &n_{1,3} \\
        n_{2,1} &n_{2,2} &n_{2,3} \\
        \cdots  &\cdots  &\cdots  \\
        n_{L,1} &n_{L,2} &n_{L,3}
    \end{bmatrix} ,
\end{equation}
with the entries in each of the rows adding up to the desired number of bosons on a corresponding site, such that
\begin{equation}
    \label{eq:constraint_nj}
    \sum_{\alpha=1}^3 n_{j,\alpha} = n_j, \qquad \text{for } j=1,\dots,L,
\end{equation}
where now $n_j$ are some given numbers, and not operators. Moreover, each integer $n_{j,\alpha}$ must comply with the bounds
\begin{equation}
    \label{eq:constraint_bounds}
    0 \leq n_{j,\alpha} \leq n_j, \qquad \forall\ j,\alpha\ .
\end{equation}

The conservation of magnetization can be implemented on the Fock basis defined above by requiring that the appropriate \emph{columns} of the $n_{j,\alpha}$ table, Eq.~\eqref{eq:n_table}, add up to the required magnetization $M$:
\begin{equation}
    \label{eq:constraint_M}
    \sum_{j=1}^L \left( n_{j,1} - n_{j,3} \right) = M.
\end{equation}
To give an example, for $L=2$ sites with $n_1=2$ particles on the first, $n_2=1$ particle on the second, and total magnetization $M=0$, the allowed Fock states are
\begin{equation}
    \begin{bmatrix}
        0 &2 &0 \\
        0 &1 &0 
    \end{bmatrix} , \quad
    \begin{bmatrix}
        1 &1 &0 \\
        0 &0 &1 
    \end{bmatrix} ,\quad
    \begin{bmatrix}
        1 &0 &1 \\
        0 &1 &0 
    \end{bmatrix} ,\quad
    \begin{bmatrix}
        0 &1 &1 \\
        1 &0 &0 
    \end{bmatrix} .
\end{equation}

The interplay of the conditions Eqs.~\eqref{eq:constraint_nj}--\eqref{eq:constraint_M} entails, at the level of the Hilbert space, the partitioning into sectors
\begin{equation}
    \label{eq:decomposition_Hilbert_noperm}
    \mathcal{H} = \bigoplus_{\{n\}} \bigoplus_{M=-n_\mathrm{tot}}^{n_\mathrm{tot}} \mathcal{H}_{\{n\},M},
\end{equation}
each identified by a set of integers $\{n\}=\{n_1,\dots,n_L\}$ and $M$. To quantify the level of Hilbert space fragmentation, a computation of the size of these sectors in the thermodynamic limit is necessary. To do so, one can use the table of integers $n_{j,\alpha}$, Eq.~\eqref{eq:n_table}, together with Eqs.~\eqref{eq:constraint_nj}--\eqref{eq:constraint_M}. The first observation is that this enumeration problem is a generalization of the ``magic square'' problem~\cite{WeissteinMagicSquare}: in the latter case, one has to determine all the square matrices of positive integers such that rows and columns add up to the same number. Unfortunately, this is a hard and yet unsolved problem in combinatorics. Even finding a closed formula for the \emph{number} of solutions, or an approximation thereof in the large $L$ limit, is an open problem. In turn, the problem of enumerating the allowed Fock states is even more challenging.  

For the reasons above, instead of looking for a closed formula for the enumeration of all the allowed states $\ket{n}$ in a given sector, it is convenient to state the problem in a form that can be solved \emph{algorithmically}. From the mathematical perspective, Eqs.~\eqref{eq:constraint_nj}--\eqref{eq:constraint_M} form a linear system of Diophantine equations that can be solved by integer linear programming~\cite{Conforti2014Integer}. The general solution of such systems of equation is known to be an NP-complete problem~\cite{Karp1972Reducibility}; therefore, it is difficult to access large systems with many particles. Nevertheless, the integer solutions $n_{j,\alpha}$ to the Eq.s~\eqref{eq:constraint_nj} and \eqref{eq:constraint_M} can be found via optimized solvers for small values of $L$ and $n_j$~\cite{ortools}.

We notice that the magnetization constraint splits each sector, identified by the occupations $\{n_j\}_{j=1,\dots,L}$, into $(2 n_\mathrm{tot} + 1)$ magnetization sectors. The fragmentation is thus \emph{weak} (the largest sector is asymptotically as big as the Hilbert space), not \emph{strong} (the largest sector is asymptotically smaller than the Hilbert space), according to the classification of Ref.~\cite{Sala2020Ergodicity}. Moreover, the fragmentation is \emph{classical}, i.e.\ it takes place in the computational basis, and not \emph{quantum} (which by definition takes place in a basis different from the computational one)~\cite{Moudgalya2022Hilbert}.

It is also interesting to notice that what is for the model under consideration a magnetization conservation, acquires the form of a \emph{local magnetic dipole} conservation if the Hamiltonian, Eq.~\eqref{eq:H}, is set on a finite-dimensional lattice~\cite{Sala2020Ergodicity}. This in turn can cause a strong fragmentation of the Hilbert space, due to the presence of frozen clusters of spin. In the present case, however, the all-to-all connectivity of the interactions prevents the formation of such clusters, and the fragmentation is less severe.

%--------------------------------------------------------------------------------------------------
%--------------------------------------------------------------------------------------------------
\section{Decomposition into sectors in presence of site-permutation symmetry}
\label{app:sec:decomposition_sectors}

Here, we detail the decomposition of the Hilbert space into irreps of the site-permutation group, i.e.\ the symmetric group $\mathrm{S}_L$. 

%--------------------------------------------------------------------------------------------------
\subsection{Actions of the symmetric and special unitary groups on the Hilbert space}
\label{app:sec:actions}

The starting point is Eq.~\eqref{eq:Hilbert_space_tensor_product}, that we rewrite here for clarity: 
\begin{equation}
    \mathcal{H} = \bigotimes_{j=1}^L \mathbb{C}^d,
\end{equation}
where $d = \binom{n+2}{2} = (n+2)(n+1)/2$ is the local Hilbert space dimension on each site. To fix ideas, we label the Fock basis for each site as $\{\ket{1}, \dots, \ket{d}\}$, having chosen an arbitrary ordering. A basis on the product space consists of the vectors $\ket{i_1,\dots,i_L} \equiv \ket{i_1} \otimes \cdots \otimes \ket{i_L}$.

A Hilbert space of the tensor product form as above admits the natural action of two groups: $\mathrm{S}_L$ and SU$(d)$. The first one is defined by the representation $\sigma$ of $\mathrm{S}_L$ that permutes sites, i.e. 
\begin{equation}
    \hat{\sigma}_\pi \ket{i_1,\dots,i_L} = \ket{i_{\pi(1)},\dots,i_{\pi(L)}},
\end{equation}
where $\pi \in \mathrm{S}_L$ is a permutation, and by $\hat \sigma_\pi$ we indicate the operators in the representation acting on the physical Hilbert space. This is the same introduced in Eq.~\eqref{eq:action_S_L} in the main text. The second one is defined by the representation $\rho$ of SU$(d)$ that acts on each site separately, i.e.\
\begin{equation}
    \hat{\rho}_U \ket{i_1,\dots,i_L} = (\hat{U}\ket{i_1}) \otimes \cdots \otimes (\hat{U}\ket{i_L})
\end{equation}
for every $\hat{U} \in \mathrm{SU}(d)$. Again, $\hat{\rho}_U$ are the operators acting on the physical Hilbert space according to the representation $\rho$.

By Maschke's theorem~\cite{Fulton1991Representation}, one can decompose
\begin{equation}
    \label{app:eq:decomposition_SU(d)_permutations_separate}
    \mathcal{H} \cong \bigoplus_\lambda \mathbb{I}_{m_\lambda} \otimes \mathcal{S}^\lambda
    \cong \bigoplus_\alpha \mathbb{I}_{\tilde{m}_\alpha} \otimes \mathcal{U}^\alpha ,
\end{equation}
where $\mathcal{S}^\lambda$ are the irreps of $\mathrm{S}_L$ and $\mathcal{U}^\alpha$ the ones of SU($d$), figuring with multiplicities $m_\lambda$ and $\tilde{m}_\alpha$, respectively. Then, since SU($d$) transformations commute with the permutations, Schur's lemma implies that $\mathrm{S}_L$ must act on the multiplicity labels of the irreps of SU($d$), and vice versa. The decomposition is thus refined to
\begin{equation}
    \label{app:eq:decomposition_SU(d)_permutations}
    \mathcal{H} \cong \bigoplus_{\lambda,\alpha} \mathbb{I}_{m_{\lambda,\alpha}} \otimes \mathcal{S}^\lambda \otimes \mathcal{U}^\alpha ,
\end{equation}
with some new, joint multiplicities $m_{\lambda,\alpha}$. 

The goal of this Appendix is to show that the above decompositions can be refined even further, by means of Schur-Weyl duality. To do so, we first describe how the irreps of both groups are structured in Apps.~\ref{app:sec:irreps_S_L} and \ref{app:sec:irreps_SU(d)}, respectively. Then, we describe the Schur-Weyl duality in App.~\ref{app:sec:Schur-Weyl}, and detail the size of the sectors in Sec.~\ref{app:sec:dimensions}.

%--------------------------------------------------------------------------------------------------
\subsection{Irreducible representations of the symmetric group}
\label{app:sec:irreps_S_L}

The theory of irreducible representations of the symmetric group is a textbook topic~\cite{Fulton1991Representation,Fulton1996Young}. We follow the presentation and notation of Ref.~\cite{Sagan2001Symmetric}. 

Irreps of the symmetric group $\mathrm{S}_L$ are labeled by partitions. A \emph{partition} of an integer $L$ is a tuple of integers $\lambda = (\lambda_1,\dots, \lambda_l)$ such that $L = \lambda_1 + \cdots + \lambda_l$ and $\lambda_1 \geq\cdots \geq \lambda_l$. It can be represented graphically as a \emph{Young diagram}; for example $\lambda = (4,2,1)$ corresponds to
\begin{equation}
    \ytableausetup{centertableaux,boxsize=1.2em}
    \ydiagram{4,2,1} \;.
\end{equation}
It is convenient to introduce also the definition of \emph{Young tableau} $t$, i.e.\ the filling of a Young diagram $\lambda$ with the numbers from 1 to $L$:
\begin{equation}
    \label{app:eq:ex_Young_tableau}
    \ytableausetup{centertableaux,boxsize=1.2em}
    \begin{ytableau} 
        1 & 3 & 7 & 6\\
        4 & 5 & \none &\none\\
        2 & \none &\none & \none 
    \end{ytableau} \;.
\end{equation}
Evidently, there are $L!$ different Young tableaux associated to a diagram. For later use, let us also define a Young tableu to be \emph{standard} if the entries in both rows and columns are increasing (e.g.\ from top to bottom and from left to right): for example, 
\begin{equation}
    \ytableausetup{boxsize=1.2em,centertableaux,notabloids}
    \begin{ytableau} 
        4 & 1 & 2 \\
        3 & 5 & \none 
    \end{ytableau}
\end{equation}
is not standard while
\begin{equation}
    \ytableausetup{boxsize=1.2em,centertableaux,notabloids}
    \begin{ytableau} 
        1 & 3 & 5 \\
        2 & 4 & \none 
    \end{ytableau}
\end{equation}
is standard. Define a \emph{Young tabloid} $\{t\}$ to be the equivalence class of Young tableaux of shape $\lambda$ under the permutation of row elements: for instance, if $\lambda = (2,1)$ and
\begin{equation}
    t = 
    \ytableausetup{boxsize=1.2em,centertableaux}
    \begin{ytableau} 
        1 & 3 \\
        2 & \none 
    \end{ytableau} \, ,
\end{equation}
then
\begin{equation}
    \{t\} = 
    \ytableausetup{boxsize=1.2em,tabloids,centertableaux}
    \begin{ytableau} 
        1 & 3 \\
        2 & \none 
    \end{ytableau}
    = 
    \ytableausetup{notabloids}
    \left\{
    \begin{ytableau} 
        1 & 3 \\
        2 & \none 
    \end{ytableau} \, ,
        \begin{ytableau} 
        3 & 1 \\
        2 & \none 
    \end{ytableau}
    \right\} \, .
\end{equation}
The graphical notation of removing the vertical bars is evocative of the permutation invariance of the rows of the tabloid. 

We are in position to introduce the \emph{permutation module} corresponding to $\lambda$: it is
\begin{equation}
    \mathcal{M}^\lambda \equiv \mathbb{C}\big[ \{t_1\}, \dots, \{t_k\} \big]
\end{equation}
where by $\mathbb{C}[v_1, \dots, v_k]$ we indicate the span $c_1 v_1 + \cdots + c_k v_k$ of the abstract objects $v_j$, considered as vectors, with complex coefficients $c_j$. Also, we assume that the list $\{t_1\}, \dots, \{t_k\}$ exhausts all possible $\lambda$-tabloids. 

It is instrumental to introduce the stabilizers of a Young tableau $t$: by definition, if $t$ has rows $R_1, \dots, R_l$ then its \emph{row stabilizer} is 
\begin{equation}
    R_t \equiv \mathrm{S}_{R_1} \times \cdots \times \mathrm{S}_{R_l},
\end{equation}
where $\mathrm{S}_Q$ is the permutation group acting on the elements of the set $Q$. Similarly, if $t$ has columns $C_1, \dots, C_m$ then its \emph{column stabilizer} is 
\begin{equation}
    C_t \equiv \mathrm{S}_{C_1} \times \cdots \times \mathrm{S}_{C_m}.
\end{equation}
An example will clarify the definition: if $t$ is given by Eq.~\eqref{app:eq:ex_Young_tableau}, then $R_t = \mathrm{S}_{\{1,3,6,7\}} \times \mathrm{S}_{\{4,5\}} \times \mathrm{S}_{\{2\}}$ and $C_t = \mathrm{S}_{\{1,2,4\}} \times \mathrm{S}_{\{3,5\}} \times \mathrm{S}_{\{7\}} \times \mathrm{S}_{\{6\}}$.

From the column stabilizer one can construct the \emph{row symmetrizer} of the tableau $t$ (of shape $\lambda$) as
\begin{equation}
    S_t \equiv \sum_{\pi \in R_t} \pi,
\end{equation}
and the \emph{column antisymmetrizer} as 
\begin{equation}
    A_t \equiv \sum_{\pi \in C_t} \mathrm{sgn}(\pi) \pi.
\end{equation}
As the name suggests, $S_t$ acts on $\mathcal{M}^\lambda$ by symmetrizing vectors according to the pattern given by the rows of $t$, and $A_t$ acts on $\mathcal{M}^\lambda$ by antisymmetrizing vectors according to the pattern given by the columns of $t$. 

We define the \emph{polytabloid} associated to $t$ as
\begin{equation}
    e_t \equiv A_t \{t\}.
\end{equation}
An example: if 
\begin{equation}
    t = 
    \ytableausetup{boxsize=1.2em,centertableaux,notabloids}
    \begin{ytableau} 
        4 & 1 & 2 \\
        3 & 5 & \none 
    \end{ytableau} \, ,
\end{equation}
then
\begin{equation}
    e_t = 
    \ytableausetup{boxsize=1.2em,centertableaux,tabloids}
    \begin{ytableau} 
        4 & 1 & 2 \\
        3 & 5 & \none 
    \end{ytableau}  -
    \begin{ytableau} 
        3 & 1 & 2 \\
        4 & 5 & \none 
    \end{ytableau}  -
    \begin{ytableau} 
        4 & 5 & 2 \\
        3 & 1 & \none 
    \end{ytableau}  +
    \begin{ytableau} 
        3 & 5 & 2 \\
        4 & 1 & \none 
    \end{ytableau} \, .
\end{equation}

At this point, we can define the irreps of $\mathrm{S}_L$, that are usually called \emph{Specht modules} and denoted as $\mathcal{S}^\lambda$: $\mathcal{S}^\lambda$ is the subspace of $\mathcal{M}^\lambda$ spanned by all the polytabloids $e_t$, with $t$ of shape $\lambda$. The central result is that $\mathcal{S}^\lambda$ form a complete list of irreps of $\mathrm{S}_L$ over $\mathbb{C}$, as $\lambda$ ranges over all the partitions of $L$.

While the description above is rigorous from the mathematical viewpoint, it might lack transparency. Let us restate the main idea: rows of a Young tableau must be symmetrized, and columns antisymmetrized, in order to find the states that compose an irrep of the permutation group. These row-symmetrized, column-antisymmetrized objects are called polytabloids in the mathematical literature; the space of all their linear combinations is the irrep, called a Specht module. In physical terms, to each polytabloid there corresponds a ket state in the Hilbert space of a permutation-symmetric system.

%--------------------------------------------------------------------------------------------------
\subsection{Irreducible representations of the special unitary group}
\label{app:sec:irreps_SU(d)}

The representation theory of the special unitary group SU$(d)$ has been object of vast research, especially in the context of high-energy physics. Here we summarize only the points relevant to us, while referring the reader to e.g.\ Refs.~\cite{Itzykson1966Unitary,Georgi2000Lie} for more details. 

The Lie group SU$(d)$ is defined as set of all $d\times d$ unitary matrices with determinant equal to one, together with the matrix multiplication as the group operation. This group plays a fundamental role in describing a $d$-level quantum system, as it represents the set of all unitary changes of basis that preserve the determinant condition. The associated Lie algebra, $\mathfrak{su}(d)$, forms a vector space over $\mathbb{C}^d$ with a dimension $d^2-1$.
In the fundamental representation, the generators of $\mathfrak{su}(d)$ are given by $d(d-1)$ off-diagonal matrices $T_{\alpha\beta}$ defined as\footnote{We reserve the hats for the operators acting on the physical Hilbert space of the quantum system.}
\begin{equation}
    \begin{split}
\left(T_{\alpha\beta}\right)_{\mu\nu}=\delta_{\alpha\mu}\delta_{\beta\nu}\ ,\qquad \alpha\neq\beta\ ,\quad \alpha,\beta=1,\dots d\ ,
    \end{split}
\end{equation}
and $d-1$ Cartan subalgebra generators 
\begin{equation}
    H_\alpha=\frac{1}{2}\left(T_{\alpha,\alpha}-T_{\alpha+1,\alpha+1}\right)\ ,\quad \alpha=1,\dots,d-1\ .
\end{equation}
The $T$-generators follow the $\mathfrak{gl}(d,\mathbb{C})$ algebra
\begin{equation}
    \comm{T_{\alpha\beta}}{T_{\mu\nu}}=\delta_{\beta\mu}T_{\alpha\nu} - \delta_{\alpha\nu}T_{\mu\beta}\ ,
\end{equation}
while the elements of the Cartan subalgebra commute $\comm{H_\alpha}{H_\beta}=0$ for all $\alpha,\beta$. Let us define the highest weight state, that all raising operators annihilate
\begin{equation}
    \label{eq:annihilation}
    T_{\alpha\beta}\ket{\mu}=0\ ,\quad \forall\ \alpha <\beta\ .
\end{equation}
Then, the highest weight state $\ket{\mu}$ in any representation $\mathcal{U}$ is a common eigenstate of the Cartan subalgebra generators
\begin{equation}
    \label{app:eq:Cartan}
    H_\alpha\ket{\mu}=h_\alpha\ket{\mu}\ .
\end{equation}
As a result, any representation $\mathcal{U}$ can be characterized by a set of $d-1$ integers $h_\alpha$.

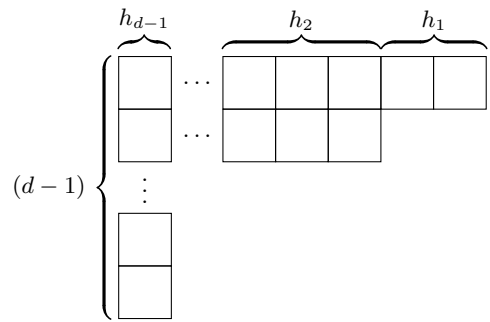
\begin{figure}
    \centering
\begin{tikzpicture}[
BC/.style = {decorate,decoration={calligraphic brace, amplitude=5pt, raise=1mm},
        very thick, pen colour={black}},
                    ]
\matrix (m) [matrix of math nodes,
             nodes={draw, minimum size=7mm, anchor=center},
             column sep=-\pgflinewidth,
             row sep=-\pgflinewidth
             ]
{
 ~~  & |[draw=none]|\dots  & ~~   &  ~~  &  ~~  &  ~~  &  ~~  \\
~~    & |[draw=none]|\dots   & ~~  &  ~~  &  ~~  &   &   \\
|[draw=none,text height=4mm]| \vdots\\
~~   &    &   &   &   &   &   \\
~~   &    &   &   &   &   &   \\
};
\draw[BC] (m-5-1.south west) -- node[left =3.0mm] {$(d-1)$} (m-1-1.north west);
\draw[BC] (m-1-1.north west) -- node[above=2.2mm] {$h_{d-1}$} (m-1-2.north west);
\draw[BC] (m-1-3.north west) -- node[above=2.2mm] {$h_2$} (m-1-6.north west);
\draw[BC] (m-1-5.north east) -- node[above=2.2mm] {$h_1$} (m-1-7.north east);
\end{tikzpicture}
    \caption{Young tableau of an example SU$(d)$ irrep characterized by a set of integers $\{h_1, h_2,\dots, h_{d-1}\}$. The value of $h_\alpha$ indicates a number of columns with exactly $\alpha$ rows.}
    \label{fig:sud}
\end{figure}

Each irrep $\mathcal{U}$ of SU$(d)$ can also be uniquely defined by a Young diagram: we denote it as $\mathcal{U}^\lambda$. The equivalence of the two definitions is encoded in the shape of the Young diagrams: the integer $h_\alpha$ characterizing the representation corresponds to $h_\alpha$ columns of the Young diagram comprising exactly $\alpha$ rows, as shown in Fig.~\ref{fig:sud}.

Often, the Young diagram representation of an irrep is more physically intuitive and, moreover, allows for an algorithmic way of combining different irreps. In the case of SU$(d)$, the permissible diagrams are constrained by the group structure. In particular, any diagram $\lambda$ can have at most $d-1$ rows. This constraint comes from the $(d-1)$-coordinate space of Cartan subalgebra generators in which at most $d-1$ coordinates can be antisymmetrized. Note the difference with the U$(d)$ representations allowing for maximum $d$ rows in the valid Young diagrams. The difference arises from the additional U$(1)$ constraint which ``removes'' the columns with $d$ rows.

%--------------------------------------------------------------------------------------------------
\subsubsection*{SU(3)}

\begin{figure}
    \centering
    \includegraphics[width=0.8\linewidth]{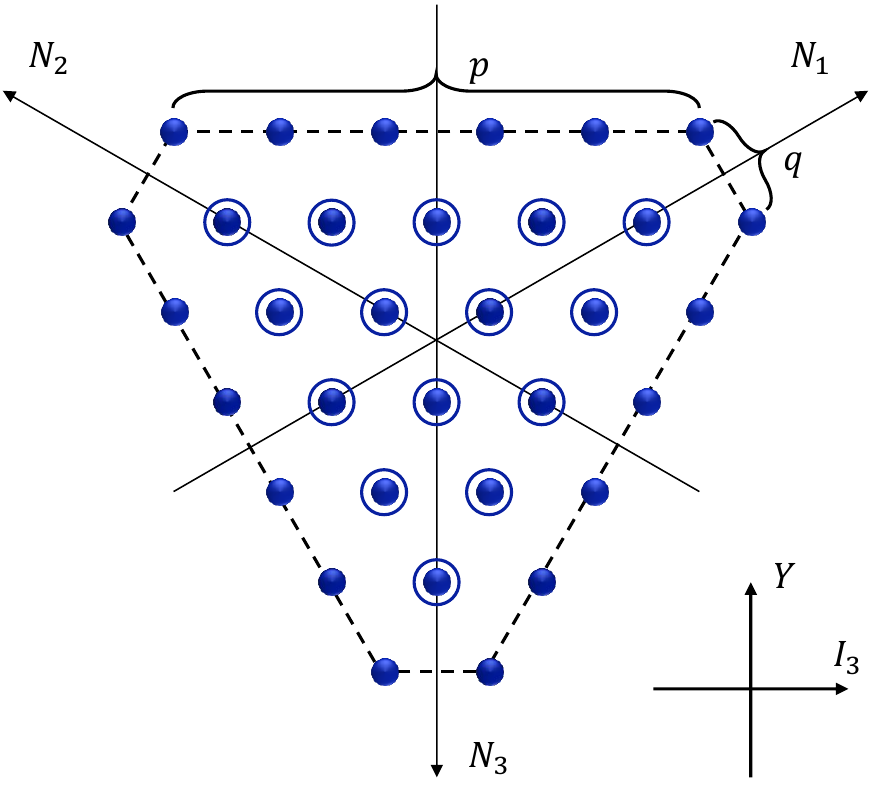}
    \caption{An example SU(3) representation for $p=5,q=1$. Each node of the diagram either represents a single state (balls) or two degenerate states (encircled balls). The representation is shown in a coordinate system of particle occupation $N_i$ of each of the three modes $i=1,2,3$. The reference isospin-hypercharge ($I_3-Y$) coordinate system is shown in the bottom right corner.}
    \label{app:fig:su3_rep}
\end{figure}

In the case of SU(3), it is customary to consider a Cartan subalgebra composed of the operators $I_3=\frac{1}{2}\left(T_{11} - T_{22}\right)$ and $Y=\frac{1}{3}\left( T_{11} + T_{22} - 2T_{33}\right)$. The SU(3) irreps are then defined by integers $(p,q)$ such that for a highest weight state $\ket{\mu}$ it holds
\begin{equation}
    I_3\ket{\mu}= \frac{p}{2}\ket{\mu}\qquad \mathrm{and}\qquad Y\ket{\mu}= \frac{p +2q}{3}\ket{\mu}\ .
\end{equation}
The highest weight state, and all the remaining states belonging to an irrep, can be represented in a coordinate system of $I_3-Y$, as shown in Fig.~\ref{app:fig:su3_rep} for $p=5, q=1$. Alternatively, the irreps can also be parametrised by the number of particles in each of the levels (also shown in Fig.~\ref{app:fig:su3_rep}), giving a more physically intuitive picture of the states involved.

%--------------------------------------------------------------------------------------------------
\subsection{Schur-Weyl duality}
\label{app:sec:Schur-Weyl}

As seen above, the representations of both the symmetric group $\mathrm{S}_L$ and the unitary group SU$(d)$ are labeled by Young diagrams. In the first case, the Young diagrams must be composed of $L$ boxes; in the second case, they must have at most $d-1$ rows.

The decomposition Eq.~\eqref{app:eq:decomposition_SU(d)_permutations} made use of Schur's lemma to combine the representations of $\mathrm{S}_L$ and SU$(d)$. A further step can be taken by noticing that SU($d$) transformations and permutations \emph{centralize each other}, i.e.\ that all the endomorphisms of $\mathcal{H}$ that commute with the unitary representations of the permutations belong to SU($d$), and vice versa. Such a mutual centralization leads to the celebrated \emph{Schur-Weyl duality}\footnote{Usually the Schur-Weyl duality is expressed for GL$(d)$ or U($d$), but for our Hamiltonian it is convenient to state it in terms of SU($d$).}:
\begin{equation}
    \label{app:eq:Schur-Weyl}
    \mathcal{H} \cong \bigoplus_{\lambda \in \mathrm{Par}(L,d)} \mathcal{S}^\lambda \otimes \mathcal{U}^{\bar\lambda} ,
\end{equation}
where the sum runs over all the partitions of $L$ into at most $d$ elements, or equivalently, the Young diagrams composed of $L$ boxes arranged in at most $d$ rows. The diagram $\bar{\lambda}$ is obtained by removing the columns of length $d$ from the diagram $\lambda$, as explained in App.~\ref{app:sec:irreps_SU(d)}. The difference wrt.\ Eq.~\eqref{app:eq:decomposition_SU(d)_permutations} is that a single index $\lambda$ is labeling the irreps of both $\mathrm{S}_L$ and SU($d$), and the multiplicities are either 0 or 1.

Equation~\eqref{app:eq:Schur-Weyl} implies that there exists a basis, sometimes called Schur basis~\cite{Bacon2005Quantum,Bacon2006Efficient,Krovi2019Efficient,Anschuetz2023Efficient}, that labels all the states in $\mathcal{H}$ as $\ket{\lambda,p_\lambda,u_\lambda} \equiv \ket{\lambda} \otimes \ket{p_\lambda} \otimes \ket{u_\lambda}$, where $\lambda$ specifies the block, permutation operators act only on $p_\lambda$ labels
\begin{equation}
    \hat{\sigma}_\pi \ket{\lambda,p_\lambda,u_\lambda} =  \ket{\lambda} \otimes \big( \hat{\sigma}_\pi \ket{p_\lambda} \big) \otimes \ket{u_\lambda},
\end{equation}
and single-site U$(d)$ operators act only on $u_\lambda$ labels: 
\begin{equation}
    \hat{\rho}_U \ket{\lambda,p_\lambda,u_\lambda} = \ket{\lambda} \otimes \ket{p_\lambda} \otimes \big(\hat{\rho}_U \ket{u_\lambda} \big).
\end{equation}
Since the Hamiltonian is composed only of products of single-site U$(d)$ operators, it follows that its sectors are specified by the labels $\lambda$ and $p_\lambda$. Full quantum chaos can be seen only at the level of $\ket{u_\lambda}$ states (once also the magnetization is taken into account).

To give a concrete example, let us consider $L=4$, i.e.\ a 4-site system. The admissible Young diagrams are
\begin{equation}
    \ytableausetup{boxsize=1.2em,centertableaux,notabloids}
    \ydiagram{4} \,, \quad
    \ydiagram{3,1} \,, \quad
    \ydiagram{2,2} \,, \quad
    \ydiagram{2,1,1} \,. 
\end{equation}
From left to right, they correspond to $(p,q,r) =(4,0,0), (2,1,0), (0,2,0)$ and $(1,0,1)$, respectively. This means that the corresponding SU(3) representations are, respectively, $D(4,0)$, $D(2,1)$, $D(0,2)$ and $D(1,0)$. Let us concentrate on the rightmost diagram. There exist 3 distinct standard Young tableaux corresponding to this diagram, and they label the irreps of $\mathrm{S}_4$:
\begin{equation}
    \label{eq:labelings}
    \ytableausetup{centertableaux} \begin{ytableau} 
    1 & 2 \\
    3 & \none\\
    4 & \none
    \end{ytableau}
    \ ,\qquad
    \begin{ytableau} 
    1 & 3 \\
    2 & \none\\
    4 & \none
    \end{ytableau}
    \ ,\qquad
    \begin{ytableau} 
    1 & 4 \\
    2 & \none\\
    3 & \none
    \end{ytableau}
    \ .  
\end{equation}
Analyzing the leftmost diagram, we follow the procedure presented in Sec.~\ref{sec:1particle}.
Firstly, define a reference state by assigning a particle of species $1$ to each block in the first row, a particle of species $2$ to each block in the second row, and a particle of species $3$ to each block in the third row. Using the vector notation in which consecutive labels correspond to the species of particles on consecutive sites, the reference state is written as $\ket{\phi}=\ket{1,1,2,3}$. The highest weight state is then appropriately symmetrised and antisymmetrised so that
\begin{align}
    \ket{\mu_1} &=A_{1,3,4}S_{1,2}\ket{\phi}\\
    &=(1-P_{1,3}-P_{1,4} - P_{3,4} \nonumber\\
    &\hspace{1cm}  + P_{1,3}P_{1,4} + P_{1,3}P_{3,4})(1+P_{1,2})\ket{\phi}\ ,
\end{align}
where $P_{a,b}\equiv(a,b)$ are two-cycles between the pairs of sites.
Explicitly, the highest-weight state is 
\begin{multline}
    \ket{\mu_1} = \ket{1,1,2,3}-\ket{1,1,3,2}-\ket{2,1,1,3}\\
    -\ket{3,1,2,1}+\ket{2,1,3,1}+\ket{3,1,1,2} .
\end{multline}

\begin{comment}
\begin{multline}
    \ket{\mu_1} = 2\ket{1,1,2,3}-2\ket{1,1,3,2}-\ket{1,2,1,3}\\
    -\ket{2,1,1,3}+\ket{1,2,3,1}+\ket{2,1,3,1}-\ket{1,3,2,1}\\
    -\ket{3,1,2,1}+\ket{1,3,1,2}+\ket{3,1,1,2} .
\end{multline}
\end{comment}

\begin{figure}
    \centering
    \includegraphics[width=0.4\linewidth]{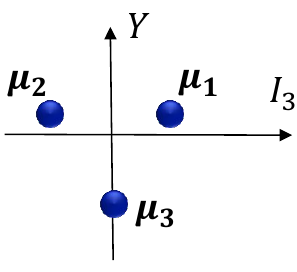}
    \caption{An example SU(3) representation for $L=4$ sites.}
    \label{fig:su3_rep_4p}
\end{figure}

The diagram analyzed corresponds to the $D(1,0)$ representation of SU(3), i.e.\ the fundamental representation. Hence, the entire representation consists of three states
\begin{equation}
    \big\{\ket{\mu_1}, \
    \ket{\mu_2}=T_{21}\ket{\mu_1}, \
    \ket{\mu_3}=T_{31}\ket{\mu_1}\big\} .
\end{equation}
Their configuration is shown in Fig.~\ref{fig:su3_rep_4p}.

%--------------------------------------------------------------------------------------------------
\subsection{Dimensions of the irreps}
\label{app:sec:dimensions}

In order to describe quantitatively the fragmentation of the Hilbert space $\mathcal{H}$ given by Eq.~\eqref{app:eq:Schur-Weyl}, it remains to quantify
\begin{enumerate}
    \item how many partitions of $L$ in at most $d$ elements there are,
    \item how large the subspaces $\mathcal{S}^\lambda$ are,
    \item how large the subspaces $\mathcal{U}^{\bar{\lambda}}$ are.
\end{enumerate}
These are known, non-trivial mathematical results, which we now summarize.

Let us start from the first point. We need the number of partitions of $L$ into at most $d$ parts, $p_d(n)$. First, we notice that if $n$ (the numbers of bosons on each site) grows at least as $L^{1/2}$, then $d \sim n^2 \gtrsim L$ and one can just use the number of \emph{unrestricted} partitions of $L$, $p(L)$. This is asymptotically given by Ramanujan's formula:
\begin{equation}
    \label{app:eq:Ramanujan}
    p(L) \sim \frac{1}{4 \sqrt{3} L} \exp( \pi \sqrt{\frac{2L}{3}}).
\end{equation}
If instead $n$ (and thus $d$) remains finite as $L$ is increased, then one needs to use $p_d(n)$. The number has been the object of numerous studies; we borrow the asymptotic formula~\cite{Agnarsson2002Sylvester,Echter2025Many}
\begin{equation}
    \frac{1}{L!} \binom{L+d-1}{L} \leq p_d(L) \leq \frac{1}{L!} \binom{L + \frac{d(d+1)}{2} - 1}{d-1}.
\end{equation}
Using Stirling's formula, the asymptotics reads
\begin{equation}
    \label{app:eq:restricted_partitions_bound}
    L^{d-1} \lesssim p_d(L) \lesssim L^d.
\end{equation}
Therefore, the total number of different Specht modules that participate in the decomposition Eq.~\eqref{app:eq:Schur-Weyl} is either a stretched exponential in $L$ if $n$ diverges as well (Eq.~\eqref{app:eq:Ramanujan}), or polynomial in $L$ if $n$ remains finite (Eq.~\eqref{app:eq:restricted_partitions_bound}).

Let us pass to point 2.\ above, i.e.\ the size of the Specht modules. The dimension of the Young diagram, $\dim \lambda$, corresponds to the number of standard Young tableaux of shape $\lambda$. This number can be found via the \emph{hook length formula}:
\begin{equation}
    \dim \lambda = \frac{L!}{\prod_{(i,j)\in \lambda} h_\lambda(i,j)}
\end{equation}
where $(i,j)$ parametrize all the boxes of the diagram, and $h_\lambda(i,j)$ is the size of the \emph{hook} starting from $(i,j)$. By definition, the hook starting from $(i,j)$ is the set of boxes that lie both to the right and below $(i,j)$, including $(i,j)$: as an example,
\begin{equation}
    \ytableausetup{centertableaux,boxsize=1.2em}
    \begin{ytableau} 
       \phantom{.} & \bullet & \times & \times\\
       \phantom{.} & \times & \none &\none\\
       \phantom{.} & \none &\none & \none 
    \end{ytableau}
\end{equation}
and the size of the hook is 4. Using the hook length formula, one can estimate the dimension of an asymptotically large Young diagram by passing to the logarithms: one finds~\cite{Cavina2024Symmetry}
\begin{equation}
    \label{app:eq:hook_asymp}
    \dim \lambda \approx e^{L s(x)}, \quad s(x) = - \sum_{i=1}^d x_i \log x_i,
\end{equation}
where $x = (\lambda_1/L, \dots, \lambda_d/L)$ is the rescaled coordinate parametrizing the length of the rows of $\lambda$. Thus, the irreps of $\mathrm{S}_L$ have a size exponentially large in $L$.

Finally, regarding point 3.,\ we recall that the dimension of the $\mathcal{U}^{\lambda}$ irrep of SU($d$) is given by the \emph{Weyl dimension formula}
\begin{equation}
    \dim \mathcal{U}^{\lambda} = \prod_{(i,j) \in \lambda} \left( 1+ \frac{\lambda_i-\lambda_j}{j-i} \right).
\end{equation}
It is interesting for our purposes to bound this number from above as~\cite{Cavina2024Symmetry}
\begin{equation}
    \dim \mathcal{U}^{\lambda} \leq L^{d(d-1)/2}.
\end{equation}
This means that, for a finite number of particles on each site, the largest dynamical sector $\mathcal{H}_{\lambda,h}$ is at most polynomial in the system size $L$, entailing a strong fragmentation. When instead $d$ is increased together with $L$, $\dim \mathcal{U}^{\lambda}$ remains exponential in $L$ for general $\lambda$, and the fragmentation is weak.

Summarizing, one can consider two separate cases. If the number $n$ of particles on each site is kept finite while $L$ is increased, then the number of different irreps, as parametrized by $\lambda$, is polynomial in $L$ (point 1.); the multiplicity of each irrep is  exponentially large in $L$ (point 2.); and the size of each sector is at most polynomial in $L$ (point 3.). If instead $n$ is increased together with $L$, then the number of different irreps is a stretched exponential in $L$ (point 1.); the multiplicity of each irrep remains exponentially large in $L$ (point 2.); and the size of each sector is exponential in $L$ (point 3.).

%--------------------------------------------------------------------------------------------------
%--------------------------------------------------------------------------------------------------
\section{Classical equations of motion}
\label{app:classical_eoms}

The classical equations of motion were computed by direct calculation and are as follows: 
\begin{widetext}
\begin{subequations}
\begin{align}
    i\der{}{t} G_{11}(t) &=i\der{}{t} G_{33}(t) = -\frac{i}{2} \der{}{t} G_{22}(t)=  g_1g_2 \left\{G_{23}(t)G_{21}(t) - G_{32}(t)G_{12}(t) \right \}\ ,\\
    i\der{}{t} G_{12}(t) &=-h G_{12}(t)+g_1\chi^+(t)\left\{G_{22}(t) - G_{11}(t)\right\} - g_2\chi^-(t)G_{13}(t)\ ,\\
    i\der{}{t} G_{23}(t) &=- h G_{23}(t) +g_2\chi^+(t)\left\{G_{33}(t) - G_{22}(t)\right\} + g_1\chi^-(t)G_{13}(t)\ ,\\
    i\der{}{t} G_{13}(t) &=-2hG_{13} (t)+\chi^+(t)\left\{g_1G_{23}(t)-g_2G_{12}(t) \right\}\ ,
\end{align}
\end{subequations}
\end{widetext}
where we defined 
\begin{equation}
    \begin{aligned}
        \chi^-(t) &\equiv g_1G^{21}(t) + g_2G^{32}(t) \ , \\
        \chi^+(t) &\equiv g_1 G^{12}(t) + g_2 G^{23}(t)
    \end{aligned}
\end{equation}
for clarity of notation. The remaining equations are obtained by Hermitian conjugation. Note that the equations of motion for the diagonal elements $G_{\alpha\alpha}$ explicitly reflect the global particle-number and magnetization conservation laws.

%--------------------------------------------------------------------------------------------------
%--------------------------------------------------------------------------------------------------
\section{Expectation values on SU(3) coherent states}\label{app:exp_vals}

We quote here the explicit formulae for the expectation values of the bosonic bilinears in the SU(3) coherent states parametrizing an irrep $D(p,q)$. The coherent states take an explicit form~\cite{Gnutzmann2000Quantum}
\begin{equation}
    \ket{\Vec{\gamma}}_{p,q} = \frac{1}{\sqrt{A_1^pA_2^q}}e^{\gamma_3 \hat{T}_{31}}e^{\gamma_1 \hat{T}_{21}}e^{\gamma_2 \hat{T}_{32}}\ket{\mu}_{p,q}\ ,
\end{equation}
where the normalization constants are
\begin{equation}
    \begin{aligned}
        A_1 &= 1 + |\gamma_1|^2 + |\gamma_3|^2\ ,\\
        A_2 &= 1 + |\gamma_2|^2 + |\gamma_3 -\gamma_1\gamma_2|^2\ . 
    \end{aligned}
\end{equation}
The objects of interest here are
\begin{equation}
    G_{\alpha\beta} = \frac{1}{L}\ev*{\hat{T}_{\alpha\beta}}{\Vec{\gamma}} = \frac{1}{L} \ev*{\sum_{j=1}^L \hat{b}^\dagger_{\alpha,j} \hat{b}_{\beta,j}^\phdagger}{\Vec{\gamma}} \ ,
\end{equation}
whose explicit expressions read~\cite{Gnutzmann2000Quantum}
\begin{widetext}
\begin{subequations}
\begin{align}
    G_{11}=&\frac{n}{3} + \frac{p}{3A_1L}\left\{2-|\gamma_1|^2-|\gamma_3|^2\right\} + \frac{q}{3A_2L}\left\{1+|\gamma_2|^2-2|\gamma_3-\gamma_1\gamma_2|^2\right\}\ ,\\
    G_{22}=&\frac{n}{3} + \frac{p}{3A_1L}\left\{-1+2|\gamma_1|^2-|\gamma_3|^2\right\} + \frac{q}{3A_2L}\left\{1-2|\gamma_2|^2+|\gamma_3-\gamma_1\gamma_2|^2\right\}\ ,\\
    G_{33}=&\frac{n}{3} + \frac{p}{3A_1L}\left\{-1-|\gamma_1|^2+2|\gamma_3|^2\right\} +\frac{q}{3A_2L}\left\{-2+|\gamma_2|^2+|\gamma_3-\gamma_1\gamma_2|^2\right\}\ ,\\
    G_{12}=&\frac{p}{A_1L}\gamma_1 - \frac{q}{A_2L}\gamma_2^*\left\{\gamma_3 - \gamma_1\gamma_2\right\}\ ,\\
    G_{13}=&\frac{p}{A_1L}\gamma_3 + \frac{q}{A_2L}\left\{\gamma_3 - \gamma_1\gamma_2\right\}\ ,\\
    G_{23}=&\frac{p}{A_1L}\gamma_1^*\gamma_3 + \frac{q}{A_2L}\gamma_2\ ,
\end{align}
\end{subequations}
\end{widetext}
where $n$ is the number of particles per site.
The remaining expectation values can be obtained via complex conjugation.

The explicit dependence of classical variables on $p/L$ and $q/L$ underlines that only representations labeled by quantum numbers that are extensive in the system size have a non-trivial classical dynamics.

%#########################################--BIBLIOGRAPHY--#########################################
\bibliography{references}

%apsrev4-2.bst 2019-01-14 (MD) hand-edited version of apsrev4-1.bst
%Control: key (0)
%Control: author (8) initials jnrlst
%Control: editor formatted (1) identically to author
%Control: production of article title (0) allowed
%Control: page (0) single
%Control: year (1) truncated
%Control: production of eprint (0) enabled
\begin{thebibliography}{73}%
\makeatletter
\providecommand \@ifxundefined [1]{%
 \@ifx{#1\undefined}
}%
\providecommand \@ifnum [1]{%
 \ifnum #1\expandafter \@firstoftwo
 \else \expandafter \@secondoftwo
 \fi
}%
\providecommand \@ifx [1]{%
 \ifx #1\expandafter \@firstoftwo
 \else \expandafter \@secondoftwo
 \fi
}%
\providecommand \natexlab [1]{#1}%
\providecommand \enquote  [1]{``#1''}%
\providecommand \bibnamefont  [1]{#1}%
\providecommand \bibfnamefont [1]{#1}%
\providecommand \citenamefont [1]{#1}%
\providecommand \href@noop [0]{\@secondoftwo}%
\providecommand \href [0]{\begingroup \@sanitize@url \@href}%
\providecommand \@href[1]{\@@startlink{#1}\@@href}%
\providecommand \@@href[1]{\endgroup#1\@@endlink}%
\providecommand \@sanitize@url [0]{\catcode `\\12\catcode `\$12\catcode
  `\&12\catcode `\#12\catcode `\^12\catcode `\_12\catcode `\%12\relax}%
\providecommand \@@startlink[1]{}%
\providecommand \@@endlink[0]{}%
\providecommand \url  [0]{\begingroup\@sanitize@url \@url }%
\providecommand \@url [1]{\endgroup\@href {#1}{\urlprefix }}%
\providecommand \urlprefix  [0]{URL }%
\providecommand \Eprint [0]{\href }%
\providecommand \doibase [0]{https://doi.org/}%
\providecommand \selectlanguage [0]{\@gobble}%
\providecommand \bibinfo  [0]{\@secondoftwo}%
\providecommand \bibfield  [0]{\@secondoftwo}%
\providecommand \translation [1]{[#1]}%
\providecommand \BibitemOpen [0]{}%
\providecommand \bibitemStop [0]{}%
\providecommand \bibitemNoStop [0]{.\EOS\space}%
\providecommand \EOS [0]{\spacefactor3000\relax}%
\providecommand \BibitemShut  [1]{\csname bibitem#1\endcsname}%
\let\auto@bib@innerbib\@empty
%</preamble>
\bibitem [{\citenamefont {Haake}(2010)}]{Haake2010Quantum}%
  \BibitemOpen
  \bibfield  {author} {\bibinfo {author} {\bibfnamefont {F.}~\bibnamefont
  {Haake}},\ }\href {https://doi.org/10.1007/978-3-642-05428-0_1} {\emph
  {\bibinfo {title} {Quantum Signatures of Chaos}}}\ (\bibinfo  {publisher}
  {Springer},\ \bibinfo {address} {Berlin, Heidelberg},\ \bibinfo {year}
  {2010})\BibitemShut {NoStop}%
\bibitem [{\citenamefont {Sciolla}\ and\ \citenamefont
  {Biroli}(2011)}]{Sciolla2011Dynamical}%
  \BibitemOpen
  \bibfield  {author} {\bibinfo {author} {\bibfnamefont {B.}~\bibnamefont
  {Sciolla}}\ and\ \bibinfo {author} {\bibfnamefont {G.}~\bibnamefont
  {Biroli}},\ }\bibfield  {title} {\bibinfo {title} {Dynamical transitions and
  quantum quenches in mean-field models},\ }\href
  {https://doi.org/10.1088/1742-5468/2011/11/P11003} {\bibfield  {journal}
  {\bibinfo  {journal} {J. Stat. Mech.: Theory Exp.}\ }\textbf {\bibinfo
  {volume} {2011}}\bibinfo  {number} { (11)},\ \bibinfo {pages}
  {P11003}}\BibitemShut {NoStop}%
\bibitem [{\citenamefont {Lewis-Swan}\ \emph {et~al.}(2019)\citenamefont
  {Lewis-Swan}, \citenamefont {Safavi-Naini}, \citenamefont {Bollinger},\ and\
  \citenamefont {Rey}}]{LewisSwan2019Unifying}%
  \BibitemOpen
\bibfield  {number} {  }\bibfield  {author} {\bibinfo {author} {\bibfnamefont
  {R.~J.}\ \bibnamefont {Lewis-Swan}}, \bibinfo {author} {\bibfnamefont
  {A.}~\bibnamefont {Safavi-Naini}}, \bibinfo {author} {\bibfnamefont {J.~J.}\
  \bibnamefont {Bollinger}},\ and\ \bibinfo {author} {\bibfnamefont {A.~M.}\
  \bibnamefont {Rey}},\ }\bibfield  {title} {\bibinfo {title} {Unifying
  scrambling, thermalization and entanglement through measurement of fidelity
  out-of-time-order correlators in the dicke model},\ }\href
  {https://doi.org/10.1038/s41467-019-09436-y} {\bibfield  {journal} {\bibinfo
  {journal} {Nat. Commun.}\ }\textbf {\bibinfo {volume} {10}},\ \bibinfo
  {pages} {1581} (\bibinfo {year} {2019})}\BibitemShut {NoStop}%
\bibitem [{\citenamefont {Walther}\ \emph {et~al.}(2006)\citenamefont
  {Walther}, \citenamefont {Varcoe}, \citenamefont {Englert},\ and\
  \citenamefont {Becker}}]{Walther2006Cavity}%
  \BibitemOpen
  \bibfield  {author} {\bibinfo {author} {\bibfnamefont {H.}~\bibnamefont
  {Walther}}, \bibinfo {author} {\bibfnamefont {B.~T.~H.}\ \bibnamefont
  {Varcoe}}, \bibinfo {author} {\bibfnamefont {B.-G.}\ \bibnamefont
  {Englert}},\ and\ \bibinfo {author} {\bibfnamefont {T.}~\bibnamefont
  {Becker}},\ }\bibfield  {title} {\bibinfo {title} {Cavity quantum
  electrodynamics},\ }\href {https://doi.org/10.1088/0034-4885/69/5/R02}
  {\bibfield  {journal} {\bibinfo  {journal} {Rep. Progr. Phys.}\ }\textbf
  {\bibinfo {volume} {69}},\ \bibinfo {pages} {1325} (\bibinfo {year}
  {2006})}\BibitemShut {NoStop}%
\bibitem [{\citenamefont {Mivehvar}\ \emph {et~al.}(2021)\citenamefont
  {Mivehvar}, \citenamefont {Piazza}, \citenamefont {Donner},\ and\
  \citenamefont {Ritsch}}]{Mivehvar2021Cavity}%
  \BibitemOpen
  \bibfield  {author} {\bibinfo {author} {\bibfnamefont {F.}~\bibnamefont
  {Mivehvar}}, \bibinfo {author} {\bibfnamefont {F.}~\bibnamefont {Piazza}},
  \bibinfo {author} {\bibfnamefont {T.}~\bibnamefont {Donner}},\ and\ \bibinfo
  {author} {\bibfnamefont {H.}~\bibnamefont {Ritsch}},\ }\bibfield  {title}
  {\bibinfo {title} {Cavity qed with quantum gases: new paradigms in many-body
  physics},\ }\href {https://doi.org/10.1080/00018732.2021.1969727} {\bibfield
  {journal} {\bibinfo  {journal} {Adv. Phys.}\ }\textbf {\bibinfo {volume}
  {70}},\ \bibinfo {pages} {1} (\bibinfo {year} {2021})}\BibitemShut {NoStop}%
\bibitem [{\citenamefont {Nu\ss{}mann}\ \emph {et~al.}(2005)\citenamefont
  {Nu\ss{}mann}, \citenamefont {Hijlkema}, \citenamefont {Weber}, \citenamefont
  {Rohde}, \citenamefont {Rempe},\ and\ \citenamefont
  {Kuhn}}]{Nussmann2005Submicron}%
  \BibitemOpen
  \bibfield  {author} {\bibinfo {author} {\bibfnamefont {S.}~\bibnamefont
  {Nu\ss{}mann}}, \bibinfo {author} {\bibfnamefont {M.}~\bibnamefont
  {Hijlkema}}, \bibinfo {author} {\bibfnamefont {B.}~\bibnamefont {Weber}},
  \bibinfo {author} {\bibfnamefont {F.}~\bibnamefont {Rohde}}, \bibinfo
  {author} {\bibfnamefont {G.}~\bibnamefont {Rempe}},\ and\ \bibinfo {author}
  {\bibfnamefont {A.}~\bibnamefont {Kuhn}},\ }\bibfield  {title} {\bibinfo
  {title} {Submicron positioning of single atoms in a microcavity},\ }\href
  {https://doi.org/10.1103/PhysRevLett.95.173602} {\bibfield  {journal}
  {\bibinfo  {journal} {Phys. Rev. Lett.}\ }\textbf {\bibinfo {volume} {95}},\
  \bibinfo {pages} {173602} (\bibinfo {year} {2005})}\BibitemShut {NoStop}%
\bibitem [{\citenamefont {Baumann}\ \emph {et~al.}(2010)\citenamefont
  {Baumann}, \citenamefont {Guerlin}, \citenamefont {Brennecke},\ and\
  \citenamefont {Esslinger}}]{Baumann2010Dicke}%
  \BibitemOpen
  \bibfield  {author} {\bibinfo {author} {\bibfnamefont {K.}~\bibnamefont
  {Baumann}}, \bibinfo {author} {\bibfnamefont {C.}~\bibnamefont {Guerlin}},
  \bibinfo {author} {\bibfnamefont {F.}~\bibnamefont {Brennecke}},\ and\
  \bibinfo {author} {\bibfnamefont {T.}~\bibnamefont {Esslinger}},\ }\bibfield
  {title} {\bibinfo {title} {Dicke quantum phase transition with a superfluid
  gas in an optical cavity},\ }\href {https://doi.org/10.1038/nature09009}
  {\bibfield  {journal} {\bibinfo  {journal} {Nature}\ }\textbf {\bibinfo
  {volume} {464}},\ \bibinfo {pages} {1301} (\bibinfo {year}
  {2010})}\BibitemShut {NoStop}%
\bibitem [{\citenamefont {Zhang}\ \emph {et~al.}(2012)\citenamefont {Zhang},
  \citenamefont {McConnell}, \citenamefont {\'{C}uk}, \citenamefont {Lin},
  \citenamefont {Schleier-Smith}, \citenamefont {Leroux},\ and\ \citenamefont
  {Vuleti\'{c}}}]{Zhang2012Collective}%
  \BibitemOpen
  \bibfield  {author} {\bibinfo {author} {\bibfnamefont {H.}~\bibnamefont
  {Zhang}}, \bibinfo {author} {\bibfnamefont {R.}~\bibnamefont {McConnell}},
  \bibinfo {author} {\bibfnamefont {S.}~\bibnamefont {\'{C}uk}}, \bibinfo
  {author} {\bibfnamefont {Q.}~\bibnamefont {Lin}}, \bibinfo {author}
  {\bibfnamefont {M.~H.}\ \bibnamefont {Schleier-Smith}}, \bibinfo {author}
  {\bibfnamefont {I.~D.}\ \bibnamefont {Leroux}},\ and\ \bibinfo {author}
  {\bibfnamefont {V.}~\bibnamefont {Vuleti\'{c}}},\ }\bibfield  {title}
  {\bibinfo {title} {Collective state measurement of mesoscopic ensembles with
  single-atom resolution},\ }\href
  {https://doi.org/10.1103/PhysRevLett.109.133603} {\bibfield  {journal}
  {\bibinfo  {journal} {Phys. Rev. Lett.}\ }\textbf {\bibinfo {volume} {109}},\
  \bibinfo {pages} {133603} (\bibinfo {year} {2012})}\BibitemShut {NoStop}%
\bibitem [{\citenamefont {Thompson}\ \emph {et~al.}(2013)\citenamefont
  {Thompson}, \citenamefont {Tiecke}, \citenamefont {de~Leon}, \citenamefont
  {Feist}, \citenamefont {Akimov}, \citenamefont {Gullans}, \citenamefont
  {Zibrov}, \citenamefont {Vuletić},\ and\ \citenamefont
  {Lukin}}]{Thompson2013Coupling}%
  \BibitemOpen
  \bibfield  {author} {\bibinfo {author} {\bibfnamefont {J.~D.}\ \bibnamefont
  {Thompson}}, \bibinfo {author} {\bibfnamefont {T.~G.}\ \bibnamefont
  {Tiecke}}, \bibinfo {author} {\bibfnamefont {N.~P.}\ \bibnamefont {de~Leon}},
  \bibinfo {author} {\bibfnamefont {J.}~\bibnamefont {Feist}}, \bibinfo
  {author} {\bibfnamefont {A.~V.}\ \bibnamefont {Akimov}}, \bibinfo {author}
  {\bibfnamefont {M.}~\bibnamefont {Gullans}}, \bibinfo {author} {\bibfnamefont
  {A.~S.}\ \bibnamefont {Zibrov}}, \bibinfo {author} {\bibfnamefont
  {V.}~\bibnamefont {Vuletić}},\ and\ \bibinfo {author} {\bibfnamefont
  {M.~D.}\ \bibnamefont {Lukin}},\ }\bibfield  {title} {\bibinfo {title}
  {Coupling a single trapped atom to a nanoscale optical cavity},\ }\href
  {https://doi.org/10.1126/science.1237125} {\bibfield  {journal} {\bibinfo
  {journal} {Science}\ }\textbf {\bibinfo {volume} {340}},\ \bibinfo {pages}
  {1202} (\bibinfo {year} {2013})}\BibitemShut {NoStop}%
\bibitem [{\citenamefont {Begley}\ \emph {et~al.}(2016)\citenamefont {Begley},
  \citenamefont {Vogt}, \citenamefont {Gulati}, \citenamefont {Takahashi},\
  and\ \citenamefont {Keller}}]{Begley2016Optimized}%
  \BibitemOpen
  \bibfield  {author} {\bibinfo {author} {\bibfnamefont {S.}~\bibnamefont
  {Begley}}, \bibinfo {author} {\bibfnamefont {M.}~\bibnamefont {Vogt}},
  \bibinfo {author} {\bibfnamefont {G.~K.}\ \bibnamefont {Gulati}}, \bibinfo
  {author} {\bibfnamefont {H.}~\bibnamefont {Takahashi}},\ and\ \bibinfo
  {author} {\bibfnamefont {M.}~\bibnamefont {Keller}},\ }\bibfield  {title}
  {\bibinfo {title} {Optimized multi-ion cavity coupling},\ }\href
  {https://doi.org/10.1103/PhysRevLett.116.223001} {\bibfield  {journal}
  {\bibinfo  {journal} {Phys. Rev. Lett.}\ }\textbf {\bibinfo {volume} {116}},\
  \bibinfo {pages} {223001} (\bibinfo {year} {2016})}\BibitemShut {NoStop}%
\bibitem [{\citenamefont {Neuzner}\ \emph {et~al.}(2016)\citenamefont
  {Neuzner}, \citenamefont {K{\"o}rber}, \citenamefont {Morin}, \citenamefont
  {Ritter},\ and\ \citenamefont {Rempe}}]{Neuzner2016Interference}%
  \BibitemOpen
  \bibfield  {author} {\bibinfo {author} {\bibfnamefont {A.}~\bibnamefont
  {Neuzner}}, \bibinfo {author} {\bibfnamefont {M.}~\bibnamefont {K{\"o}rber}},
  \bibinfo {author} {\bibfnamefont {O.}~\bibnamefont {Morin}}, \bibinfo
  {author} {\bibfnamefont {S.}~\bibnamefont {Ritter}},\ and\ \bibinfo {author}
  {\bibfnamefont {G.}~\bibnamefont {Rempe}},\ }\bibfield  {title} {\bibinfo
  {title} {Interference and dynamics of light from a distance-controlled atom
  pair in an optical cavity},\ }\href {https://doi.org/10.1038/nphoton.2016.19}
  {\bibfield  {journal} {\bibinfo  {journal} {Nat. Photonics}\ }\textbf
  {\bibinfo {volume} {10}},\ \bibinfo {pages} {303} (\bibinfo {year}
  {2016})}\BibitemShut {NoStop}%
\bibitem [{\citenamefont {Davis}\ \emph {et~al.}(2018)\citenamefont {Davis},
  \citenamefont {Wang}, \citenamefont {Safavi-Naeini},\ and\ \citenamefont
  {Schleier-Smith}}]{Davis2018Painting}%
  \BibitemOpen
  \bibfield  {author} {\bibinfo {author} {\bibfnamefont {E.~J.}\ \bibnamefont
  {Davis}}, \bibinfo {author} {\bibfnamefont {Z.}~\bibnamefont {Wang}},
  \bibinfo {author} {\bibfnamefont {A.~H.}\ \bibnamefont {Safavi-Naeini}},\
  and\ \bibinfo {author} {\bibfnamefont {M.~H.}\ \bibnamefont
  {Schleier-Smith}},\ }\bibfield  {title} {\bibinfo {title} {Painting
  nonclassical states of spin or motion with shaped single photons},\ }\href
  {https://doi.org/10.1103/PhysRevLett.121.123602} {\bibfield  {journal}
  {\bibinfo  {journal} {Phys. Rev. Lett.}\ }\textbf {\bibinfo {volume} {121}},\
  \bibinfo {pages} {123602} (\bibinfo {year} {2018})}\BibitemShut {NoStop}%
\bibitem [{\citenamefont {Norcia}\ \emph {et~al.}(2018)\citenamefont {Norcia},
  \citenamefont {Lewis-Swan}, \citenamefont {Cline}, \citenamefont {Zhu},
  \citenamefont {Rey},\ and\ \citenamefont {Thompson}}]{Norcia2018Cavity}%
  \BibitemOpen
  \bibfield  {author} {\bibinfo {author} {\bibfnamefont {M.~A.}\ \bibnamefont
  {Norcia}}, \bibinfo {author} {\bibfnamefont {R.~J.}\ \bibnamefont
  {Lewis-Swan}}, \bibinfo {author} {\bibfnamefont {J.~R.~K.}\ \bibnamefont
  {Cline}}, \bibinfo {author} {\bibfnamefont {B.}~\bibnamefont {Zhu}}, \bibinfo
  {author} {\bibfnamefont {A.~M.}\ \bibnamefont {Rey}},\ and\ \bibinfo {author}
  {\bibfnamefont {J.~K.}\ \bibnamefont {Thompson}},\ }\bibfield  {title}
  {\bibinfo {title} {Cavity-mediated collective spin-exchange interactions in a
  strontium superradiant laser},\ }\href
  {https://doi.org/10.1126/science.aar3102} {\bibfield  {journal} {\bibinfo
  {journal} {Science}\ }\textbf {\bibinfo {volume} {361}},\ \bibinfo {pages}
  {259} (\bibinfo {year} {2018})}\BibitemShut {NoStop}%
\bibitem [{\citenamefont {Davis}\ \emph {et~al.}(2019)\citenamefont {Davis},
  \citenamefont {Bentsen}, \citenamefont {Homeier}, \citenamefont {Li},\ and\
  \citenamefont {Schleier-Smith}}]{Davis2019Photon}%
  \BibitemOpen
  \bibfield  {author} {\bibinfo {author} {\bibfnamefont {E.~J.}\ \bibnamefont
  {Davis}}, \bibinfo {author} {\bibfnamefont {G.}~\bibnamefont {Bentsen}},
  \bibinfo {author} {\bibfnamefont {L.}~\bibnamefont {Homeier}}, \bibinfo
  {author} {\bibfnamefont {T.}~\bibnamefont {Li}},\ and\ \bibinfo {author}
  {\bibfnamefont {M.~H.}\ \bibnamefont {Schleier-Smith}},\ }\bibfield  {title}
  {\bibinfo {title} {{Photon-Mediated Spin-Exchange Dynamics of Spin-1
  Atoms}},\ }\href {https://doi.org/10.1103/PhysRevLett.122.010405} {\bibfield
  {journal} {\bibinfo  {journal} {Phys. Rev. Lett.}\ }\textbf {\bibinfo
  {volume} {122}},\ \bibinfo {pages} {010405} (\bibinfo {year}
  {2019})}\BibitemShut {NoStop}%
\bibitem [{\citenamefont {Davis}\ \emph {et~al.}(2020)\citenamefont {Davis},
  \citenamefont {Periwal}, \citenamefont {Cooper}, \citenamefont {Bentsen},
  \citenamefont {Evered}, \citenamefont {Van~Kirk},\ and\ \citenamefont
  {Schleier-Smith}}]{Davis2020Protecting}%
  \BibitemOpen
  \bibfield  {author} {\bibinfo {author} {\bibfnamefont {E.~J.}\ \bibnamefont
  {Davis}}, \bibinfo {author} {\bibfnamefont {A.}~\bibnamefont {Periwal}},
  \bibinfo {author} {\bibfnamefont {E.~S.}\ \bibnamefont {Cooper}}, \bibinfo
  {author} {\bibfnamefont {G.}~\bibnamefont {Bentsen}}, \bibinfo {author}
  {\bibfnamefont {S.~J.}\ \bibnamefont {Evered}}, \bibinfo {author}
  {\bibfnamefont {K.}~\bibnamefont {Van~Kirk}},\ and\ \bibinfo {author}
  {\bibfnamefont {M.~H.}\ \bibnamefont {Schleier-Smith}},\ }\bibfield  {title}
  {\bibinfo {title} {Protecting spin coherence in a tunable heisenberg model},\
  }\href {https://doi.org/10.1103/PhysRevLett.125.060402} {\bibfield  {journal}
  {\bibinfo  {journal} {Phys. Rev. Lett.}\ }\textbf {\bibinfo {volume} {125}},\
  \bibinfo {pages} {060402} (\bibinfo {year} {2020})}\BibitemShut {NoStop}%
\bibitem [{\citenamefont {Yan}\ \emph {et~al.}(2023)\citenamefont {Yan},
  \citenamefont {Ho}, \citenamefont {Lu}, \citenamefont {Masson}, \citenamefont
  {Asenjo-Garcia},\ and\ \citenamefont {Stamper-Kurn}}]{Yan2023Superradiant}%
  \BibitemOpen
  \bibfield  {author} {\bibinfo {author} {\bibfnamefont {Z.}~\bibnamefont
  {Yan}}, \bibinfo {author} {\bibfnamefont {J.}~\bibnamefont {Ho}}, \bibinfo
  {author} {\bibfnamefont {Y.-H.}\ \bibnamefont {Lu}}, \bibinfo {author}
  {\bibfnamefont {S.~J.}\ \bibnamefont {Masson}}, \bibinfo {author}
  {\bibfnamefont {A.}~\bibnamefont {Asenjo-Garcia}},\ and\ \bibinfo {author}
  {\bibfnamefont {D.~M.}\ \bibnamefont {Stamper-Kurn}},\ }\bibfield  {title}
  {\bibinfo {title} {Superradiant and subradiant cavity scattering by atom
  arrays},\ }\href {https://doi.org/10.1103/PhysRevLett.131.253603} {\bibfield
  {journal} {\bibinfo  {journal} {Phys. Rev. Lett.}\ }\textbf {\bibinfo
  {volume} {131}},\ \bibinfo {pages} {253603} (\bibinfo {year}
  {2023})}\BibitemShut {NoStop}%
\bibitem [{\citenamefont {Sauerwein}\ \emph {et~al.}(2023)\citenamefont
  {Sauerwein}, \citenamefont {Orsi}, \citenamefont {Uhrich}, \citenamefont
  {Bandyopadhyay}, \citenamefont {Mattiotti}, \citenamefont {Cantat-Moltrecht},
  \citenamefont {Pupillo}, \citenamefont {Hauke},\ and\ \citenamefont
  {Brantut}}]{Sauerwein2023Engineering}%
  \BibitemOpen
  \bibfield  {author} {\bibinfo {author} {\bibfnamefont {N.}~\bibnamefont
  {Sauerwein}}, \bibinfo {author} {\bibfnamefont {F.}~\bibnamefont {Orsi}},
  \bibinfo {author} {\bibfnamefont {P.}~\bibnamefont {Uhrich}}, \bibinfo
  {author} {\bibfnamefont {S.}~\bibnamefont {Bandyopadhyay}}, \bibinfo {author}
  {\bibfnamefont {F.}~\bibnamefont {Mattiotti}}, \bibinfo {author}
  {\bibfnamefont {T.}~\bibnamefont {Cantat-Moltrecht}}, \bibinfo {author}
  {\bibfnamefont {G.}~\bibnamefont {Pupillo}}, \bibinfo {author} {\bibfnamefont
  {P.}~\bibnamefont {Hauke}},\ and\ \bibinfo {author} {\bibfnamefont {J.-P.}\
  \bibnamefont {Brantut}},\ }\bibfield  {title} {\bibinfo {title} {Engineering
  random spin models with atoms in a high-finesse cavity},\ }\href
  {https://doi.org/10.1038/s41567-023-02033-3} {\bibfield  {journal} {\bibinfo
  {journal} {Nat. Phys.}\ }\textbf {\bibinfo {volume} {19}},\ \bibinfo {pages}
  {1128} (\bibinfo {year} {2023})}\BibitemShut {NoStop}%
\bibitem [{\citenamefont {Orsi}\ \emph {et~al.}(2024)\citenamefont {Orsi},
  \citenamefont {Sauerwein}, \citenamefont {Bhatt}, \citenamefont {Faltinath},
  \citenamefont {Fedotova}, \citenamefont {Reiter}, \citenamefont
  {Cantat-Moltrecht},\ and\ \citenamefont {Brantut}}]{Orsi2024Cavity}%
  \BibitemOpen
  \bibfield  {author} {\bibinfo {author} {\bibfnamefont {F.}~\bibnamefont
  {Orsi}}, \bibinfo {author} {\bibfnamefont {N.}~\bibnamefont {Sauerwein}},
  \bibinfo {author} {\bibfnamefont {R.~P.}\ \bibnamefont {Bhatt}}, \bibinfo
  {author} {\bibfnamefont {J.}~\bibnamefont {Faltinath}}, \bibinfo {author}
  {\bibfnamefont {E.}~\bibnamefont {Fedotova}}, \bibinfo {author}
  {\bibfnamefont {N.}~\bibnamefont {Reiter}}, \bibinfo {author} {\bibfnamefont
  {T.}~\bibnamefont {Cantat-Moltrecht}},\ and\ \bibinfo {author} {\bibfnamefont
  {J.-P.}\ \bibnamefont {Brantut}},\ }\bibfield  {title} {\bibinfo {title}
  {Cavity microscope for micrometer-scale control of atom-photon
  interactions},\ }\href {https://doi.org/10.1103/PRXQuantum.5.040333}
  {\bibfield  {journal} {\bibinfo  {journal} {PRX Quantum}\ }\textbf {\bibinfo
  {volume} {5}},\ \bibinfo {pages} {040333} (\bibinfo {year}
  {2024})}\BibitemShut {NoStop}%
\bibitem [{\citenamefont {Leroux}\ \emph {et~al.}(2010)\citenamefont {Leroux},
  \citenamefont {Schleier-Smith},\ and\ \citenamefont
  {Vuleti\'{c}}}]{Leroux2010Implementation}%
  \BibitemOpen
  \bibfield  {author} {\bibinfo {author} {\bibfnamefont {I.~D.}\ \bibnamefont
  {Leroux}}, \bibinfo {author} {\bibfnamefont {M.~H.}\ \bibnamefont
  {Schleier-Smith}},\ and\ \bibinfo {author} {\bibfnamefont {V.}~\bibnamefont
  {Vuleti\'{c}}},\ }\bibfield  {title} {\bibinfo {title} {Implementation of
  cavity squeezing of a collective atomic spin},\ }\href
  {https://doi.org/10.1103/PhysRevLett.104.073602} {\bibfield  {journal}
  {\bibinfo  {journal} {Phys. Rev. Lett.}\ }\textbf {\bibinfo {volume} {104}},\
  \bibinfo {pages} {073602} (\bibinfo {year} {2010})}\BibitemShut {NoStop}%
\bibitem [{\citenamefont {Emary}\ and\ \citenamefont
  {Brandes}(2003{\natexlab{a}})}]{Emary2003Quantum}%
  \BibitemOpen
  \bibfield  {author} {\bibinfo {author} {\bibfnamefont {C.}~\bibnamefont
  {Emary}}\ and\ \bibinfo {author} {\bibfnamefont {T.}~\bibnamefont
  {Brandes}},\ }\bibfield  {title} {\bibinfo {title} {Quantum chaos triggered
  by precursors of a quantum phase transition: The dicke model},\ }\href
  {https://doi.org/10.1103/PhysRevLett.90.044101} {\bibfield  {journal}
  {\bibinfo  {journal} {Phys. Rev. Lett.}\ }\textbf {\bibinfo {volume} {90}},\
  \bibinfo {pages} {044101} (\bibinfo {year} {2003}{\natexlab{a}})}\BibitemShut
  {NoStop}%
\bibitem [{\citenamefont {Emary}\ and\ \citenamefont
  {Brandes}(2003{\natexlab{b}})}]{Emary2003Chaos}%
  \BibitemOpen
  \bibfield  {author} {\bibinfo {author} {\bibfnamefont {C.}~\bibnamefont
  {Emary}}\ and\ \bibinfo {author} {\bibfnamefont {T.}~\bibnamefont
  {Brandes}},\ }\bibfield  {title} {\bibinfo {title} {Chaos and the quantum
  phase transition in the dicke model},\ }\href
  {https://doi.org/10.1103/PhysRevE.67.066203} {\bibfield  {journal} {\bibinfo
  {journal} {Phys. Rev. E}\ }\textbf {\bibinfo {volume} {67}},\ \bibinfo
  {pages} {066203} (\bibinfo {year} {2003}{\natexlab{b}})}\BibitemShut
  {NoStop}%
\bibitem [{\citenamefont {Bogoliubov}\ \emph {et~al.}(1996)\citenamefont
  {Bogoliubov}, \citenamefont {Bullough},\ and\ \citenamefont
  {Timonen}}]{Bogoliubov1996Exact}%
  \BibitemOpen
  \bibfield  {author} {\bibinfo {author} {\bibfnamefont {N.~M.}\ \bibnamefont
  {Bogoliubov}}, \bibinfo {author} {\bibfnamefont {R.~K.}\ \bibnamefont
  {Bullough}},\ and\ \bibinfo {author} {\bibfnamefont {J.}~\bibnamefont
  {Timonen}},\ }\bibfield  {title} {\bibinfo {title} {Exact solution of
  generalized tavis - cummings models in quantum optics},\ }\href
  {https://doi.org/10.1088/0305-4470/29/19/015} {\bibfield  {journal} {\bibinfo
   {journal} {J. Phys. A}\ }\textbf {\bibinfo {volume} {29}},\ \bibinfo {pages}
  {6305} (\bibinfo {year} {1996})}\BibitemShut {NoStop}%
\bibitem [{\citenamefont {Ziolkowska}\ \emph {et~al.}(tion)\citenamefont
  {Ziolkowska}, \citenamefont {Hosseinabadi}, \citenamefont {Marino},\ and\
  \citenamefont {Mikheev}}]{my2PI}%
  \BibitemOpen
  \bibfield  {author} {\bibinfo {author} {\bibfnamefont {A.~A.}\ \bibnamefont
  {Ziolkowska}}, \bibinfo {author} {\bibfnamefont {H.}~\bibnamefont
  {Hosseinabadi}}, \bibinfo {author} {\bibfnamefont {J.}~\bibnamefont
  {Marino}},\ and\ \bibinfo {author} {\bibfnamefont {A.~N.}\ \bibnamefont
  {Mikheev}},\ }\href@noop {} {\bibinfo {title} {Quantum fluctuations and chaos
  in fully connected spin models: A {2PI} approach}} (\bibinfo {year} {in
  preparation})\BibitemShut {NoStop}%
\bibitem [{\citenamefont {Qi}\ \emph {et~al.}(2023)\citenamefont {Qi},
  \citenamefont {Scaffidi},\ and\ \citenamefont {Cao}}]{Qi2023Surprises}%
  \BibitemOpen
  \bibfield  {author} {\bibinfo {author} {\bibfnamefont {Z.}~\bibnamefont
  {Qi}}, \bibinfo {author} {\bibfnamefont {T.}~\bibnamefont {Scaffidi}},\ and\
  \bibinfo {author} {\bibfnamefont {X.}~\bibnamefont {Cao}},\ }\bibfield
  {title} {\bibinfo {title} {Surprises in the deep hilbert space of all-to-all
  systems: From superexponential scrambling to slow entanglement growth},\
  }\href {https://doi.org/10.1103/PhysRevB.108.054301} {\bibfield  {journal}
  {\bibinfo  {journal} {Phys. Rev. B}\ }\textbf {\bibinfo {volume} {108}},\
  \bibinfo {pages} {054301} (\bibinfo {year} {2023})}\BibitemShut {NoStop}%
\bibitem [{\citenamefont {Pi\~neiro Orioli}\ \emph {et~al.}(2022)\citenamefont
  {Pi\~neiro Orioli}, \citenamefont {Thompson},\ and\ \citenamefont
  {Rey}}]{Pineiro2022Emergent}%
  \BibitemOpen
  \bibfield  {author} {\bibinfo {author} {\bibfnamefont {A.}~\bibnamefont
  {Pi\~neiro Orioli}}, \bibinfo {author} {\bibfnamefont {J.~K.}\ \bibnamefont
  {Thompson}},\ and\ \bibinfo {author} {\bibfnamefont {A.~M.}\ \bibnamefont
  {Rey}},\ }\bibfield  {title} {\bibinfo {title} {Emergent dark states from
  superradiant dynamics in multilevel atoms in a cavity},\ }\href
  {https://doi.org/10.1103/PhysRevX.12.011054} {\bibfield  {journal} {\bibinfo
  {journal} {Phys. Rev. X}\ }\textbf {\bibinfo {volume} {12}},\ \bibinfo
  {pages} {011054} (\bibinfo {year} {2022})}\BibitemShut {NoStop}%
\bibitem [{\citenamefont {Chu}\ \emph {et~al.}(2023)\citenamefont {Chu},
  \citenamefont {Orioli}, \citenamefont {Barberena}, \citenamefont {Thompson},\
  and\ \citenamefont {Rey}}]{Chu2023Photon}%
  \BibitemOpen
  \bibfield  {author} {\bibinfo {author} {\bibfnamefont {A.}~\bibnamefont
  {Chu}}, \bibinfo {author} {\bibfnamefont {A.~P.~n.}\ \bibnamefont {Orioli}},
  \bibinfo {author} {\bibfnamefont {D.}~\bibnamefont {Barberena}}, \bibinfo
  {author} {\bibfnamefont {J.~K.}\ \bibnamefont {Thompson}},\ and\ \bibinfo
  {author} {\bibfnamefont {A.~M.}\ \bibnamefont {Rey}},\ }\bibfield  {title}
  {\bibinfo {title} {Photon-mediated correlated hopping in a synthetic
  ladder},\ }\href {https://doi.org/10.1103/PhysRevResearch.5.L022034}
  {\bibfield  {journal} {\bibinfo  {journal} {Phys. Rev. Res.}\ }\textbf
  {\bibinfo {volume} {5}},\ \bibinfo {pages} {L022034} (\bibinfo {year}
  {2023})}\BibitemShut {NoStop}%
\bibitem [{\citenamefont {Valencia-Tortora}\ \emph {et~al.}(2023)\citenamefont
  {Valencia-Tortora}, \citenamefont {Kelly}, \citenamefont {Donner},
  \citenamefont {Morigi}, \citenamefont {Fazio},\ and\ \citenamefont
  {Marino}}]{ValenciaTortora2023Crafting}%
  \BibitemOpen
  \bibfield  {author} {\bibinfo {author} {\bibfnamefont {R.~J.}\ \bibnamefont
  {Valencia-Tortora}}, \bibinfo {author} {\bibfnamefont {S.~P.}\ \bibnamefont
  {Kelly}}, \bibinfo {author} {\bibfnamefont {T.}~\bibnamefont {Donner}},
  \bibinfo {author} {\bibfnamefont {G.}~\bibnamefont {Morigi}}, \bibinfo
  {author} {\bibfnamefont {R.}~\bibnamefont {Fazio}},\ and\ \bibinfo {author}
  {\bibfnamefont {J.}~\bibnamefont {Marino}},\ }\bibfield  {title} {\bibinfo
  {title} {{Crafting the dynamical structure of synchronization by harnessing
  bosonic multilevel cavity QED}},\ }\href
  {https://doi.org/10.1103/PhysRevResearch.5.023112} {\bibfield  {journal}
  {\bibinfo  {journal} {Phys. Rev. Res.}\ }\textbf {\bibinfo {volume} {5}},\
  \bibinfo {pages} {023112} (\bibinfo {year} {2023})}\BibitemShut {NoStop}%
\bibitem [{\citenamefont {Fulton}\ and\ \citenamefont
  {Harris}(1991)}]{Fulton1991Representation}%
  \BibitemOpen
  \bibfield  {author} {\bibinfo {author} {\bibfnamefont {W.}~\bibnamefont
  {Fulton}}\ and\ \bibinfo {author} {\bibfnamefont {J.}~\bibnamefont
  {Harris}},\ }\href {https://doi.org/10.1007/978-1-4612-0979-9} {\emph
  {\bibinfo {title} {{Representation Theory: A First Course}}}},\ Graduate
  Texts in Mathematics\ (\bibinfo  {publisher} {Springer},\ \bibinfo {year}
  {1991})\BibitemShut {NoStop}%
\bibitem [{\citenamefont {Sala}\ \emph {et~al.}(2020)\citenamefont {Sala},
  \citenamefont {Rakovszky}, \citenamefont {Verresen}, \citenamefont {Knap},\
  and\ \citenamefont {Pollmann}}]{Sala2020Ergodicity}%
  \BibitemOpen
  \bibfield  {author} {\bibinfo {author} {\bibfnamefont {P.}~\bibnamefont
  {Sala}}, \bibinfo {author} {\bibfnamefont {T.}~\bibnamefont {Rakovszky}},
  \bibinfo {author} {\bibfnamefont {R.}~\bibnamefont {Verresen}}, \bibinfo
  {author} {\bibfnamefont {M.}~\bibnamefont {Knap}},\ and\ \bibinfo {author}
  {\bibfnamefont {F.}~\bibnamefont {Pollmann}},\ }\bibfield  {title} {\bibinfo
  {title} {{Ergodicity Breaking Arising from Hilbert Space Fragmentation in
  Dipole-Conserving Hamiltonians}},\ }\href
  {https://doi.org/10.1103/PhysRevX.10.011047} {\bibfield  {journal} {\bibinfo
  {journal} {Phys. Rev. X}\ }\textbf {\bibinfo {volume} {10}},\ \bibinfo
  {pages} {011047} (\bibinfo {year} {2020})}\BibitemShut {NoStop}%
\bibitem [{\citenamefont {Moudgalya}\ \emph {et~al.}(2022)\citenamefont
  {Moudgalya}, \citenamefont {Bernevig},\ and\ \citenamefont
  {Regnault}}]{Moudgalya2022Quantum}%
  \BibitemOpen
  \bibfield  {author} {\bibinfo {author} {\bibfnamefont {S.}~\bibnamefont
  {Moudgalya}}, \bibinfo {author} {\bibfnamefont {B.~A.}\ \bibnamefont
  {Bernevig}},\ and\ \bibinfo {author} {\bibfnamefont {N.}~\bibnamefont
  {Regnault}},\ }\bibfield  {title} {\bibinfo {title} {{Quantum many-body scars
  and Hilbert space fragmentation: a review of exact results}},\ }\href
  {https://doi.org/10.1088/1361-6633/ac73a0} {\bibfield  {journal} {\bibinfo
  {journal} {Rep. Progr. Phys.}\ }\textbf {\bibinfo {volume} {85}},\ \bibinfo
  {pages} {086501} (\bibinfo {year} {2022})}\BibitemShut {NoStop}%
\bibitem [{\citenamefont {Gnutzmann}\ \emph {et~al.}(2000)\citenamefont
  {Gnutzmann}, \citenamefont {Haake},\ and\ \citenamefont
  {Kus}}]{Gnutzmann2000Quantum}%
  \BibitemOpen
  \bibfield  {author} {\bibinfo {author} {\bibfnamefont {S.}~\bibnamefont
  {Gnutzmann}}, \bibinfo {author} {\bibfnamefont {F.}~\bibnamefont {Haake}},\
  and\ \bibinfo {author} {\bibfnamefont {M.}~\bibnamefont {Kus}},\ }\bibfield
  {title} {\bibinfo {title} {{Quantum chaos of SU3 observables}},\ }\href
  {https://doi.org/10.1088/0305-4470/33/1/309} {\bibfield  {journal} {\bibinfo
  {journal} {Journal of Physics A: Mathematical and General}\ }\textbf
  {\bibinfo {volume} {33}},\ \bibinfo {pages} {143} (\bibinfo {year}
  {2000})}\BibitemShut {NoStop}%
\bibitem [{\citenamefont {Moudgalya}\ and\ \citenamefont
  {Motrunich}(2022)}]{Moudgalya2022Hilbert}%
  \BibitemOpen
  \bibfield  {author} {\bibinfo {author} {\bibfnamefont {S.}~\bibnamefont
  {Moudgalya}}\ and\ \bibinfo {author} {\bibfnamefont {O.~I.}\ \bibnamefont
  {Motrunich}},\ }\bibfield  {title} {\bibinfo {title} {{Hilbert Space
  Fragmentation and Commutant Algebras}},\ }\href
  {https://doi.org/10.1103/PhysRevX.12.011050} {\bibfield  {journal} {\bibinfo
  {journal} {Phys. Rev. X}\ }\textbf {\bibinfo {volume} {12}},\ \bibinfo
  {pages} {011050} (\bibinfo {year} {2022})}\BibitemShut {NoStop}%
\bibitem [{\citenamefont {Li}\ \emph {et~al.}(2025)\citenamefont {Li},
  \citenamefont {Pollmann}, \citenamefont {Read},\ and\ \citenamefont
  {Sala}}]{Li2025Highly}%
  \BibitemOpen
  \bibfield  {author} {\bibinfo {author} {\bibfnamefont {Y.}~\bibnamefont
  {Li}}, \bibinfo {author} {\bibfnamefont {F.}~\bibnamefont {Pollmann}},
  \bibinfo {author} {\bibfnamefont {N.}~\bibnamefont {Read}},\ and\ \bibinfo
  {author} {\bibfnamefont {P.}~\bibnamefont {Sala}},\ }\bibfield  {title}
  {\bibinfo {title} {Highly entangled stationary states from strong
  symmetries},\ }\href {https://doi.org/10.1103/PhysRevX.15.011068} {\bibfield
  {journal} {\bibinfo  {journal} {Phys. Rev. X}\ }\textbf {\bibinfo {volume}
  {15}},\ \bibinfo {pages} {011068} (\bibinfo {year} {2025})}\BibitemShut
  {NoStop}%
\bibitem [{\citenamefont {Tavis}\ and\ \citenamefont
  {Cummings}(1968)}]{Tavis1968Exact}%
  \BibitemOpen
  \bibfield  {author} {\bibinfo {author} {\bibfnamefont {M.}~\bibnamefont
  {Tavis}}\ and\ \bibinfo {author} {\bibfnamefont {F.~W.}\ \bibnamefont
  {Cummings}},\ }\bibfield  {title} {\bibinfo {title} {Exact solution for an
  $n$-molecule---radiation-field hamiltonian},\ }\href
  {https://doi.org/10.1103/PhysRev.170.379} {\bibfield  {journal} {\bibinfo
  {journal} {Phys. Rev.}\ }\textbf {\bibinfo {volume} {170}},\ \bibinfo {pages}
  {379} (\bibinfo {year} {1968})}\BibitemShut {NoStop}%
\bibitem [{\citenamefont {Dicke}(1954)}]{Dicke1054Coherence}%
  \BibitemOpen
  \bibfield  {author} {\bibinfo {author} {\bibfnamefont {R.~H.}\ \bibnamefont
  {Dicke}},\ }\bibfield  {title} {\bibinfo {title} {Coherence in spontaneous
  radiation processes},\ }\href {https://doi.org/10.1103/PhysRev.93.99}
  {\bibfield  {journal} {\bibinfo  {journal} {Phys. Rev.}\ }\textbf {\bibinfo
  {volume} {93}},\ \bibinfo {pages} {99} (\bibinfo {year} {1954})}\BibitemShut
  {NoStop}%
\bibitem [{\citenamefont {Kirton}\ \emph {et~al.}(2019)\citenamefont {Kirton},
  \citenamefont {Roses}, \citenamefont {Keeling},\ and\ \citenamefont
  {Dalla~Torre}}]{Kirton2019Introduction}%
  \BibitemOpen
  \bibfield  {author} {\bibinfo {author} {\bibfnamefont {P.}~\bibnamefont
  {Kirton}}, \bibinfo {author} {\bibfnamefont {M.~M.}\ \bibnamefont {Roses}},
  \bibinfo {author} {\bibfnamefont {J.}~\bibnamefont {Keeling}},\ and\ \bibinfo
  {author} {\bibfnamefont {E.~G.}\ \bibnamefont {Dalla~Torre}},\ }\bibfield
  {title} {\bibinfo {title} {Introduction to the dicke model: From equilibrium
  to nonequilibrium, and vice versa},\ }\href
  {https://doi.org/10.1002/qute.201800043} {\bibfield  {journal} {\bibinfo
  {journal} {Adv. Quantum Tech.}\ }\textbf {\bibinfo {volume} {2}},\ \bibinfo
  {pages} {1800043} (\bibinfo {year} {2019})}\BibitemShut {NoStop}%
\bibitem [{\citenamefont {Meshkov}\ \emph {et~al.}(1965)\citenamefont
  {Meshkov}, \citenamefont {Glick},\ and\ \citenamefont
  {Lipkin}}]{Meshkov1965Validity}%
  \BibitemOpen
  \bibfield  {author} {\bibinfo {author} {\bibfnamefont {N.}~\bibnamefont
  {Meshkov}}, \bibinfo {author} {\bibfnamefont {A.}~\bibnamefont {Glick}},\
  and\ \bibinfo {author} {\bibfnamefont {H.}~\bibnamefont {Lipkin}},\
  }\bibfield  {title} {\bibinfo {title} {Validity of many-body approximation
  methods for a solvable model: (ii). linearization procedures},\ }\href
  {https://doi.org/10.1016/0029-5582(65)90863-1} {\bibfield  {journal}
  {\bibinfo  {journal} {Nucl. Phys.}\ }\textbf {\bibinfo {volume} {62}},\
  \bibinfo {pages} {199} (\bibinfo {year} {1965})}\BibitemShut {NoStop}%
\bibitem [{\citenamefont {Schuckert}\ \emph {et~al.}(2018)\citenamefont
  {Schuckert}, \citenamefont {Pi\~neiro Orioli},\ and\ \citenamefont
  {Berges}}]{Schuckert2018Nonequilibrium}%
  \BibitemOpen
  \bibfield  {author} {\bibinfo {author} {\bibfnamefont {A.}~\bibnamefont
  {Schuckert}}, \bibinfo {author} {\bibfnamefont {A.}~\bibnamefont {Pi\~neiro
  Orioli}},\ and\ \bibinfo {author} {\bibfnamefont {J.}~\bibnamefont
  {Berges}},\ }\bibfield  {title} {\bibinfo {title} {Nonequilibrium quantum
  spin dynamics from two-particle irreducible functional integral techniques in
  the schwinger boson representation},\ }\href
  {https://doi.org/10.1103/PhysRevB.98.224304} {\bibfield  {journal} {\bibinfo
  {journal} {Phys. Rev. B}\ }\textbf {\bibinfo {volume} {98}},\ \bibinfo
  {pages} {224304} (\bibinfo {year} {2018})}\BibitemShut {NoStop}%
\bibitem [{\citenamefont {Halimeh}\ \emph {et~al.}(2018)\citenamefont
  {Halimeh}, \citenamefont {Punk},\ and\ \citenamefont
  {Piazza}}]{Halimeh2018Aging}%
  \BibitemOpen
  \bibfield  {author} {\bibinfo {author} {\bibfnamefont {J.~C.}\ \bibnamefont
  {Halimeh}}, \bibinfo {author} {\bibfnamefont {M.}~\bibnamefont {Punk}},\ and\
  \bibinfo {author} {\bibfnamefont {F.}~\bibnamefont {Piazza}},\ }\bibfield
  {title} {\bibinfo {title} {Aging dynamics in quenched noisy long-range
  quantum ising models},\ }\href {https://doi.org/10.1103/PhysRevB.98.045111}
  {\bibfield  {journal} {\bibinfo  {journal} {Phys. Rev. B}\ }\textbf {\bibinfo
  {volume} {98}},\ \bibinfo {pages} {045111} (\bibinfo {year}
  {2018})}\BibitemShut {NoStop}%
\bibitem [{\citenamefont {Lambert}\ \emph {et~al.}(2009)\citenamefont
  {Lambert}, \citenamefont {Chen}, \citenamefont {Johansson},\ and\
  \citenamefont {Nori}}]{Lambert2009Quantum}%
  \BibitemOpen
  \bibfield  {author} {\bibinfo {author} {\bibfnamefont {N.}~\bibnamefont
  {Lambert}}, \bibinfo {author} {\bibfnamefont {Y.-n.}\ \bibnamefont {Chen}},
  \bibinfo {author} {\bibfnamefont {R.}~\bibnamefont {Johansson}},\ and\
  \bibinfo {author} {\bibfnamefont {F.}~\bibnamefont {Nori}},\ }\bibfield
  {title} {\bibinfo {title} {Quantum chaos and critical behavior on a chip},\
  }\href {https://doi.org/10.1103/PhysRevB.80.165308} {\bibfield  {journal}
  {\bibinfo  {journal} {Phys. Rev. B}\ }\textbf {\bibinfo {volume} {80}},\
  \bibinfo {pages} {165308} (\bibinfo {year} {2009})}\BibitemShut {NoStop}%
\bibitem [{\citenamefont {Altland}\ and\ \citenamefont
  {Haake}(2012)}]{Altland2012Quantum}%
  \BibitemOpen
  \bibfield  {author} {\bibinfo {author} {\bibfnamefont {A.}~\bibnamefont
  {Altland}}\ and\ \bibinfo {author} {\bibfnamefont {F.}~\bibnamefont
  {Haake}},\ }\bibfield  {title} {\bibinfo {title} {Quantum chaos and effective
  thermalization},\ }\href {https://doi.org/10.1103/PhysRevLett.108.073601}
  {\bibfield  {journal} {\bibinfo  {journal} {Phys. Rev. Lett.}\ }\textbf
  {\bibinfo {volume} {108}},\ \bibinfo {pages} {073601} (\bibinfo {year}
  {2012})}\BibitemShut {NoStop}%
\bibitem [{\citenamefont {Bakemeier}\ \emph {et~al.}(2013)\citenamefont
  {Bakemeier}, \citenamefont {Alvermann},\ and\ \citenamefont
  {Fehske}}]{Bakemeier2013Dynamics}%
  \BibitemOpen
  \bibfield  {author} {\bibinfo {author} {\bibfnamefont {L.}~\bibnamefont
  {Bakemeier}}, \bibinfo {author} {\bibfnamefont {A.}~\bibnamefont
  {Alvermann}},\ and\ \bibinfo {author} {\bibfnamefont {H.}~\bibnamefont
  {Fehske}},\ }\bibfield  {title} {\bibinfo {title} {Dynamics of the dicke
  model close to the classical limit},\ }\href
  {https://doi.org/10.1103/PhysRevA.88.043835} {\bibfield  {journal} {\bibinfo
  {journal} {Phys. Rev. A}\ }\textbf {\bibinfo {volume} {88}},\ \bibinfo
  {pages} {043835} (\bibinfo {year} {2013})}\BibitemShut {NoStop}%
\bibitem [{\citenamefont {Bentsen}\ \emph {et~al.}(2019)\citenamefont
  {Bentsen}, \citenamefont {Potirniche}, \citenamefont {Bulchandani},
  \citenamefont {Scaffidi}, \citenamefont {Cao}, \citenamefont {Qi},
  \citenamefont {Schleier-Smith},\ and\ \citenamefont
  {Altman}}]{Bentsen2019Integrable}%
  \BibitemOpen
  \bibfield  {author} {\bibinfo {author} {\bibfnamefont {G.}~\bibnamefont
  {Bentsen}}, \bibinfo {author} {\bibfnamefont {I.-D.}\ \bibnamefont
  {Potirniche}}, \bibinfo {author} {\bibfnamefont {V.~B.}\ \bibnamefont
  {Bulchandani}}, \bibinfo {author} {\bibfnamefont {T.}~\bibnamefont
  {Scaffidi}}, \bibinfo {author} {\bibfnamefont {X.}~\bibnamefont {Cao}},
  \bibinfo {author} {\bibfnamefont {X.-L.}\ \bibnamefont {Qi}}, \bibinfo
  {author} {\bibfnamefont {M.}~\bibnamefont {Schleier-Smith}},\ and\ \bibinfo
  {author} {\bibfnamefont {E.}~\bibnamefont {Altman}},\ }\bibfield  {title}
  {\bibinfo {title} {Integrable and chaotic dynamics of spins coupled to an
  optical cavity},\ }\href {https://doi.org/10.1103/PhysRevX.9.041011}
  {\bibfield  {journal} {\bibinfo  {journal} {Phys. Rev. X}\ }\textbf {\bibinfo
  {volume} {9}},\ \bibinfo {pages} {041011} (\bibinfo {year}
  {2019})}\BibitemShut {NoStop}%
\bibitem [{\citenamefont {Periwal}\ \emph {et~al.}(2021)\citenamefont
  {Periwal}, \citenamefont {Cooper}, \citenamefont {Kunkel}, \citenamefont
  {Wienand}, \citenamefont {Davis},\ and\ \citenamefont
  {Schleier-Smith}}]{Periwal2021Programmable}%
  \BibitemOpen
  \bibfield  {author} {\bibinfo {author} {\bibfnamefont {A.}~\bibnamefont
  {Periwal}}, \bibinfo {author} {\bibfnamefont {E.~S.}\ \bibnamefont {Cooper}},
  \bibinfo {author} {\bibfnamefont {P.}~\bibnamefont {Kunkel}}, \bibinfo
  {author} {\bibfnamefont {J.~F.}\ \bibnamefont {Wienand}}, \bibinfo {author}
  {\bibfnamefont {E.~J.}\ \bibnamefont {Davis}},\ and\ \bibinfo {author}
  {\bibfnamefont {M.}~\bibnamefont {Schleier-Smith}},\ }\bibfield  {title}
  {\bibinfo {title} {Programmable interactions and emergent geometry in an
  array of atom clouds},\ }\href {https://doi.org/10.1038/s41586-021-04156-0}
  {\bibfield  {journal} {\bibinfo  {journal} {Nature}\ }\textbf {\bibinfo
  {volume} {600}},\ \bibinfo {pages} {630} (\bibinfo {year}
  {2021})}\BibitemShut {NoStop}%
\bibitem [{\citenamefont {Sawicki}\ and\ \citenamefont
  {Kuś}(2010)}]{Sawicki2010Classical}%
  \BibitemOpen
  \bibfield  {author} {\bibinfo {author} {\bibfnamefont {A.}~\bibnamefont
  {Sawicki}}\ and\ \bibinfo {author} {\bibfnamefont {M.}~\bibnamefont {Kuś}},\
  }\bibfield  {title} {\bibinfo {title} {{Classical nonintegrability of a
  quantum chaotic Hamiltonian system}},\ }\href
  {https://doi.org/10.1016/j.physd.2010.02.005} {\bibfield  {journal} {\bibinfo
   {journal} {Physica D: Nonlinear Phenomena}\ }\textbf {\bibinfo {volume}
  {239}},\ \bibinfo {pages} {719–726} (\bibinfo {year} {2010})}\BibitemShut
  {NoStop}%
\bibitem [{\citenamefont {Fradkin}(1965)}]{Fradkin1965Three}%
  \BibitemOpen
  \bibfield  {author} {\bibinfo {author} {\bibfnamefont {D.~M.}\ \bibnamefont
  {Fradkin}},\ }\bibfield  {title} {\bibinfo {title} {{Three-Dimensional
  Isotropic Harmonic Oscillator and SU3}},\ }\href
  {https://doi.org/10.1119/1.1971373} {\bibfield  {journal} {\bibinfo
  {journal} {American Journal of Physics}\ }\textbf {\bibinfo {volume} {33}},\
  \bibinfo {pages} {207} (\bibinfo {year} {1965})}\BibitemShut {NoStop}%
\bibitem [{\citenamefont {Mukherjee}\ \emph {et~al.}(2021)\citenamefont
  {Mukherjee}, \citenamefont {Banerjee}, \citenamefont {Sengupta},\ and\
  \citenamefont {Sen}}]{Mukherjee2021Minimal}%
  \BibitemOpen
  \bibfield  {author} {\bibinfo {author} {\bibfnamefont {B.}~\bibnamefont
  {Mukherjee}}, \bibinfo {author} {\bibfnamefont {D.}~\bibnamefont {Banerjee}},
  \bibinfo {author} {\bibfnamefont {K.}~\bibnamefont {Sengupta}},\ and\
  \bibinfo {author} {\bibfnamefont {A.}~\bibnamefont {Sen}},\ }\bibfield
  {title} {\bibinfo {title} {Minimal model for hilbert space fragmentation with
  local constraints},\ }\href {https://doi.org/10.1103/PhysRevB.104.155117}
  {\bibfield  {journal} {\bibinfo  {journal} {Phys. Rev. B}\ }\textbf {\bibinfo
  {volume} {104}},\ \bibinfo {pages} {155117} (\bibinfo {year}
  {2021})}\BibitemShut {NoStop}%
\bibitem [{\citenamefont {Brighi}\ \emph {et~al.}(2023)\citenamefont {Brighi},
  \citenamefont {Ljubotina},\ and\ \citenamefont {Serbyn}}]{Brighi2023Hilbert}%
  \BibitemOpen
  \bibfield  {author} {\bibinfo {author} {\bibfnamefont {P.}~\bibnamefont
  {Brighi}}, \bibinfo {author} {\bibfnamefont {M.}~\bibnamefont {Ljubotina}},\
  and\ \bibinfo {author} {\bibfnamefont {M.}~\bibnamefont {Serbyn}},\
  }\bibfield  {title} {\bibinfo {title} {{Hilbert space fragmentation and slow
  dynamics in particle-conserving quantum East models}},\ }\href
  {https://doi.org/10.21468/SciPostPhys.15.3.093} {\bibfield  {journal}
  {\bibinfo  {journal} {SciPost Phys.}\ }\textbf {\bibinfo {volume} {15}},\
  \bibinfo {pages} {093} (\bibinfo {year} {2023})}\BibitemShut {NoStop}%
\bibitem [{\citenamefont {Bohigas}\ \emph {et~al.}(1984)\citenamefont
  {Bohigas}, \citenamefont {Giannoni},\ and\ \citenamefont
  {Schmit}}]{Bohigas1984Characterization}%
  \BibitemOpen
  \bibfield  {author} {\bibinfo {author} {\bibfnamefont {O.}~\bibnamefont
  {Bohigas}}, \bibinfo {author} {\bibfnamefont {M.~J.}\ \bibnamefont
  {Giannoni}},\ and\ \bibinfo {author} {\bibfnamefont {C.}~\bibnamefont
  {Schmit}},\ }\bibfield  {title} {\bibinfo {title} {{Characterization of
  Chaotic Quantum Spectra and Universality of Level Fluctuation Laws}},\ }\href
  {https://doi.org/10.1103/PhysRevLett.52.1} {\bibfield  {journal} {\bibinfo
  {journal} {Phys. Rev. Lett.}\ }\textbf {\bibinfo {volume} {52}},\ \bibinfo
  {pages} {1} (\bibinfo {year} {1984})}\BibitemShut {NoStop}%
\bibitem [{\citenamefont {Berry}\ and\ \citenamefont
  {Tabor}(1977)}]{Berry1977Level}%
  \BibitemOpen
  \bibfield  {author} {\bibinfo {author} {\bibfnamefont {M.~V.}\ \bibnamefont
  {Berry}}\ and\ \bibinfo {author} {\bibfnamefont {M.}~\bibnamefont {Tabor}},\
  }\bibfield  {title} {\bibinfo {title} {{Level Clustering in the Regular
  Spectrum}},\ }\href@noop {} {\bibfield  {journal} {\bibinfo  {journal} {Proc.
  R. Soc. London A}\ }\textbf {\bibinfo {volume} {356}},\ \bibinfo {pages}
  {375} (\bibinfo {year} {1977})}\BibitemShut {NoStop}%
\bibitem [{\citenamefont {Dahlbom}\ \emph {et~al.}(2022)\citenamefont
  {Dahlbom}, \citenamefont {Miles}, \citenamefont {Zhang}, \citenamefont
  {Batista},\ and\ \citenamefont {Barros}}]{Dahlbom2022Langevin}%
  \BibitemOpen
  \bibfield  {author} {\bibinfo {author} {\bibfnamefont {D.}~\bibnamefont
  {Dahlbom}}, \bibinfo {author} {\bibfnamefont {C.}~\bibnamefont {Miles}},
  \bibinfo {author} {\bibfnamefont {H.}~\bibnamefont {Zhang}}, \bibinfo
  {author} {\bibfnamefont {C.~D.}\ \bibnamefont {Batista}},\ and\ \bibinfo
  {author} {\bibfnamefont {K.}~\bibnamefont {Barros}},\ }\bibfield  {title}
  {\bibinfo {title} {{Langevin dynamics of generalized spins as SU($N$)
  coherent states}},\ }\href {https://doi.org/10.1103/PhysRevB.106.235154}
  {\bibfield  {journal} {\bibinfo  {journal} {Phys. Rev. B}\ }\textbf {\bibinfo
  {volume} {106}},\ \bibinfo {pages} {235154} (\bibinfo {year}
  {2022})}\BibitemShut {NoStop}%
\bibitem [{\citenamefont {Zhang}\ and\ \citenamefont
  {Batista}(2021)}]{Zhang2021Classical}%
  \BibitemOpen
  \bibfield  {author} {\bibinfo {author} {\bibfnamefont {H.}~\bibnamefont
  {Zhang}}\ and\ \bibinfo {author} {\bibfnamefont {C.~D.}\ \bibnamefont
  {Batista}},\ }\bibfield  {title} {\bibinfo {title} {{Classical spin dynamics
  based on $\mathrm{SU}(N)$ coherent states}},\ }\href
  {https://doi.org/10.1103/PhysRevB.104.104409} {\bibfield  {journal} {\bibinfo
   {journal} {Phys. Rev. B}\ }\textbf {\bibinfo {volume} {104}},\ \bibinfo
  {pages} {104409} (\bibinfo {year} {2021})}\BibitemShut {NoStop}%
\bibitem [{\citenamefont {Gross}\ \emph {et~al.}(2010)\citenamefont {Gross},
  \citenamefont {Zibold}, \citenamefont {Nicklas}, \citenamefont {Estève},\
  and\ \citenamefont {Oberthaler}}]{Gross2010Nonlinear}%
  \BibitemOpen
  \bibfield  {author} {\bibinfo {author} {\bibfnamefont {C.}~\bibnamefont
  {Gross}}, \bibinfo {author} {\bibfnamefont {T.}~\bibnamefont {Zibold}},
  \bibinfo {author} {\bibfnamefont {E.}~\bibnamefont {Nicklas}}, \bibinfo
  {author} {\bibfnamefont {J.}~\bibnamefont {Estève}},\ and\ \bibinfo {author}
  {\bibfnamefont {M.~K.}\ \bibnamefont {Oberthaler}},\ }\bibfield  {title}
  {\bibinfo {title} {Nonlinear atom interferometer surpasses classical
  precision limit},\ }\href {https://doi.org/10.1038/nature08919} {\bibfield
  {journal} {\bibinfo  {journal} {Nature}\ }\textbf {\bibinfo {volume} {464}},\
  \bibinfo {pages} {1165–1169} (\bibinfo {year} {2010})}\BibitemShut
  {NoStop}%
\bibitem [{\citenamefont {Fernholz}\ \emph {et~al.}(2008)\citenamefont
  {Fernholz}, \citenamefont {Krauter}, \citenamefont {Jensen}, \citenamefont
  {Sherson}, \citenamefont {S\o{}rensen},\ and\ \citenamefont
  {Polzik}}]{Fernholz2008Spin}%
  \BibitemOpen
  \bibfield  {author} {\bibinfo {author} {\bibfnamefont {T.}~\bibnamefont
  {Fernholz}}, \bibinfo {author} {\bibfnamefont {H.}~\bibnamefont {Krauter}},
  \bibinfo {author} {\bibfnamefont {K.}~\bibnamefont {Jensen}}, \bibinfo
  {author} {\bibfnamefont {J.~F.}\ \bibnamefont {Sherson}}, \bibinfo {author}
  {\bibfnamefont {A.~S.}\ \bibnamefont {S\o{}rensen}},\ and\ \bibinfo {author}
  {\bibfnamefont {E.~S.}\ \bibnamefont {Polzik}},\ }\bibfield  {title}
  {\bibinfo {title} {Spin squeezing of atomic ensembles via nuclear-electronic
  spin entanglement},\ }\href {https://doi.org/10.1103/PhysRevLett.101.073601}
  {\bibfield  {journal} {\bibinfo  {journal} {Phys. Rev. Lett.}\ }\textbf
  {\bibinfo {volume} {101}},\ \bibinfo {pages} {073601} (\bibinfo {year}
  {2008})}\BibitemShut {NoStop}%
\bibitem [{\citenamefont {Hamley}\ \emph {et~al.}(2012)\citenamefont {Hamley},
  \citenamefont {Gerving}, \citenamefont {Hoang}, \citenamefont {Bookjans},\
  and\ \citenamefont {Chapman}}]{Hamley2012Spin}%
  \BibitemOpen
  \bibfield  {author} {\bibinfo {author} {\bibfnamefont {C.~D.}\ \bibnamefont
  {Hamley}}, \bibinfo {author} {\bibfnamefont {C.~S.}\ \bibnamefont {Gerving}},
  \bibinfo {author} {\bibfnamefont {T.~M.}\ \bibnamefont {Hoang}}, \bibinfo
  {author} {\bibfnamefont {E.~M.}\ \bibnamefont {Bookjans}},\ and\ \bibinfo
  {author} {\bibfnamefont {M.~S.}\ \bibnamefont {Chapman}},\ }\bibfield
  {title} {\bibinfo {title} {Spin-nematic squeezed vacuum in a quantum gas},\
  }\href {https://doi.org/10.1038/nphys2245} {\bibfield  {journal} {\bibinfo
  {journal} {Nature Physics}\ }\textbf {\bibinfo {volume} {8}},\ \bibinfo
  {pages} {305–308} (\bibinfo {year} {2012})}\BibitemShut {NoStop}%
\bibitem [{\citenamefont {Corre}\ \emph {et~al.}(2015)\citenamefont {Corre},
  \citenamefont {Zibold}, \citenamefont {Frapolli}, \citenamefont {Shao},
  \citenamefont {Dalibard},\ and\ \citenamefont {Gerbier}}]{Corre2015Spin}%
  \BibitemOpen
  \bibfield  {author} {\bibinfo {author} {\bibfnamefont {V.}~\bibnamefont
  {Corre}}, \bibinfo {author} {\bibfnamefont {T.}~\bibnamefont {Zibold}},
  \bibinfo {author} {\bibfnamefont {C.}~\bibnamefont {Frapolli}}, \bibinfo
  {author} {\bibfnamefont {L.}~\bibnamefont {Shao}}, \bibinfo {author}
  {\bibfnamefont {J.}~\bibnamefont {Dalibard}},\ and\ \bibinfo {author}
  {\bibfnamefont {F.}~\bibnamefont {Gerbier}},\ }\bibfield  {title} {\bibinfo
  {title} {Spin-1 condensates at thermal equilibrium: A su(3) coherent state
  approach},\ }\href {https://doi.org/10.1209/0295-5075/110/26001} {\bibfield
  {journal} {\bibinfo  {journal} {Europhys. Lett.}\ }\textbf {\bibinfo {volume}
  {110}},\ \bibinfo {pages} {26001} (\bibinfo {year} {2015})}\BibitemShut
  {NoStop}%
\bibitem [{\citenamefont {Gnutzmann}\ and\ \citenamefont
  {Kus}(1998)}]{Gnutzmann1998Coherent}%
  \BibitemOpen
  \bibfield  {author} {\bibinfo {author} {\bibfnamefont {S.}~\bibnamefont
  {Gnutzmann}}\ and\ \bibinfo {author} {\bibfnamefont {M.}~\bibnamefont
  {Kus}},\ }\bibfield  {title} {\bibinfo {title} {{Coherent states and the
  classical limit on irreducible representations}},\ }\href
  {https://doi.org/10.1088/0305-4470/31/49/011} {\bibfield  {journal} {\bibinfo
   {journal} {Journal of Physics A: Mathematical and General}\ }\textbf
  {\bibinfo {volume} {31}},\ \bibinfo {pages} {9871} (\bibinfo {year}
  {1998})}\BibitemShut {NoStop}%
\bibitem [{\citenamefont {Classen-Howes}\ \emph {et~al.}(2024)\citenamefont
  {Classen-Howes}, \citenamefont {Fendley}, \citenamefont {Pandey},\ and\
  \citenamefont {Parameswaran}}]{ClassenHowes2024Bipartite}%
  \BibitemOpen
  \bibfield  {author} {\bibinfo {author} {\bibfnamefont {J.}~\bibnamefont
  {Classen-Howes}}, \bibinfo {author} {\bibfnamefont {P.}~\bibnamefont
  {Fendley}}, \bibinfo {author} {\bibfnamefont {A.}~\bibnamefont {Pandey}},\
  and\ \bibinfo {author} {\bibfnamefont {S.~A.}\ \bibnamefont {Parameswaran}},\
  }\bibfield  {title} {\bibinfo {title} {Bipartite sachdev-ye models with
  read-saleur symmetries},\ }\href
  {https://doi.org/10.1103/PhysRevB.110.125140} {\bibfield  {journal} {\bibinfo
   {journal} {Phys. Rev. B}\ }\textbf {\bibinfo {volume} {110}},\ \bibinfo
  {pages} {125140} (\bibinfo {year} {2024})}\BibitemShut {NoStop}%
\bibitem [{\citenamefont {Weisstein}()}]{WeissteinMagicSquare}%
  \BibitemOpen
  \bibfield  {author} {\bibinfo {author} {\bibfnamefont {E.~W.}\ \bibnamefont
  {Weisstein}},\ }\href@noop {} {\bibinfo {title} {{Magic Square. From
  MathWorld--A Wolfram Web Resource}}}\BibitemShut {NoStop}%
\bibitem [{\citenamefont {Conforti}\ \emph {et~al.}(2014)\citenamefont
  {Conforti}, \citenamefont {Cornu{\'e}jols},\ and\ \citenamefont
  {Zambelli}}]{Conforti2014Integer}%
  \BibitemOpen
  \bibfield  {author} {\bibinfo {author} {\bibfnamefont {M.}~\bibnamefont
  {Conforti}}, \bibinfo {author} {\bibfnamefont {G.}~\bibnamefont
  {Cornu{\'e}jols}},\ and\ \bibinfo {author} {\bibfnamefont {G.}~\bibnamefont
  {Zambelli}},\ }\href {https://doi.org/10.1007/978-3-319-11008-0} {\emph
  {\bibinfo {title} {{Integer Programming}}}}\ (\bibinfo  {publisher}
  {Springer},\ \bibinfo {address} {Cham},\ \bibinfo {year} {2014})\BibitemShut
  {NoStop}%
\bibitem [{\citenamefont {Karp}(1972)}]{Karp1972Reducibility}%
  \BibitemOpen
  \bibfield  {author} {\bibinfo {author} {\bibfnamefont {R.~M.}\ \bibnamefont
  {Karp}},\ }\bibinfo {title} {{Reducibility among Combinatorial Problems}},\
  in\ \href {https://doi.org/10.1007/978-1-4684-2001-2_9} {\emph {\bibinfo
  {booktitle} {Complexity of Computer Computations}}},\ \bibinfo {series and
  number} {The IBM Research Symposia Series},\ \bibinfo {editor} {edited by\
  \bibinfo {editor} {\bibfnamefont {R.~E.}\ \bibnamefont {Miller}}, \bibinfo
  {editor} {\bibfnamefont {J.~W.}\ \bibnamefont {Thatcher}},\ and\ \bibinfo
  {editor} {\bibfnamefont {J.~D.}\ \bibnamefont {Bohlinger}}}\ (\bibinfo
  {publisher} {Springer},\ \bibinfo {address} {Boston, MA},\ \bibinfo {year}
  {1972})\ pp.\ \bibinfo {pages} {85--103}\BibitemShut {NoStop}%
\bibitem [{\citenamefont {Perron}\ and\ \citenamefont {Furnon}()}]{ortools}%
  \BibitemOpen
  \bibfield  {author} {\bibinfo {author} {\bibfnamefont {L.}~\bibnamefont
  {Perron}}\ and\ \bibinfo {author} {\bibfnamefont {V.}~\bibnamefont
  {Furnon}},\ }\href@noop {} {\bibinfo {title} {{OR-Tools}}}\BibitemShut
  {NoStop}%
\bibitem [{\citenamefont {Fulton}(1996)}]{Fulton1996Young}%
  \BibitemOpen
  \bibfield  {author} {\bibinfo {author} {\bibfnamefont {W.}~\bibnamefont
  {Fulton}},\ }\href@noop {} {\emph {\bibinfo {title} {{Young Tableaux: With
  Applications to Representation Theory and Geometry}}}}\ (\bibinfo
  {publisher} {Cambridge University Press},\ \bibinfo {year}
  {1996})\BibitemShut {NoStop}%
\bibitem [{\citenamefont {Sagan}(2001)}]{Sagan2001Symmetric}%
  \BibitemOpen
  \bibfield  {author} {\bibinfo {author} {\bibfnamefont {B.~E.}\ \bibnamefont
  {Sagan}},\ }\href {https://doi.org/10.1007/978-1-4757-6804-6} {\emph
  {\bibinfo {title} {{The Symmetric Group: Representations, Combinatorial
  Algorithms, and Symmetric Functions}}}}\ (\bibinfo  {publisher} {Springer},\
  \bibinfo {year} {2001})\BibitemShut {NoStop}%
\bibitem [{\citenamefont {Itzykson}\ and\ \citenamefont
  {Nauenberg}(1966)}]{Itzykson1966Unitary}%
  \BibitemOpen
  \bibfield  {author} {\bibinfo {author} {\bibfnamefont {C.}~\bibnamefont
  {Itzykson}}\ and\ \bibinfo {author} {\bibfnamefont {M.}~\bibnamefont
  {Nauenberg}},\ }\bibfield  {title} {\bibinfo {title} {{Unitary Groups:
  Representations and Decompositions}},\ }\href
  {https://doi.org/10.1103/RevModPhys.38.95} {\bibfield  {journal} {\bibinfo
  {journal} {Rev. Mod. Phys.}\ }\textbf {\bibinfo {volume} {38}},\ \bibinfo
  {pages} {95} (\bibinfo {year} {1966})}\BibitemShut {NoStop}%
\bibitem [{\citenamefont {Georgi}(2000)}]{Georgi2000Lie}%
  \BibitemOpen
  \bibfield  {author} {\bibinfo {author} {\bibfnamefont {H.}~\bibnamefont
  {Georgi}},\ }\href {https://doi.org/10.1201/9780429499210} {\emph {\bibinfo
  {title} {{Lie algebras in particle physics: from isospin to unified
  theories}}}}\ (\bibinfo  {publisher} {Taylor \& Francis},\ \bibinfo {year}
  {2000})\BibitemShut {NoStop}%
\bibitem [{\citenamefont {Bacon}\ \emph {et~al.}(2005)\citenamefont {Bacon},
  \citenamefont {Chuang},\ and\ \citenamefont {Harrow}}]{Bacon2005Quantum}%
  \BibitemOpen
  \bibfield  {author} {\bibinfo {author} {\bibfnamefont {D.}~\bibnamefont
  {Bacon}}, \bibinfo {author} {\bibfnamefont {I.~L.}\ \bibnamefont {Chuang}},\
  and\ \bibinfo {author} {\bibfnamefont {A.~W.}\ \bibnamefont {Harrow}},\
  }\href@noop {} {\bibinfo {title} {{The Quantum Schur Transform: I. Efficient
  Qudit Circuits}}} (\bibinfo {year} {2005}),\ \Eprint
  {https://arxiv.org/abs/quant-ph/0601001} {arXiv:quant-ph/0601001 [quant-ph]}
  \BibitemShut {NoStop}%
\bibitem [{\citenamefont {Bacon}\ \emph {et~al.}(2006)\citenamefont {Bacon},
  \citenamefont {Chuang},\ and\ \citenamefont {Harrow}}]{Bacon2006Efficient}%
  \BibitemOpen
  \bibfield  {author} {\bibinfo {author} {\bibfnamefont {D.}~\bibnamefont
  {Bacon}}, \bibinfo {author} {\bibfnamefont {I.~L.}\ \bibnamefont {Chuang}},\
  and\ \bibinfo {author} {\bibfnamefont {A.~W.}\ \bibnamefont {Harrow}},\
  }\bibfield  {title} {\bibinfo {title} {{Efficient Quantum Circuits for Schur
  and Clebsch-Gordan Transforms}},\ }\href
  {https://doi.org/10.1103/PhysRevLett.97.170502} {\bibfield  {journal}
  {\bibinfo  {journal} {Phys. Rev. Lett.}\ }\textbf {\bibinfo {volume} {97}},\
  \bibinfo {pages} {170502} (\bibinfo {year} {2006})}\BibitemShut {NoStop}%
\bibitem [{\citenamefont {Krovi}(2019)}]{Krovi2019Efficient}%
  \BibitemOpen
  \bibfield  {author} {\bibinfo {author} {\bibfnamefont {H.}~\bibnamefont
  {Krovi}},\ }\bibfield  {title} {\bibinfo {title} {{An efficient high
  dimensional quantum {S}chur transform}},\ }\href
  {https://doi.org/10.22331/q-2019-02-14-122} {\bibfield  {journal} {\bibinfo
  {journal} {{Quantum}}\ }\textbf {\bibinfo {volume} {3}},\ \bibinfo {pages}
  {122} (\bibinfo {year} {2019})}\BibitemShut {NoStop}%
\bibitem [{\citenamefont {Anschuetz}\ \emph {et~al.}(2023)\citenamefont
  {Anschuetz}, \citenamefont {Bauer}, \citenamefont {Kiani},\ and\
  \citenamefont {Lloyd}}]{Anschuetz2023Efficient}%
  \BibitemOpen
  \bibfield  {author} {\bibinfo {author} {\bibfnamefont {E.~R.}\ \bibnamefont
  {Anschuetz}}, \bibinfo {author} {\bibfnamefont {A.}~\bibnamefont {Bauer}},
  \bibinfo {author} {\bibfnamefont {B.~T.}\ \bibnamefont {Kiani}},\ and\
  \bibinfo {author} {\bibfnamefont {S.}~\bibnamefont {Lloyd}},\ }\bibfield
  {title} {\bibinfo {title} {{Efficient classical algorithms for simulating
  symmetric quantum systems}},\ }\href
  {https://doi.org/10.22331/q-2023-11-28-1189} {\bibfield  {journal} {\bibinfo
  {journal} {{Quantum}}\ }\textbf {\bibinfo {volume} {7}},\ \bibinfo {pages}
  {1189} (\bibinfo {year} {2023})}\BibitemShut {NoStop}%
\bibitem [{\citenamefont {Agnarsson}(2002)}]{Agnarsson2002Sylvester}%
  \BibitemOpen
  \bibfield  {author} {\bibinfo {author} {\bibfnamefont {G.}~\bibnamefont
  {Agnarsson}},\ }\bibfield  {title} {\bibinfo {title} {{On the Sylvester
  denumerants for general restricted partitions}},\ }\href@noop {} {\bibfield
  {journal} {\bibinfo  {journal} {Congressus numerantium}\ ,\ \bibinfo {pages}
  {49}} (\bibinfo {year} {2002})}\BibitemShut {NoStop}%
\bibitem [{\citenamefont {Echter}\ \emph {et~al.}(2025)\citenamefont {Echter},
  \citenamefont {Maier}, \citenamefont {Urbina}, \citenamefont {Lewenkopf},\
  and\ \citenamefont {Richter}}]{Echter2025Many}%
  \BibitemOpen
  \bibfield  {author} {\bibinfo {author} {\bibfnamefont {C.}~\bibnamefont
  {Echter}}, \bibinfo {author} {\bibfnamefont {G.}~\bibnamefont {Maier}},
  \bibinfo {author} {\bibfnamefont {J.-D.}\ \bibnamefont {Urbina}}, \bibinfo
  {author} {\bibfnamefont {C.}~\bibnamefont {Lewenkopf}},\ and\ \bibinfo
  {author} {\bibfnamefont {K.}~\bibnamefont {Richter}},\ }\bibfield  {title}
  {\bibinfo {title} {{Many-body density of states of bosonic and fermionic
  gases: a combinatorial approach}},\ }\href@noop {} {\bibfield  {journal}
  {\bibinfo  {journal} {J. Phys. A: Math. Theor.}\ } (\bibinfo {year}
  {2025})}\BibitemShut {NoStop}%
\bibitem [{\citenamefont {Cavina}\ \emph {et~al.}(2024)\citenamefont {Cavina},
  \citenamefont {Soret}, \citenamefont {Aslyamov}, \citenamefont
  {Ptaszy\ifmmode~\acute{n}\else \'{n}\fi{}ski},\ and\ \citenamefont
  {Esposito}}]{Cavina2024Symmetry}%
  \BibitemOpen
  \bibfield  {author} {\bibinfo {author} {\bibfnamefont {V.}~\bibnamefont
  {Cavina}}, \bibinfo {author} {\bibfnamefont {A.}~\bibnamefont {Soret}},
  \bibinfo {author} {\bibfnamefont {T.}~\bibnamefont {Aslyamov}}, \bibinfo
  {author} {\bibfnamefont {K.}~\bibnamefont {Ptaszy\ifmmode~\acute{n}\else
  \'{n}\fi{}ski}},\ and\ \bibinfo {author} {\bibfnamefont {M.}~\bibnamefont
  {Esposito}},\ }\bibfield  {title} {\bibinfo {title} {{Symmetry Shapes
  Thermodynamics of Macroscopic Quantum Systems}},\ }\href
  {https://doi.org/10.1103/PhysRevLett.133.130401} {\bibfield  {journal}
  {\bibinfo  {journal} {Phys. Rev. Lett.}\ }\textbf {\bibinfo {volume} {133}},\
  \bibinfo {pages} {130401} (\bibinfo {year} {2024})}\BibitemShut {NoStop}%
\end{thebibliography}%

\end{document}